\documentclass[apj]{emulateapj}
\usepackage{subfigure}
\usepackage{graphics}

\def\spose#1{\hbox to 0pt{#1\hss}}
\def\simlt{\mathrel{\spose{\lower 3pt\hbox{$\mathchar"218$}}
    \raise 2.0pt\hbox{$\mathchar"13C$}}}
\def\simgt{\mathrel{\spose{\lower 3pt\hbox{$\mathchar"218$}}
    \raise 2.0pt\hbox{$\mathchar"13E$}}}

\newcommand{\oiiin}{\mbox{[\ion{O}{3}]}}
\newcommand{\oiii}{\mbox{[\ion{O}{3}]} $\,$}
\newcommand{\oiiiw}{\mbox{[\ion{O}{3}] $\lambda$5007} $\,$}
\newcommand{\oiiiwn}{\mbox{[\ion{O}{3}] $\lambda$5007}}

\newcommand{\ha}{\mbox{H$\alpha$} $\,$}

\slugcomment{\it Accepted for publication in ApJ}
\shortauthors{Comerford et al.}
\shorttitle{Merger-driven Fueling of Active Galactic Nuclei}

\begin{document}

\title{Merger-driven Fueling of Active Galactic Nuclei: Six Dual and Offset Active Galactic Nuclei Discovered with {\it Chandra} and {\it Hubble Space Telescope} Observations}

\author{Julia M. Comerford\altaffilmark{1}, David Pooley\altaffilmark{2}, R. Scott Barrows\altaffilmark{1}, Jenny E. Greene\altaffilmark{3}, Nadia L. Zakamska\altaffilmark{4}, \\ Greg M. Madejski\altaffilmark{5}, and Michael C. Cooper\altaffilmark{6}}

\affil{$^1$Department of Astrophysical and Planetary Sciences, University of Colorado, Boulder, CO 80309, USA}
\affil{$^2$Department of Physics, Sam Houston State University, Huntsville, TX 77341, USA; Eureka Scientific, Inc., 2452 Delmer Street Suite 100, Oakland, CA 94602, USA}
\affil{$^3$Department of Astrophysical Sciences, Princeton University, Princeton, NJ 08544, USA}
\affil{$^4$Department of Physics and Astronomy, Johns Hopkins University, Bloomberg Center, 3400 N. Charles St., Baltimore, MD 21218}
\affil{$^5$Kavli Institute for Particle Astrophysics and Cosmology,  M/S 29, Stanford Linear Accelerator Center, \\ 2575 Sand Hill Rd., Menlo Park, CA 94725}
\affil{$^6$Center for Galaxy Evolution, Department of Physics and Astronomy, University of California, Irvine, \\ 4129 Frederick Reines Hall, Irvine, CA 92697}
  
\begin{abstract}
Dual active galactic nuclei (AGNs) and offset AGNs are kpc-scale separation supermassive black holes pairs created during galaxy mergers, where both or one of the black holes are AGNs, respectively.  These dual and offset AGNs are valuable probes of the link between mergers and AGNs but are challenging to identify.  Here we present {\it Chandra}/ACIS observations of 12 optically-selected dual AGN candidates at $z < 0.34$, where we use the X-rays to identify AGNs.  We also present {\it HST}/WFC3 observations of 10 of these candidates, which reveal any stellar bulges accompanying the AGNs.  We discover a dual AGN system with separation $\Delta x = 2.2$ kpc, where the two stellar bulges have coincident \oiiiw and X-ray sources. This system is an extremely minor merger (460:1) that may include a dwarf galaxy hosting an intermediate mass black hole.  We also find six single AGNs, and five systems that are either dual or offset AGNs with separations $\Delta x < 10$ kpc. Four of the six dual AGNs and dual/offset AGNs are in ongoing major mergers, and these AGNs are 10 times more luminous, on average, than the single AGNs in our sample.  This hints that major mergers may preferentially trigger higher luminosity AGNs.  Further, we find that confirmed dual AGNs have hard X-ray luminosities that are half of those of single AGNs at fixed \oiiiw luminosity, on average.  This could be explained by high densities of gas funneled to galaxy centers during mergers, and emphasizes the need for deeper X-ray observations of dual AGN candidates.
\end{abstract}


\section{Introduction}
\label{intro}

Dual supermassive black holes (SMBHs) are $<10$ kpc separation SMBH pairs that are created by galaxy mergers.  The dual stage lasts for $\sim100$ Myr \citep{BE80.1}, until tidal forces drag the SMBHs into sub-pc separation binary SMBH orbits that are expected to culminate in the coalescence of the two SMBHs.  The galaxy merger itself may funnel gas onto one or both of the dual SMBHs (e.g., \citealt{BA91.1,SP05.2,HO09.1,VA12.1,BL13.1}), causing them to be observable as offset active galactic nuclei (AGNs; \citealt{CO14.1}) or dual AGNs \citep{GE07.1,CO09.1}, respectively.  

Dual AGNs and offset AGNs are uniquely well suited to studies of the link between AGN activity and galaxy mergers, as they are actively accreting SMBHs in the midst of mergers.  For example, AGNs can be triggered by major mergers of galaxies, secular processes including accretion from galactic halos, and cosmological accretion along filaments (e.g., \citealt{NO83.1,HE84.1,SA88.2,MA99.3,SP05.2,KO04.2,HO06.1,LI08.3,DE09.1}), and the relative roles of these triggering mechanisms may depend on AGN properties.  Galaxy merger simulations have predicted that major mergers preferentially trigger the most luminous AGNs (e.g., \citealt{HO09.1}), but the observational evidence for this is still conflicted because of the difficulty in building a clean sample of AGNs in mergers (e.g., \citealt{KO12.2,TR12.1,VI14.1}).  Dual and offset AGNs enable new and robust observational determinations of these sorts of connections between AGNs and galaxy mergers. 

Although searches have turned up many dual AGN candidates (e.g., \citealt{CO09.3,XU09.1,BA12.1,GE12.1,BA13.1,CO13.1,FU14.1,HU14.1,IM14.1,SH14.1,WO14.1}), there are still relatively few confirmed cases of dual AGNs.  The observational challenge lies in confirming dual AGN systems, which requires spatial resolution of the two individual AGNs.  

\begin{deluxetable*}{llllllll}
\tabletypesize{\scriptsize}
\tablewidth{0pt}
\tablecolumns{8}
\tablecaption{Summary of Chandra and HST Observations} 
\tablehead{
\colhead{SDSS Designation} &
\colhead{{\it Chandra}/} & 
\colhead{Counts} &
\colhead{{\it Chandra}/ACIS} & 
\colhead{{\it HST}/WFC3} &
\colhead{{\it HST}/WFC3} &
\colhead{{\it HST}/WFC3} & 
\colhead{{\it HST}/WFC3} \\ 
 & ACIS exp. & (0.5-8 keV) & obs. date (UT) & F160W & F814W & F438W & obs. date (UT) \\
 & time (s) & & & exp. time (s) & exp. time (s) & exp. time (s) & 
}
\startdata
\scriptsize{SDSS J014209.01$-$005050.0} & 19804 & $691.5 \pm 26.3$ & 2012-09-16 & 147 & 909 & 1050 & 2011-12-19 \\    
\scriptsize{SDSS J075223.35+273643.1} & 29650 & $15.4 \pm 4.0$ & 2011-12-21 & n/a & n/a & n/a & n/a \\
\scriptsize{SDSS J084135.09+010156.2} & 19801 & $216.7 \pm 15.0$ & 2012-02-25 & 147 & 900 & 966 & 2012-03-12 \\
\scriptsize{SDSS J085416.76+502632.0} & 20078 & $7.5 \pm 2.8$ & 2012-02-10 & 147 & 1026 & 1050 & 2011-11-01 \\
\scriptsize{SDSS J095207.62+255257.2} & 19807 & $43.6 \pm 6.6$ & 2012-10-02 & 147 & 900 & 990 & 2012-11-21 \\
\scriptsize{SDSS J100654.20+464717.2} & 19783 & $11.5 \pm 3.5$ & 2013-01-14 & 147 & 972 & 1050 & 2012-03-30 \\
\scriptsize{SDSS J112659.54+294442.8} & 19798 & $27.5 \pm 5.3$ & 2012-02-17 & 147 & 900 & 990 & 2012-05-12 \\ 
\scriptsize{SDSS J123915.40+531414.6} & 19804 & $902.4 \pm 30.1$ & 2013-01-13 & 147 & 1026 & 1050 & 2012-09-12 \\      
\scriptsize{SDSS J132231.86+263159.1} & 19807 & $31.3 \pm 5.7$ & 2012-11-13 & 147 & 900 & 990 & 2012-02-13 \\ 
\scriptsize{SDSS J135646.11+102609.1} & 19804 & $72.5 \pm 8.5$ & 2012-03-31 & 147 & 900 & 978 & 2012-05-19 \\    
\scriptsize{SDSS J144804.17+182537.9} & 19807 & $15.6 \pm 4.0$ & 2012-12-05 & 147 & 924 & 1050 & 2012-05-19 \\     
\scriptsize{SDSS J160436.21+500958.1} & 29582 & $60.3 \pm 7.8$ & 2011-12-02 & n/a & n/a & n/a & n/a
\enddata
\tablecomments{Column 2 shows the exposure time for the {\it Chandra}/ACIS observation.  Column 3 shows the total number of counts observed in 0.5-8 keV. Column 4 shows the UT date of the {\it Chandra}/ACIS observation.  Columns 5, 6, and 7 show the exposure times for the {\it HST}/WFC3 F160W, F814W, and F438W observations.  Column 8 shows the UT date of the {\it HST}/WFC3 observations.}
\label{tbl-1}
\end{deluxetable*}

\begin{figure}
\begin{center}
\includegraphics[width=3.5in]{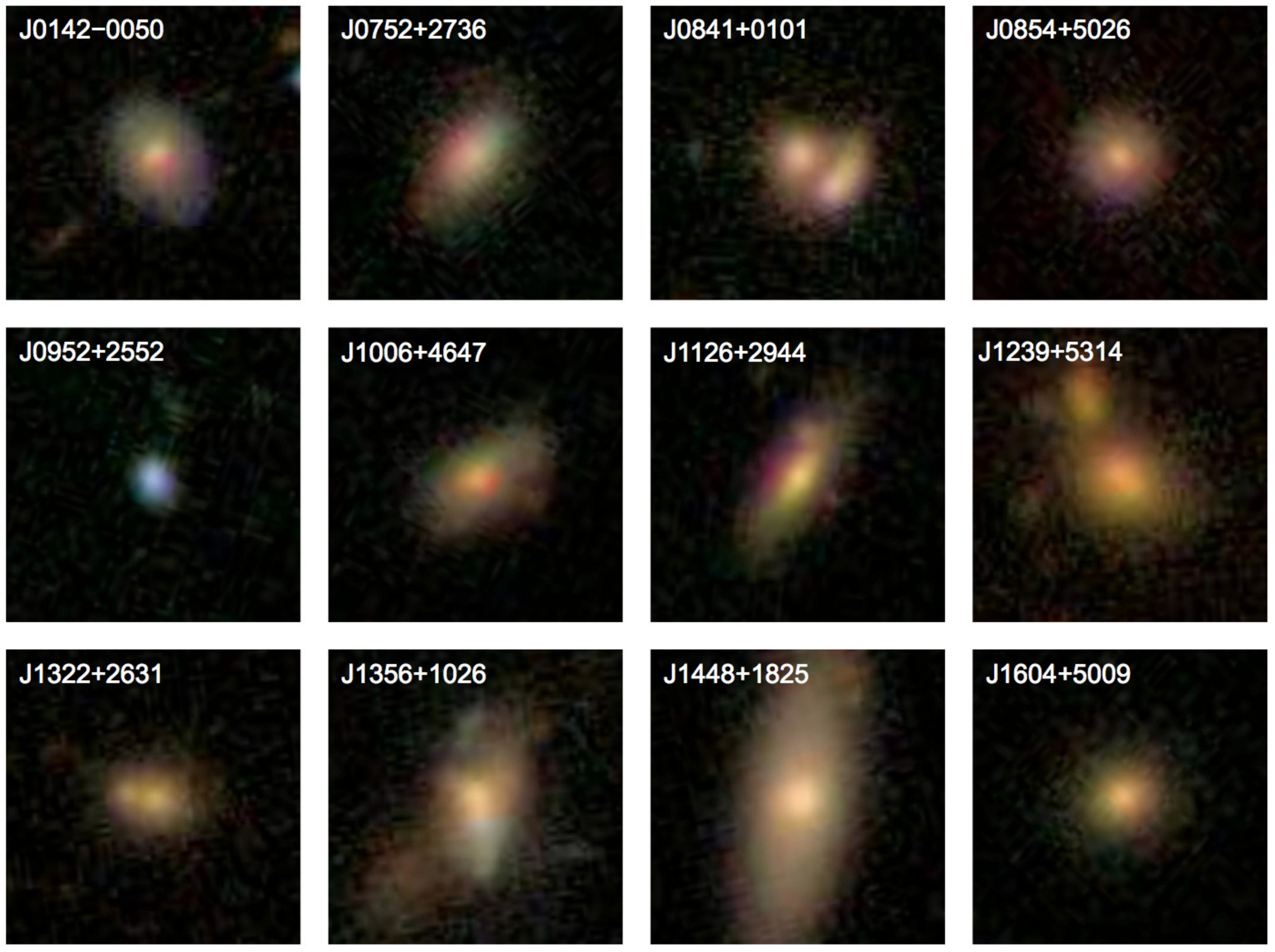}
\end{center}
\vspace{-0.1in}
\caption{$25^{\prime\prime} \times 25^{\prime\prime}$ SDSS $gri$ color composite images of the 12 dual AGN candidates, centered on the coordinates of each SDSS spectrum. In each image, North is up and East is to the left.}
\label{fig:sdss}
\end{figure}

Some wavelength regimes are not, by themselves, well suited to making these confirmations.  The detection of two AGN emission components in spatially resolved optical spectroscopy \citep{MC11.1,GR11.1,SH11.1,CO12.1,FU12.1} is not alone proof of dual AGNs, as kinematics of the narrow-line region (NLR) around a single AGN can also produce multiple emission components.  Optical and near-infrared imaging, especially when assisted by adaptive optics, can detect two stellar components with small separations \citep{MC11.1,RO11.1,GR11.1,SH11.1,FU12.1} but cannot decipher whether both stellar components host AGNs.  Even cases where two optical AGN emission components are observed to be coincident with two stellar bulges can be explained by, e.g., a single AGN that is located between the two stellar bulges and that has extended emission.

Confirmation of dual AGNs requires unambiguous spatial resolution of two individual AGNs, which can be accomplished with radio \citep{RO10.1,FU11.3,TI11.1,DE14.1,GA14.1,WR14.1,WR14.2} or X-ray \citep{CO11.2,KO11.1,KO12.1,TE12.1,LI13.1} observations.  In this paper we focus on X-ray observations of 12 dual AGN candidates with the {\it Chandra X-ray Observatory} Advanced CCD Imaging Spectrometer ({\it Chandra}/ACIS), which is well suited to making dual AGN discoveries since it has the highest spatial resolution of any X-ray telescope.  We also show multiband {\it Hubble Space Telescope} Wide Field Camera 3 ({\it HST}/WFC3) observations of 10 of the dual AGN candidates, to identify stellar bulges and place the dual AGNs in the larger context of their host galaxies.  

The 12 dual AGN candidates are AGNs that exhibit double-peaked narrow emission lines in the Sloan Digital Sky Survey (SDSS), where follow-up optical long-slit spectroscopy show two AGN-fueled \oiiiw emission components with angular separations ($> 0\farcs75$) that are larger than {\it Chandra}'s angular resolution limit. Our purpose is to use the {\it Chandra} and {\it HST} observations of these candidates (Table~\ref{tbl-1}) to identify dual AGNs and determine what types of galaxy mergers produce these systems.

We assume a Hubble constant $H_0 =70$ km s$^{-1}$ Mpc$^{-1}$, $\Omega_m=0.3$, and $\Omega_\Lambda=0.7$ throughout, and all distances are given in physical (not comoving) units.

\begin{deluxetable*}{lllllllllllll}
\tabletypesize{\scriptsize}
\tablewidth{0pt}
\tablecolumns{13}
\tablecaption{Detections of X-ray Sources Corresponding to Optical Emission Sources} 
\tablehead{
\colhead{SDSS Name} &
\colhead{$z$} & 
\colhead{$\Delta x_{\oiiin}$} &
\colhead{$\Delta x_{\oiiin}$} &
\colhead{$\theta_{\oiiin}$} &
\colhead{$X_A+X_B$} &
\colhead{$X_A$} &
\colhead{$X_B$} &
\colhead{$X_A$} &
\colhead{$X_B$} &
\colhead{$X_A$} &
\colhead{$X_B$} &
\colhead{Ref.} \\
 & & ($\prime\prime$) & (kpc) & ($^\circ$) & counts & id. & id. & amplitude & amplitude & sig. & sig. & \\
  & & & & & ($0.3-8$ keV) & & & & & & &
}
\startdata  
J0142$-$0050 & 0.133 & $0.78 \pm 0.07$ & $1.84 \pm 0.16$ & $169.4 \pm 9.2$ & 731.6 & NW & SE & $0.98^{+0.02}_{-0.08}$ & $0.02^{+0.01}_{-0.01}$ & $>$5$\sigma$ &  1.5$\sigma$ & 1 \\    
J0752+2736 & 0.069 & $0.81 \pm 0.07$ & $1.07 \pm 0.09$ & $170.6 \pm 2.4$ & 26.4 & NW & SE & $0.94^{+0.06}_{-0.19}$ & $0.06^{+0.12}_{-0.06}$ &  0.8$\sigma$ &  0.8$\sigma$ & 1 \\
J0841+0101 & 0.111 & $3.60 \pm 0.29$ & $7.28 \pm 0.59$ & $47.3 \pm 3.1$ & 237.2 & NE & SW & $1.00^{+0.00}_{-0.06}$ & $0.00^{+0.01}_{-0.00}$ &   $>$5$\sigma$ & 0$\sigma$ & 2 \\
J0854+5026 & 0.096 & $0.76 \pm 0.02$ & $1.35 \pm 0.04$ & $33.9 \pm 2.2$ & 8.3 & NE & SW &$1.00^{+0.00}_{-0.39}$ & $0.00^{+0.10}_{-0.00}$ & 4.4$\sigma$ &    0$\sigma$ & 1 \\
J0952+2552  & 0.339 & $1.00 \pm 0.05$ & $4.85 \pm 0.24$ & $3.8 \pm 2.5$ & 44.6 & NE & SW & $0.98^{+0.02}_{-0.29}$ & $0.02^{+0.05}_{-0.02}$ & $>$5$\sigma$ &  0.5$\sigma$ & 1 \\
J1006+4647 & 0.123 & $0.83 \pm 0.04$ & $1.82 \pm 0.09$ & $91.2 \pm 1.2$ & 10.2 & W & E & $1.00^{+0.00}_{-1.00}$ & $0.00^{+0.56}_{-0.00}$ &  1.7$\sigma$ &   0$\sigma$ & 1 \\
J1126+2944 & 0.102 & $0.94 \pm 0.10$ & $1.76 \pm 0.19$ & $133.2 \pm 2.3$ & 26.6 & NW & SE & $0.89^{+0.11}_{-0.27}$ & $0.11^{+0.11}_{-0.06}$ & $>$5$\sigma$ &  2.3$\sigma$ & 1 \\ 
J1239+5314 & 0.201 & $0.39 \pm 0.05^a$ & $1.29 \pm 0.18$ & $40.6 \pm 3.5$ & 908.2 & NE & SW & $0.98^{+0.02}_{-0.08}$ & $0.02^{+0.02}_{-0.02}$ & $>$5$\sigma$ & 1.0$\sigma$ &  1 \\    
J1322+2631 & 0.144 & $2.35 \pm 0.02^b$ & $5.94 \pm 0.05$ & $79.7 \pm 1.1$ &  29.8 & SW & NE & $1.00^{+0.00}_{-0.17}$ & $0.00^{+0.02}_{-0.00}$ & $>$5$\sigma$ & 0$\sigma$ & 3 \\
J1356+1026 & 0.123 & $1.33 \pm 0.04^c$ & \bf $2.94 \pm 0.09$ & $5.1 \pm 1.6$ & 78.6 & NE & SW & $0.81^{+0.12}_{-0.10}$ &  $0.19^{+0.07}_{-0.06}$ & $>$5$\sigma$ & 4.4$\sigma$ &  3 \\   
J1448+1825 & 0.038 & $0.84 \pm 0.08$ & $0.63 \pm 0.06$ & $174.6 \pm 2.1$ & 15.8 & SE & NW & $0.64^{+0.36}_{-0.30}$ &  $0.36^{+0.38}_{-0.20}$ & 4.8$\sigma$ & 2.9$\sigma$ &  1 \\     
J1604+5009 & 0.146 & $1.05 \pm 0.08$ & $2.69 \pm 0.20$ & $98.2 \pm 5.4$ & 63.6 & NW & SE & $1.00^{+0.00}_{-0.23}$ & $0.00^{+0.01}_{-0.00}$ & $>$5$\sigma$ &   0$\sigma$ & 1      
\enddata
\tablecomments{Column 3 shows the angular projected spatial separation between the two \oiiiw emission components, as measured from long-slit spectroscopy.  Column 4 shows the corresponding physical projected spatial separation. Column 5 shows the position angle between the two \oiiiw emission features on the sky, as measured from long-slit spectroscopy in degrees East of North. Column 6 shows the total number of $0.3-8$ keV counts from the best-fit models to the stronger X-ray source (source $X_A$) and weaker X-ray source (source $X_B$) combined.  Columns 7 and 8 show the identifiers for $X_A$ and $X_B$, based on their relative cardinal directions from one another.  Column 9 shows the amplitude of the fit to source $X_A$.  Column 10 shows the amplitude of a fit to source $X_B$, required to be at a position relative to the primary X-ray source such that the two X-ray sources have the same separation and position angle as the two \oiiiw emission components.  Column 11 shows the significance of the detection of source $X_A$, and column 12 shows the significance of the detection of source $X_B$. Column 13 shows references for the long-slit spectroscopy: (1) \cite{CO12.1}, (2) \cite{GR11.1}, (3) this paper.}
\tablenotetext{a}{We also detect another \oiiiw emission component at a separation of $1\farcs20$; see Section~\ref{1239}.}
\tablenotetext{b}{Because the \oiiiw emission was only observed at one position angle (separation $2\farcs1$ at position angle $79^\circ$ East of North; \citealt{SH11.1}), it provides only a lower limit on the true separation of the emission components.  To determine the full separation of emission components on the sky, we measured the spatial separation between \ha components in the {\it HST}/F814W observations (Section~\ref{hst}).}
\tablenotetext{c}{Because this galaxy also contains a large-scale outflow, the \oiiiw emission is very spatially extended and it is difficult to define a spatial separation $\Delta x$ between \oiiiw emission components.  Since much of the \oiiiw emission coincides with two stellar bulges, we instead report the $\Delta x$ between the two stellar bulges in the {\it HST} F160W image (Section~\ref{hst}).}
\label{tbl-2}
\end{deluxetable*}

\section{The Sample}

The galaxies presented here were chosen from the parent sample of 340 unique AGNs with double-peaked narrow emission lines identified in SDSS at $0.01 < z < 0.7$ \citep{WA09.1,SM10.1,LI10.1}.  Such double-peaked profiles can be produced by the relative motions of dual AGNs, but also by a single AGN with outflows, jets, or a rotating gaseous disk (e.g., \citealt{HE81.1,CR00.1,VE01.1,ER03.2,WH04.1,DA05.1,CR10.1,FI11.1}).  Follow-up observations are required to distinguish between these scenarios.

For our sample, we selected the double-peaked AGNs where the follow-up long-slit spectroscopy showed two \oiiiw emission components separated by $\Delta x_{[\mathrm{O \,III}]} > 0\farcs75$ \citep{CO12.1,GR11.1,SH11.1}.  The separation threshold was set so that {\it Chandra} could spatially resolve X-ray sources corresponding to the two \oiiiw emission components without difficulty.  We then used the \oiiiw fluxes measured in Section~\ref{sdss} and the \oiiiw to hard X-ray flux scaling relations measured by \cite{HE05.1} for single AGNs to estimate the $2-10$ keV X-ray flux of each \oiiiw peak. We required that the estimated $2-10$ keV fluxes of each peak be $F_{X,2-10\mathrm{keV}} > 8 \times 10^{-15}$ erg cm$^{-2}$ s$^{-1}$, so that each X-ray source could be detected with $< 30$ ks exposures with {\it Chandra}.   As a result, each target in our sample has an \oiiiw flux $>8 \times 10^{-16}$ erg cm$^{-2}$ s$^{-1}$ in each \oiiiw peak.

The 13 galaxies that we selected in this way were observed over two separate programs for {\it Chandra}, and for 10 of the galaxies we also obtained multiband {\it HST}/WFC3 imaging to examine the host galaxies.  One of the galaxies was presented in \cite{CO11.2}, and we analyze the 12 remaining galaxies here.  These galaxies are located at redshifts $0.04 < z < 0.34$, and two of them are classified as Type 1 AGNs by their SDSS spectra while the others are classified as Type 2 AGNs. SDSS images of the galaxies are shown in Figure~\ref{fig:sdss}.

\section{Observations and Analysis}

\subsection{Optical SDSS Spectra}
\label{sdss}

Using the SDSS spectrum of each target, we measured the \oiiiw flux of each component of the double-peaked \oiiiw emission line and the line-of-sight velocity separation between them.  We made an exception for SDSS J1239+5314, since our analysis shows that the two X-ray sources correspond not to the main double peaks, but to the blueshifted main peak and a third \oiiiw emission component located between the main double peaks (Section~\ref{1239}).  Consequently, for SDSS J1239+5314 we measured the \oiiiw fluxes for the blueshifted \oiiiw emission component (corresponding to J1239+5314NE) and for the third \oiiiw emission component (corresponding to J1239+531SW) and the velocity separation between them.

First, we used a gas and absorption line fitting code ({\tt gandalf}; \citealt{SA06.1}) to fit a stellar continuum model to each spectrum.  After subtracting off the continuum models, we fit two Gaussians to the double-peaked \oiiiw emission line.  Each Gaussian returned the flux of one of the peaks in \oiiiwn, and we then converted the fluxes to luminosities.  Then, we used the central wavelengths of the two Gaussians to measure the line-of-sight velocity separation between them.  The errors on the luminosities and velocities are propagated from the errors on the Gaussian fitting parameters.    

We did not apply reddening corrections to the fluxes, since these corrections use oversimplified dust models that introduce large uncertainties (e.g., \citealt{RE08.1}), and because ultimately we will compare the observed \oiiiw luminosities to the observed X-ray luminosities (Section~\ref{Lx-Loiii}).

\begin{figure*}
\begin{center}
\includegraphics[height=2.5in]{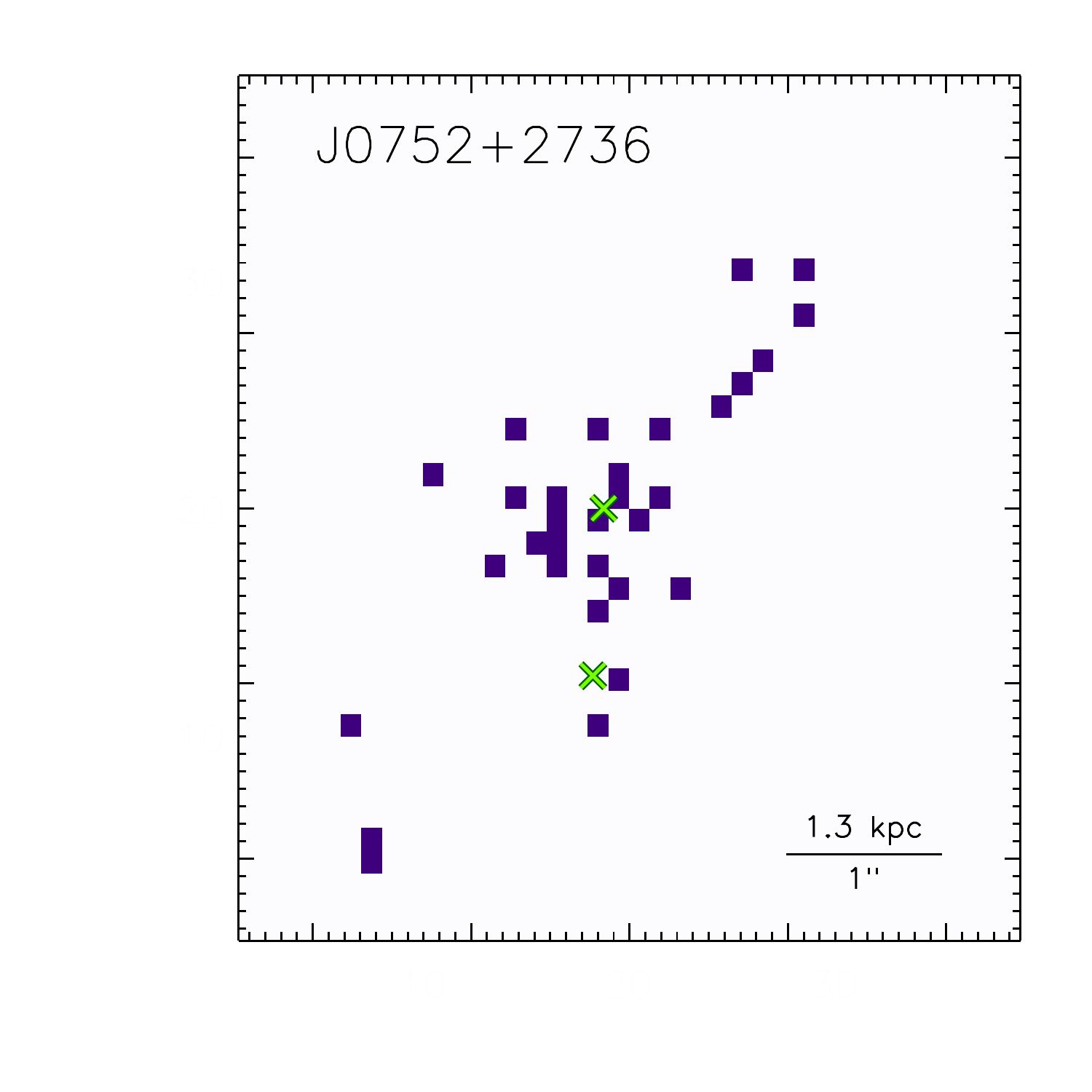}
\hspace{-.7in}
\includegraphics[height=2.5in]{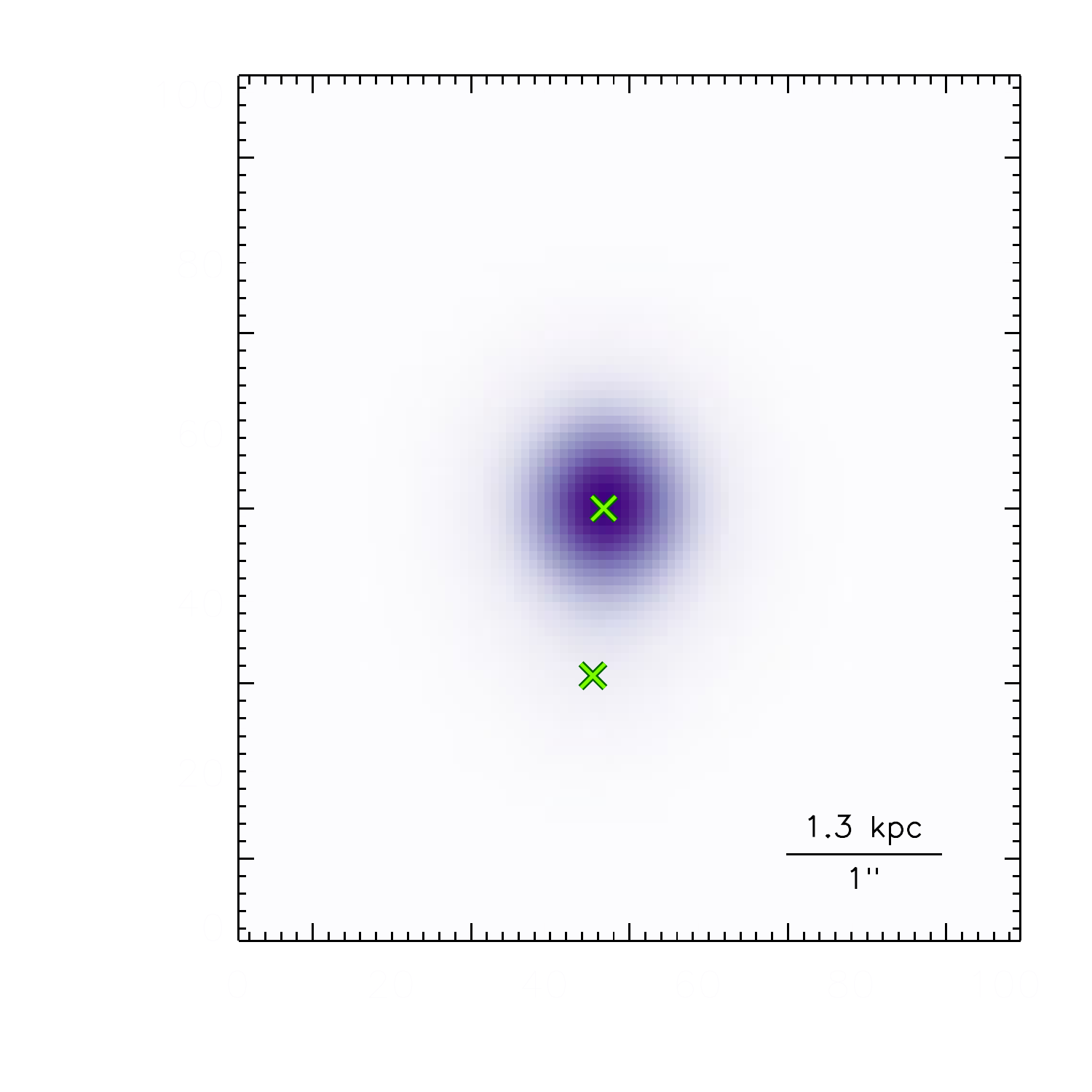}
\hspace{-.12in}
\raisebox{0.17\height}{\includegraphics[height=2.in]{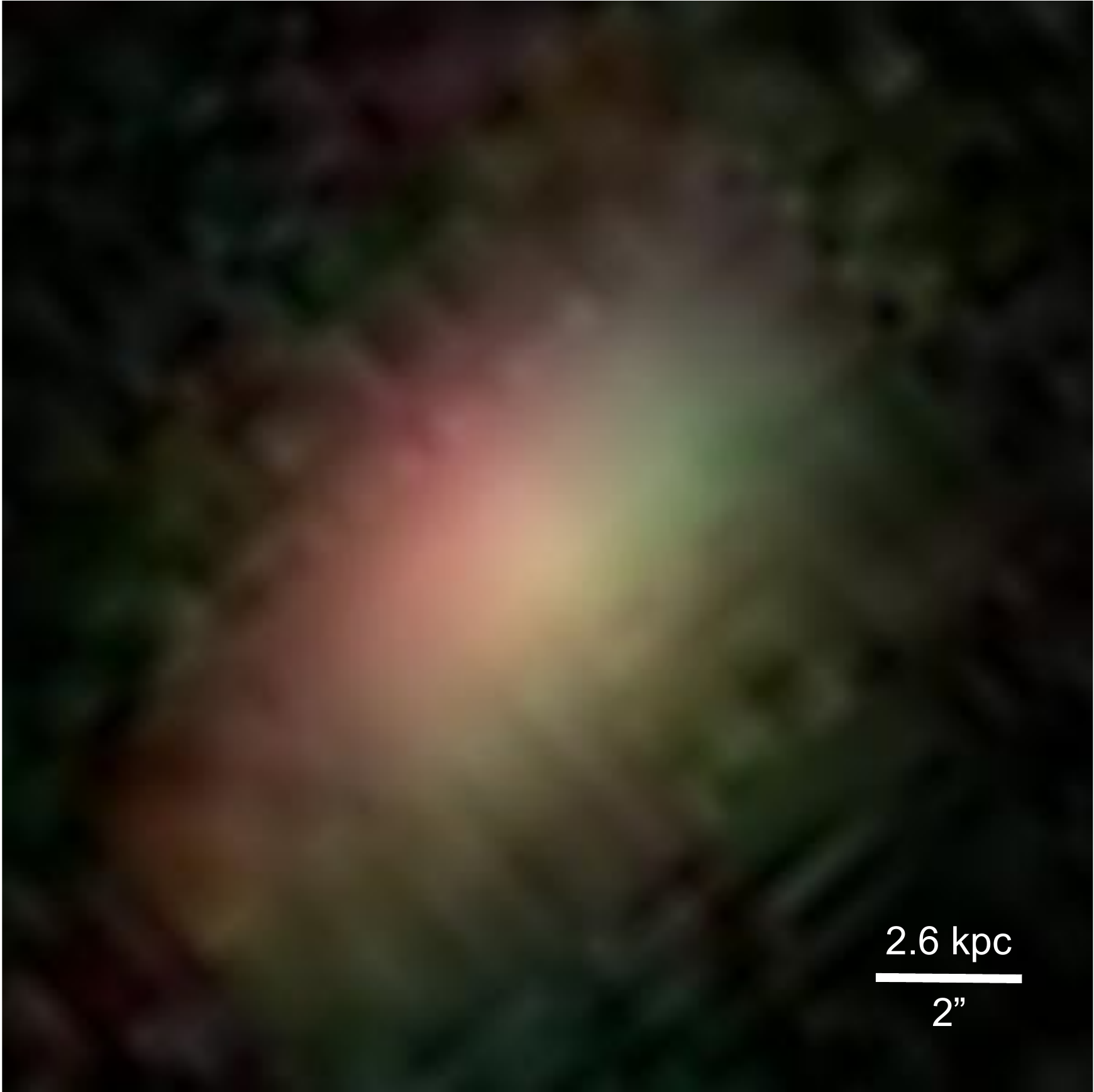}} \\
\vspace{-.4in}
\includegraphics[height=2.5in]{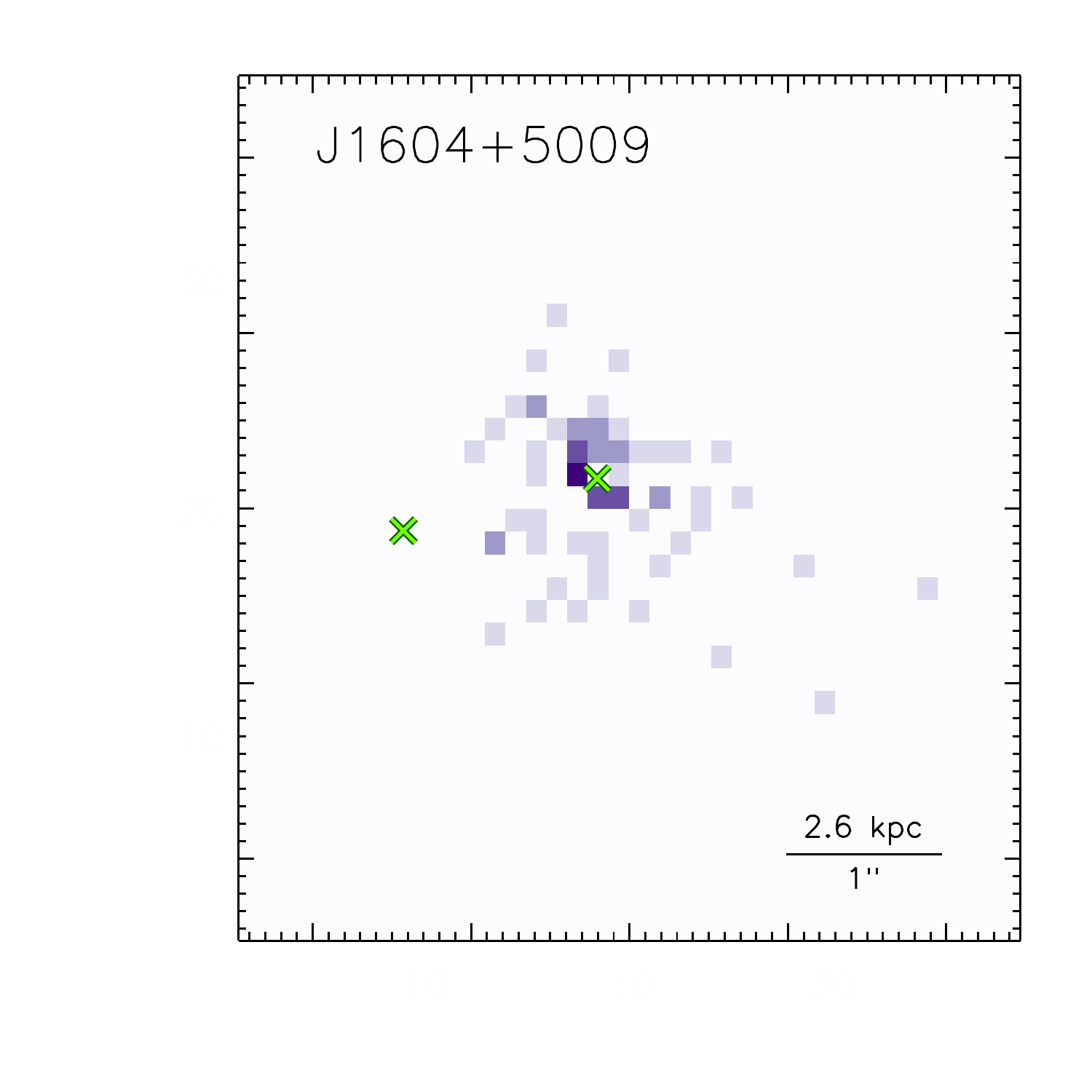}
\hspace{-.7in}
\includegraphics[height=2.5in]{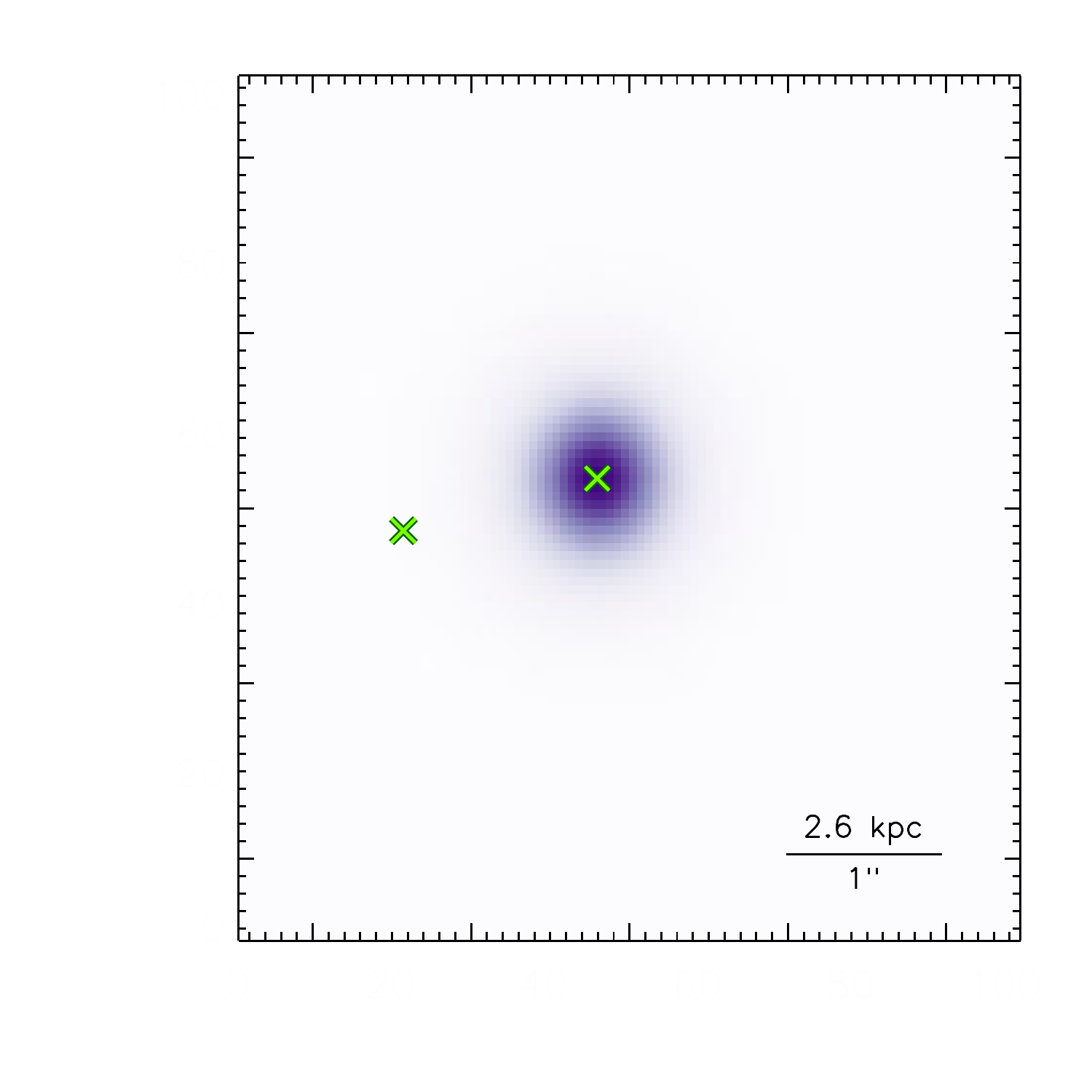}
\hspace{-.12in}
\raisebox{0.17\height}{\includegraphics[height=2.in]{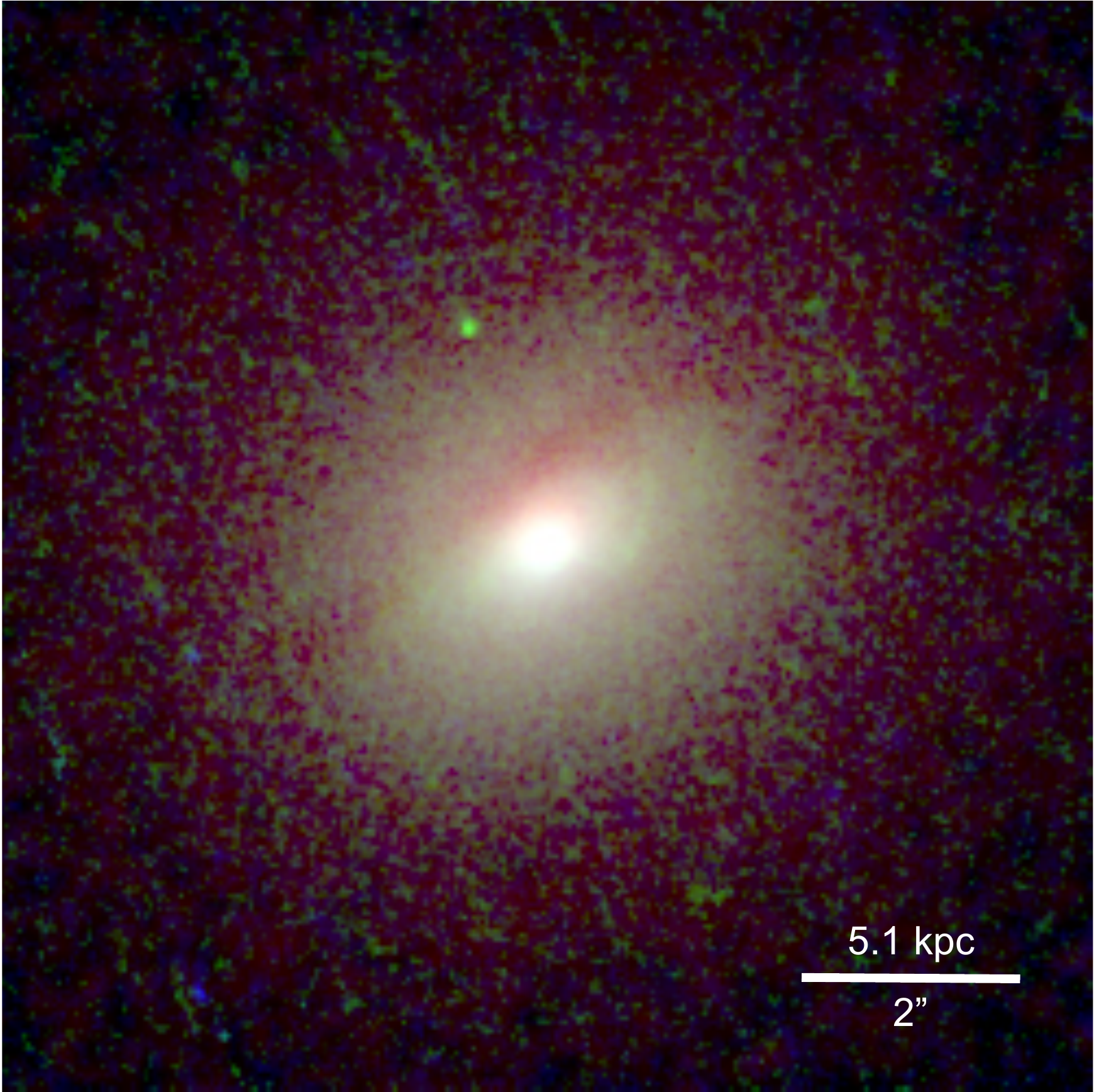}} \\
\end{center}
\vspace{-0.4in}
\caption{{\it Chandra} $0.3-8$ keV observations (left), model to the {\it Chandra} observations (middle), and imaging (right) for the two dual AGN candidates that were not a part of our {\it HST} program.  In all panels, North is up and East is to the left.  The left and middle panels are $5^{\prime\prime} \times 5^{\prime\prime}$ images centered on the coordinates of each SDSS spectrum.  The left panels show one-fourth size {\it Chandra} pixels (purple) and best-fit locations of two X-ray sources (green crosses) that coincide within $3\sigma$ to the locations of two observed \oiiiw components. The middle panels show the model fits to the two X-ray sources (purple) and the locations of two X-ray sources (green crosses).  The right panels show an SDSS $gri$ color composite image (top) and an archival {\it HST} image (red: F105W; green: F621M; blue: F547M; GO 12521, PI: Liu; bottom).}
\label{fig:nohst}
\end{figure*}

\subsection{Optical Long-slit Spectra}
\label{longslit}

All of our targets were observed with optical long-slit spectroscopy as follow-up to determine the spatial profiles of their emission.  The targets were observed using a combination of the Kast Spectrograph on the Lick 3 m telescope, the Dual Imaging Spectrograph on the Apache Point Observatory 3.5 m telescope, the Double Spectrograph on the Palomar 5 m telescope, the Blue Channel Spectrograph on the MMT 6.5 m telescope, and the Low-Dispersion Survey Spectrograph on the Magellan II 6.5 m telescope \citep{GR11.1,SH11.1,CO12.1}.  All of the galaxies (with the exception of SDSS J1322+2631) were observed twice, with the slit oriented along two different position angles on the sky.  For each galaxy the spatial separations between \oiiiw emission components were measured at each slit orientation, and then combined to yield the true spatial separation of the two \oiiiw emission components on the sky $\Delta x_{[\mathrm{O \,III}]}$ and the true position angle of the two \oiiiw emission components on the sky $\theta_{[\mathrm{O \,III}]}$ (Table~\ref{tbl-2}). 

SDSS J1322+2631 was only observed once, with the slit oriented along the major axis of the galaxy.  The long-slit spectrum, taken at a position angle of $79^\circ$ East of North, revealed two \oiiiw emission components separated by $2\farcs1$, or 5.3 kpc \citep{SH11.1}.  This measurement provides a lower limit on the full spatial separation of the emission components on the sky.  To determine the true spatial separation and position angle on the sky of the two emission sources, we measure the positions of the two \ha emission components visible in the {\it HST}/F814W observations of SDSS J1322+2631 (Section~\ref{hst}; Table~\ref{tbl-2}). 

The long-slit spectra of SDSS J1356+1026 revealed ionized line emission out to $\simgt20$ kpc, indicating the presence of a galaxy-wide outflow \citep{GR11.1,GR12.2}.  Because of the extreme spatial extent of the \oiiiw emission in this galaxy, pinpointing centroids of the \oiiiw emission from which to measure a spatial separation is not feasible.  Instead, we note that much of the \oiiiw emission spatially coincides with the locations of two stellar bulges visible in the {\it HST} F160W observations.  We take the separation and position angle between the stellar bulges on the sky (Section~\ref{hst}) to be representative of that of the \oiiiw emission components (Table~\ref{tbl-2}).

\begin{figure*}
\begin{center}
\includegraphics[height=2.5in]{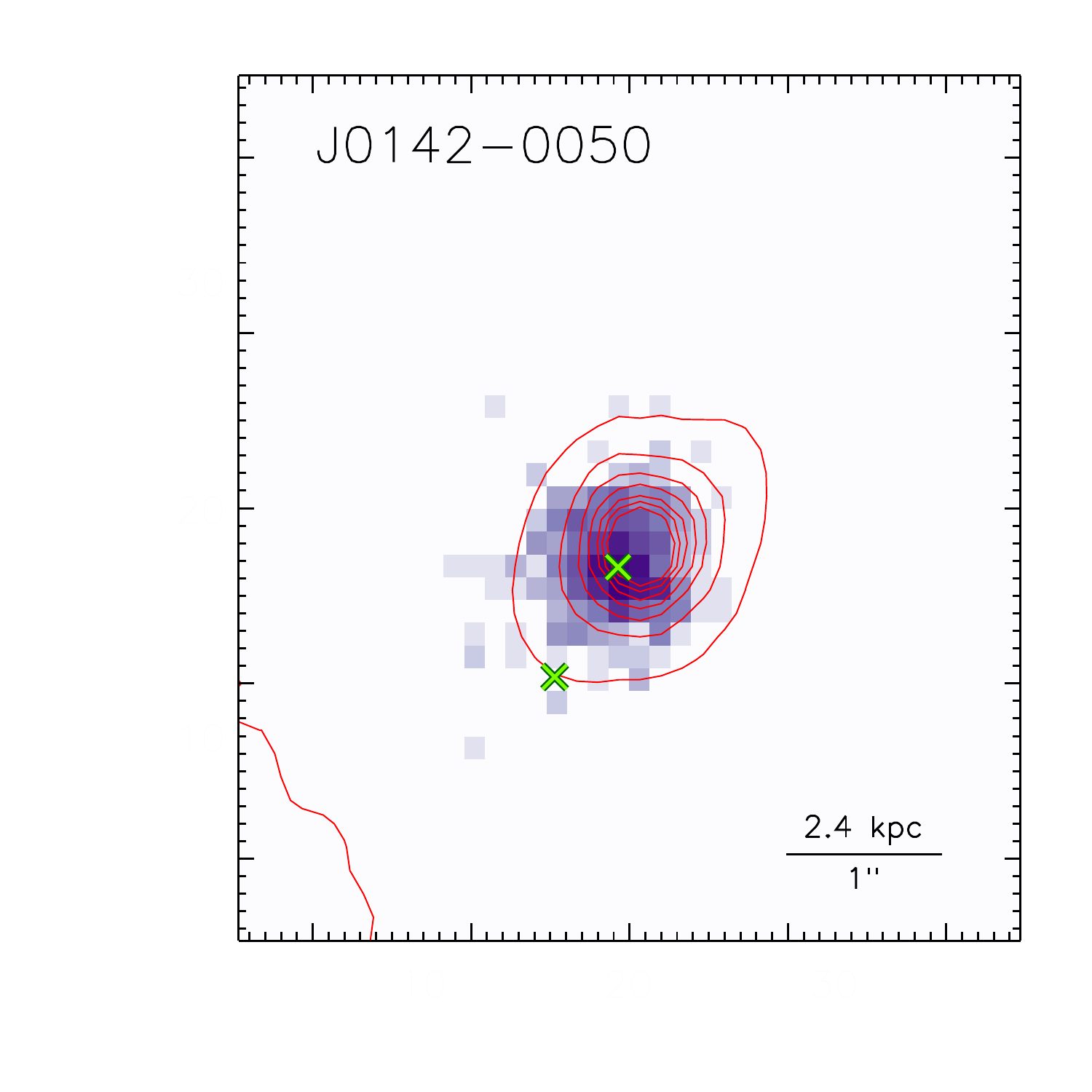}
\hspace{-.7in}
\includegraphics[height=2.5in]{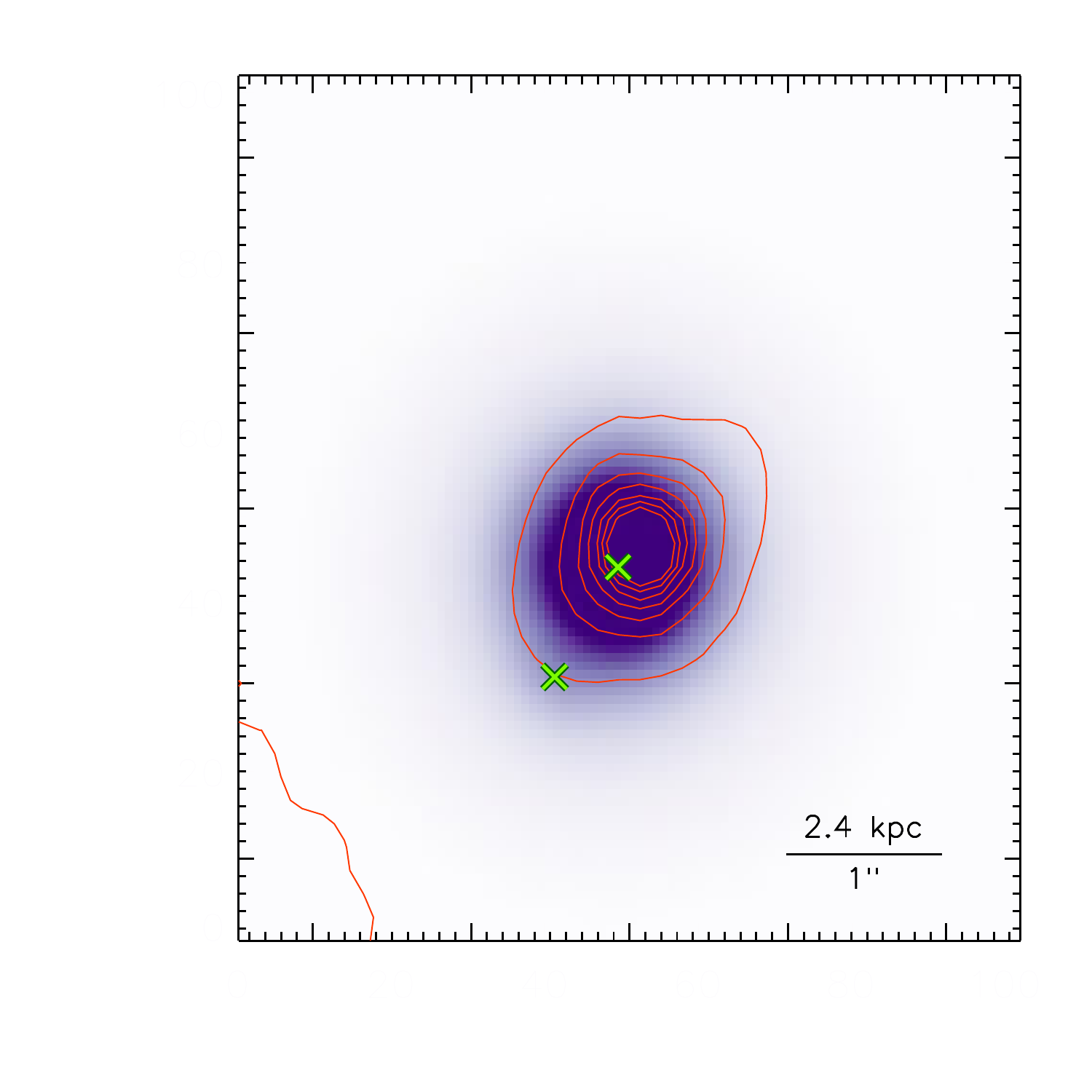}
\hspace{-.12in}
\raisebox{0.17\height}{\includegraphics[height=2.in]{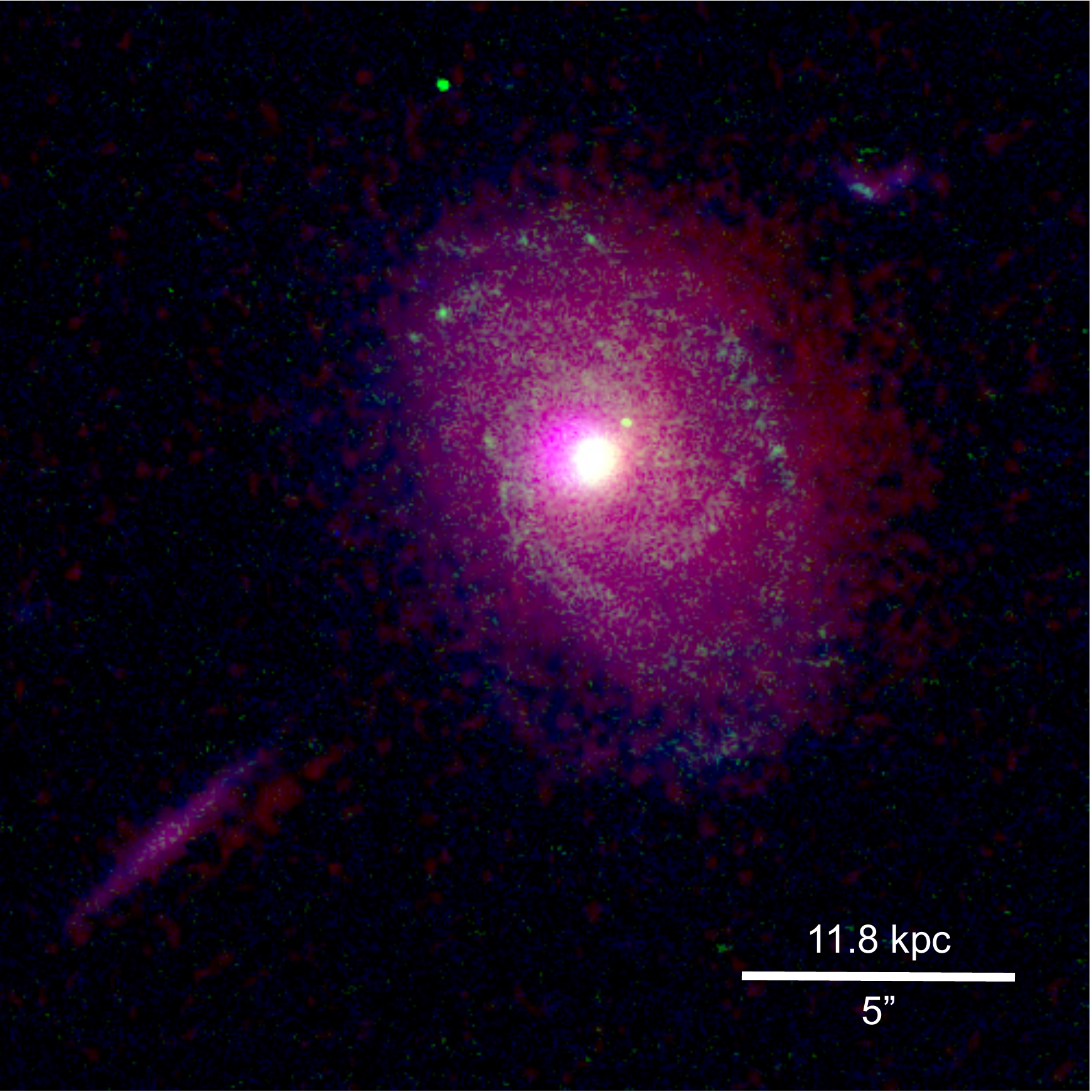}} \\
\vspace{-.4in}
\includegraphics[height=2.5in]{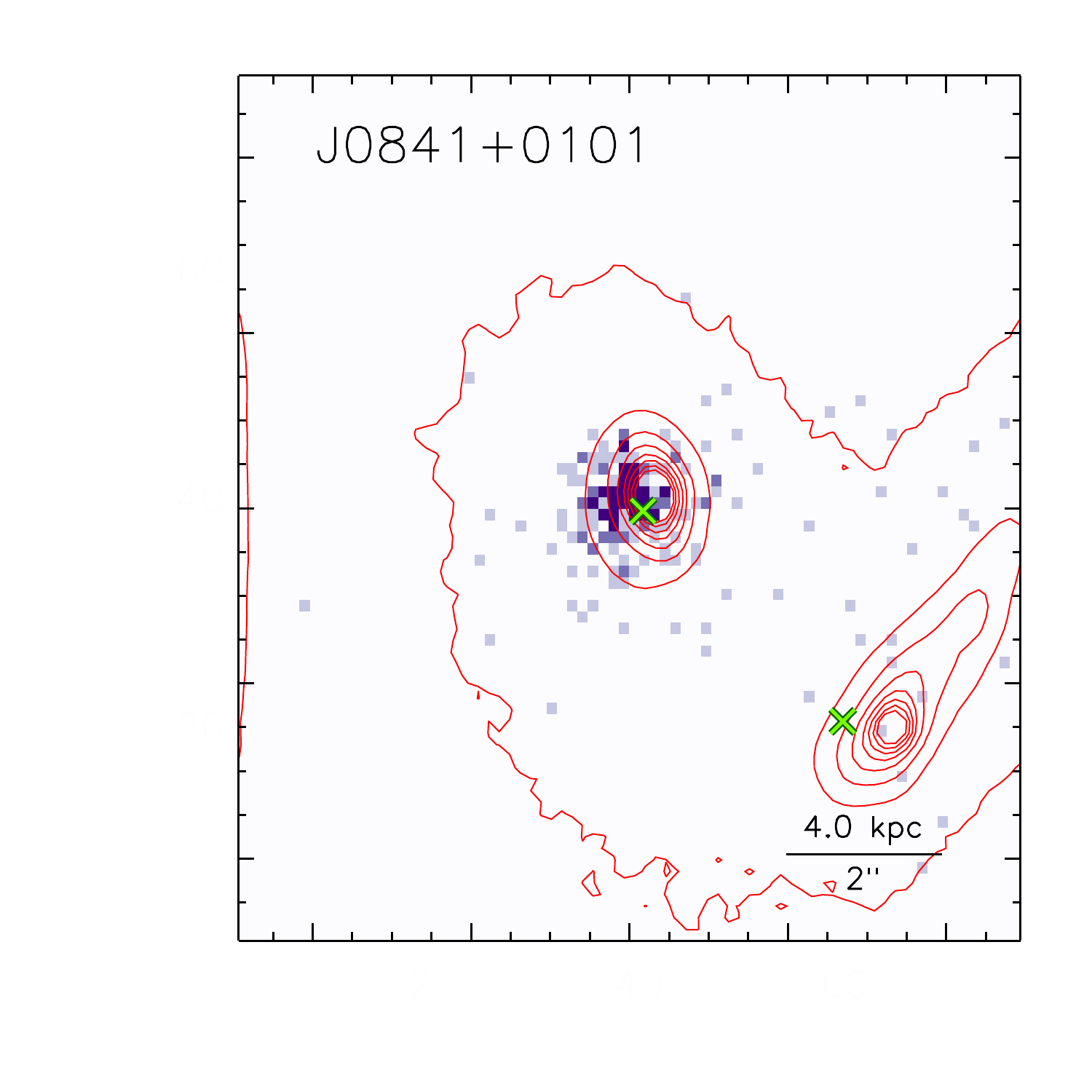}
\hspace{-.7in}
\includegraphics[height=2.5in]{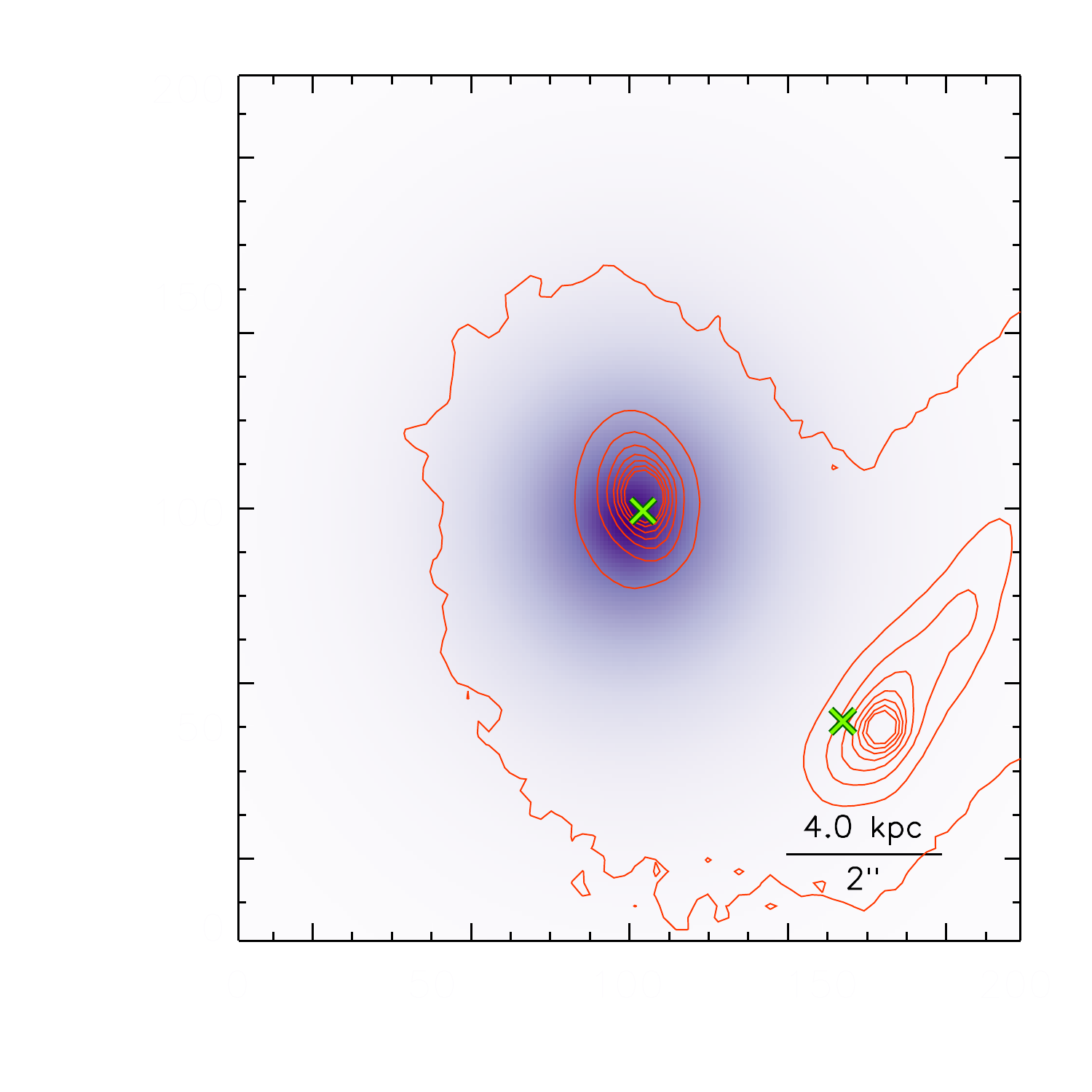}
\hspace{-.12in}
\raisebox{0.17\height}{\includegraphics[height=2.in]{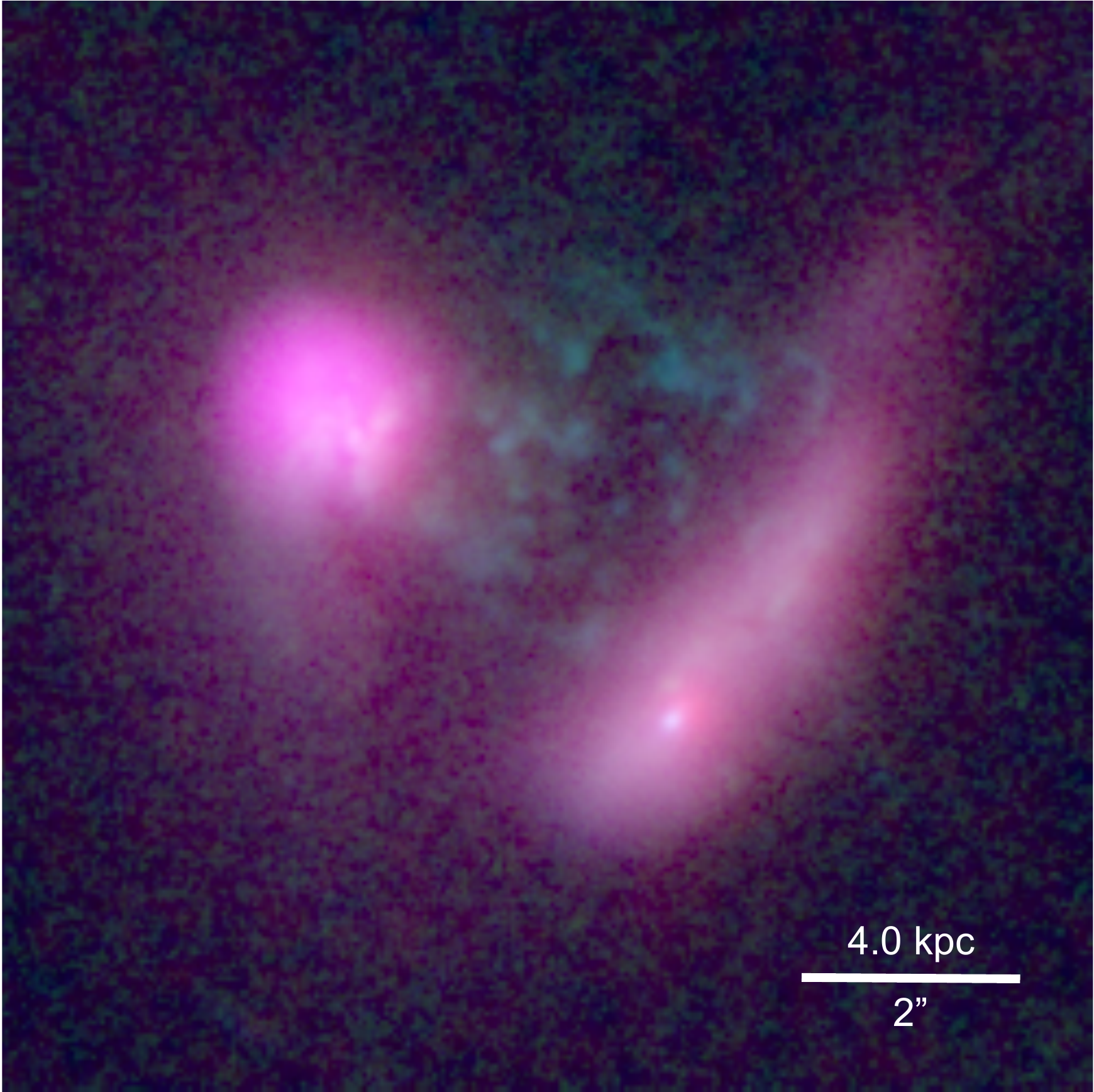}} \\
\vspace{-.4in}
\includegraphics[height=2.5in]{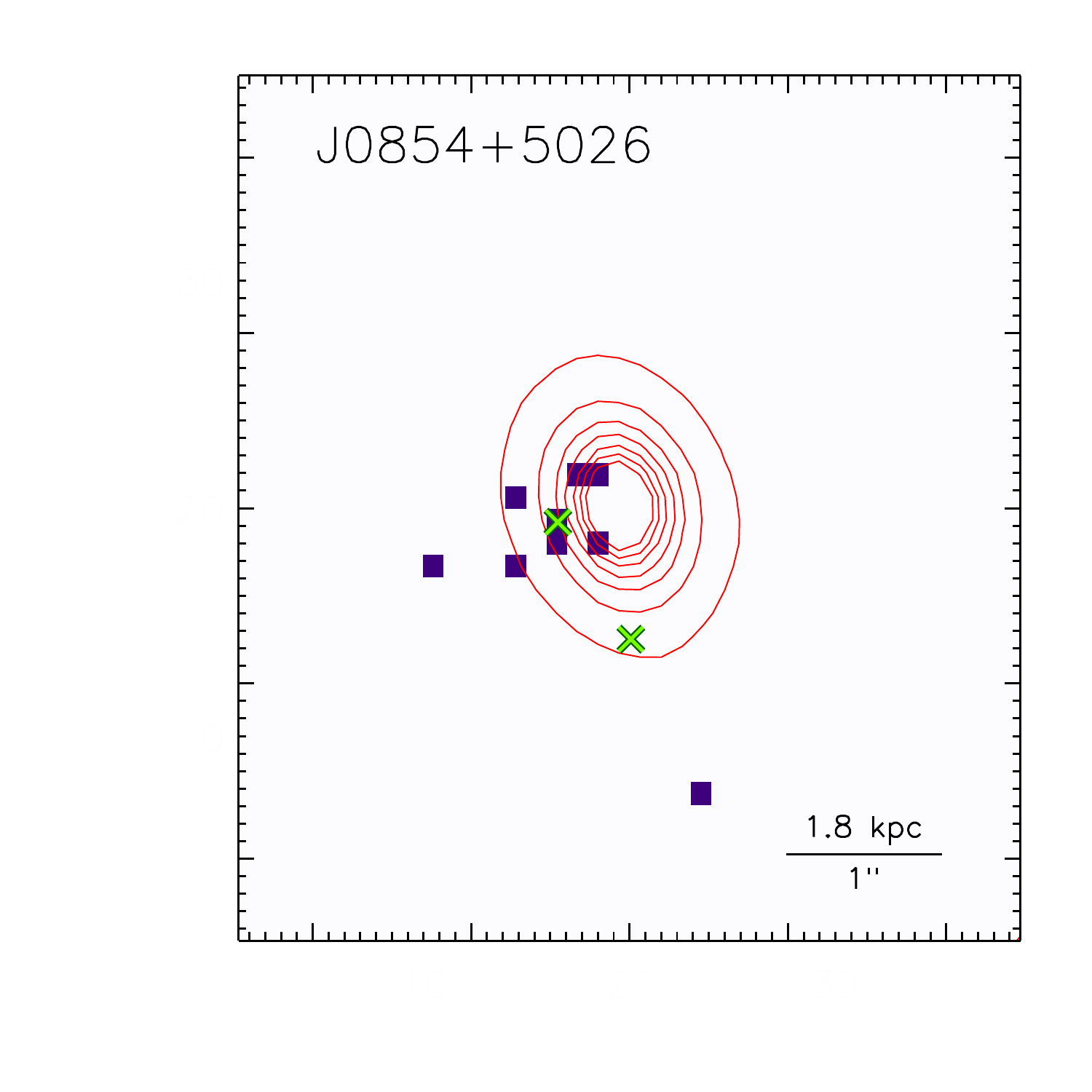}
\hspace{-.7in}
\includegraphics[height=2.5in]{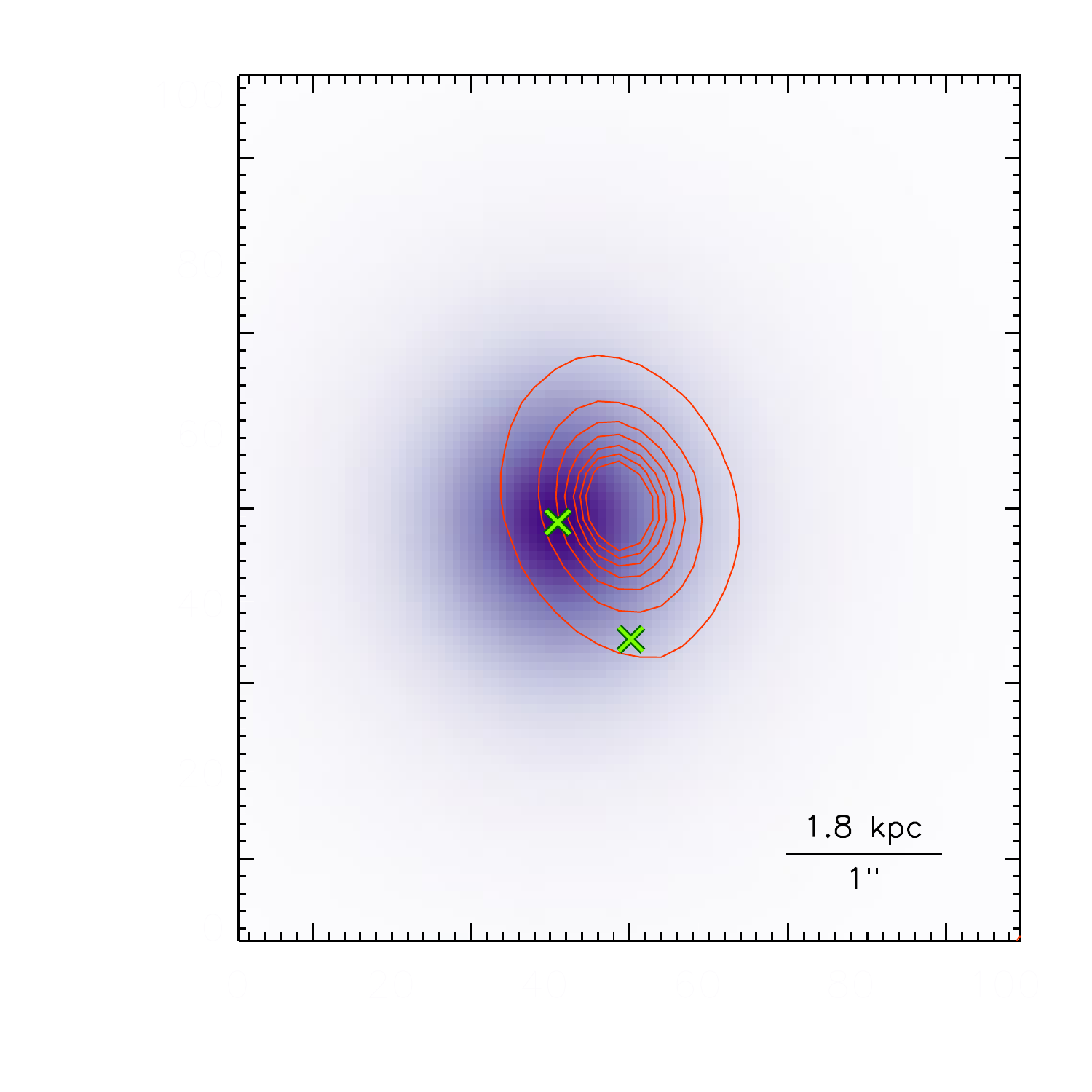}
\hspace{-.12in}
\raisebox{0.17\height}{\includegraphics[height=2.in]{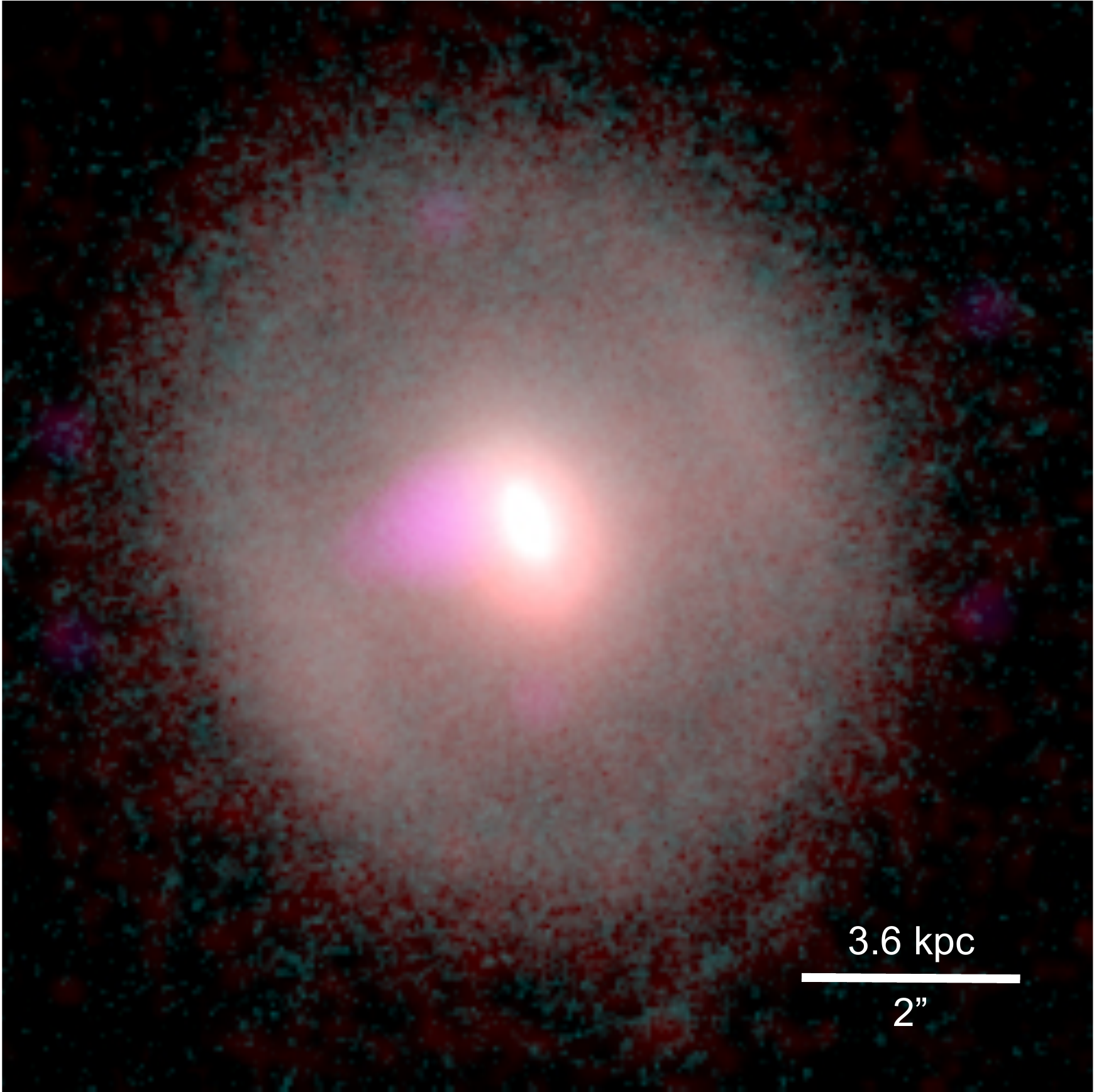}} \\
\vspace{-.4in}
\includegraphics[height=2.5in]{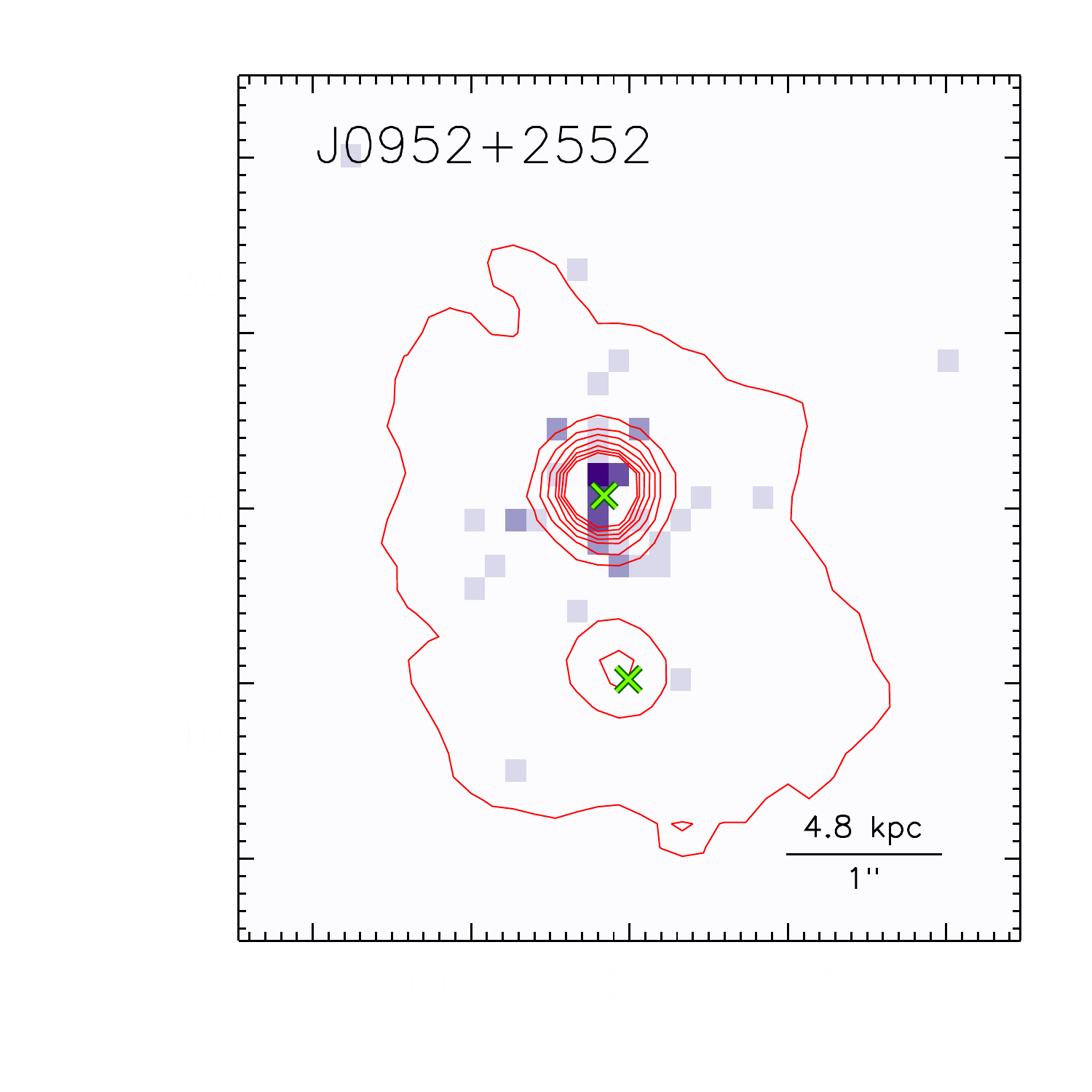}
\hspace{-.7in}
\includegraphics[height=2.5in]{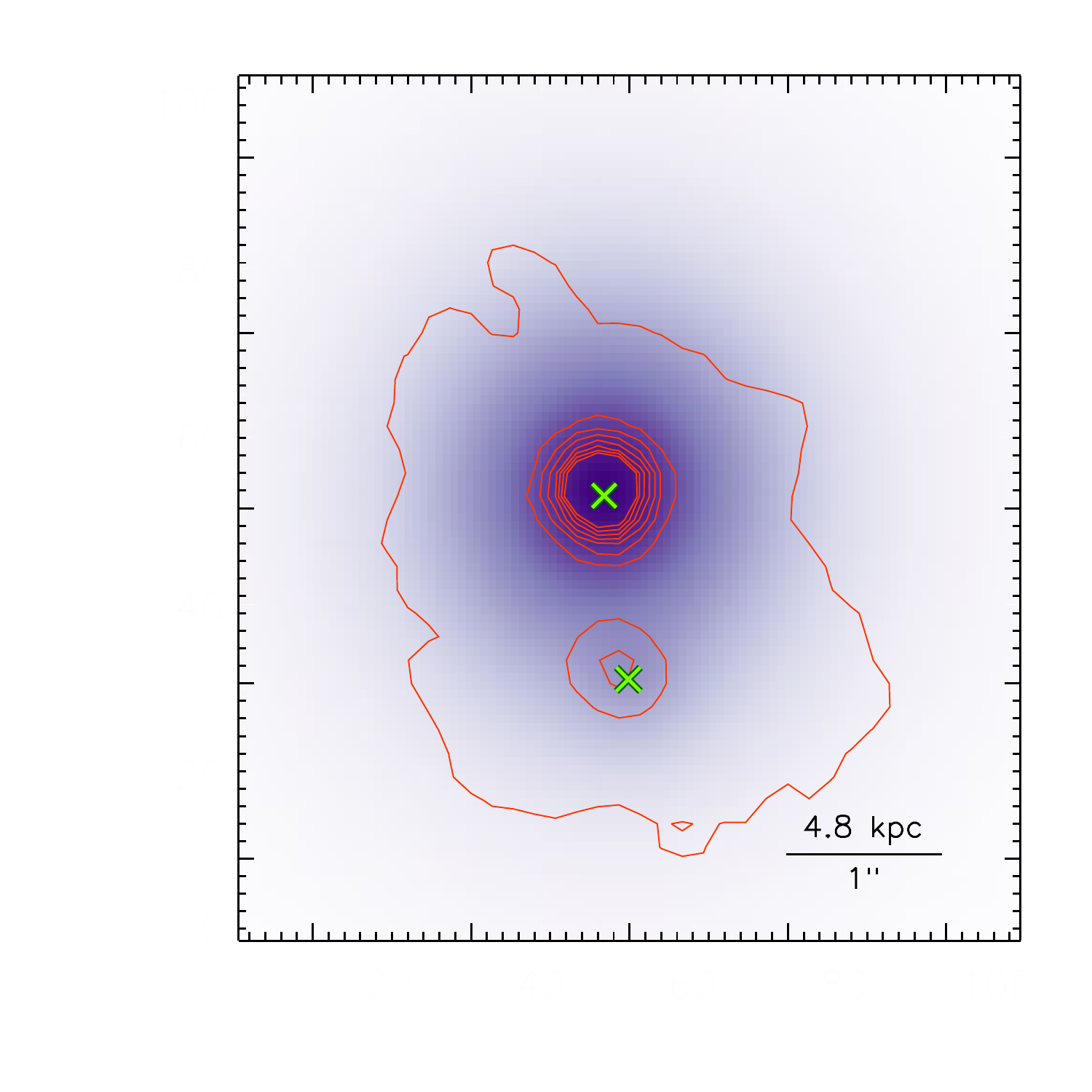}
\hspace{-.12in}
\raisebox{0.17\height}{\includegraphics[height=2.in]{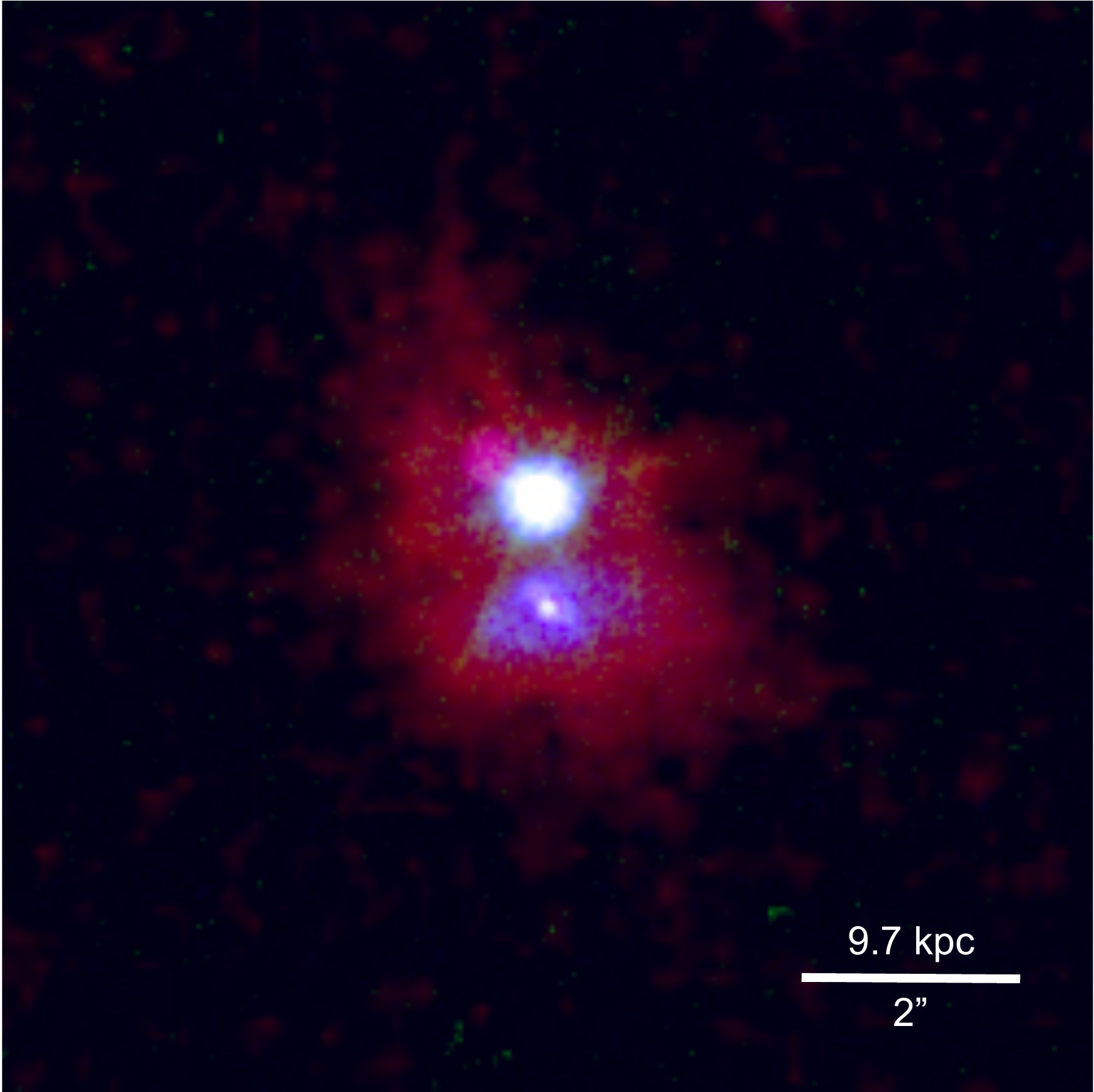}}
\end{center}
\end{figure*}

\begin{figure*}
\begin{center}
\includegraphics[height=2.5in]{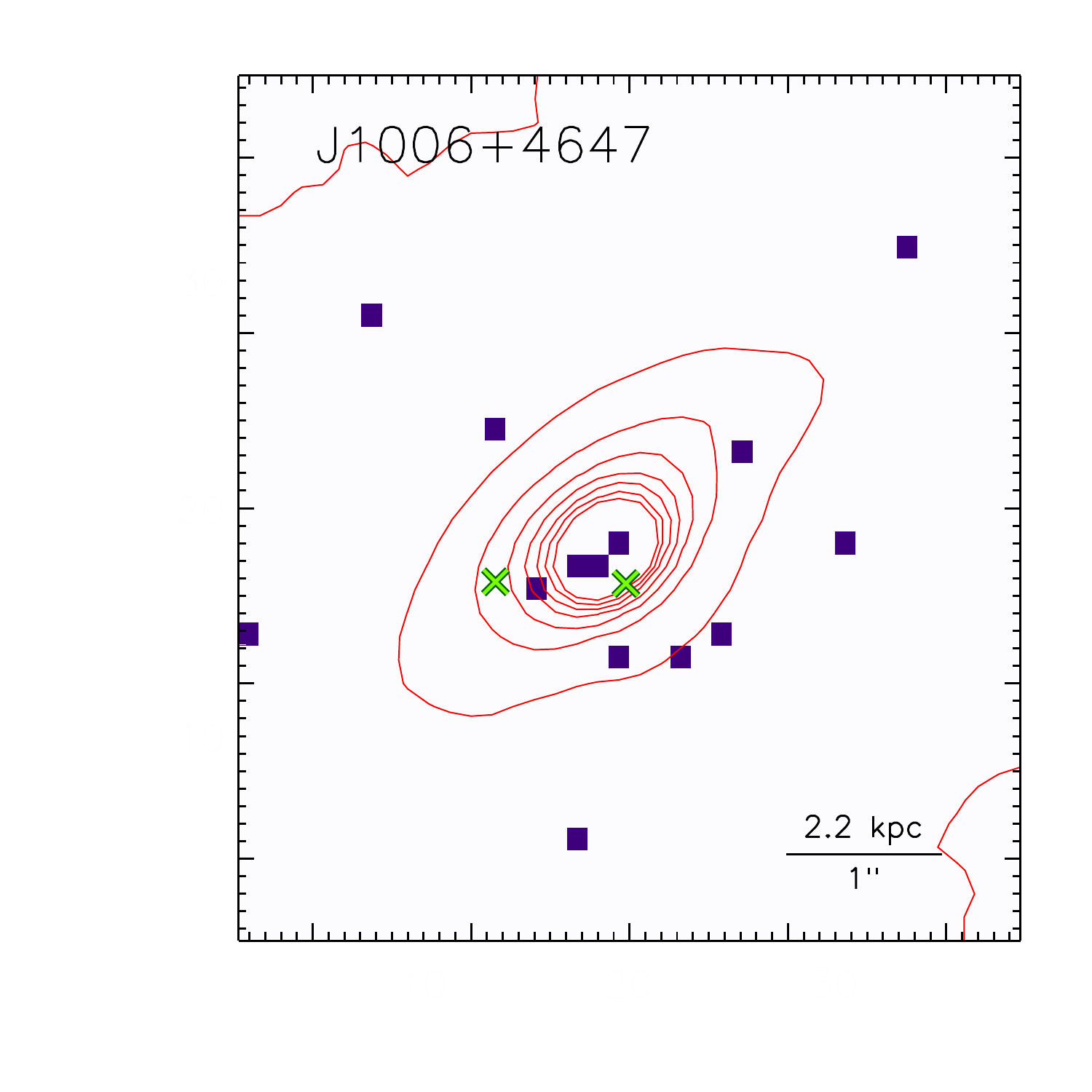}
\hspace{-.7in}
\includegraphics[height=2.5in]{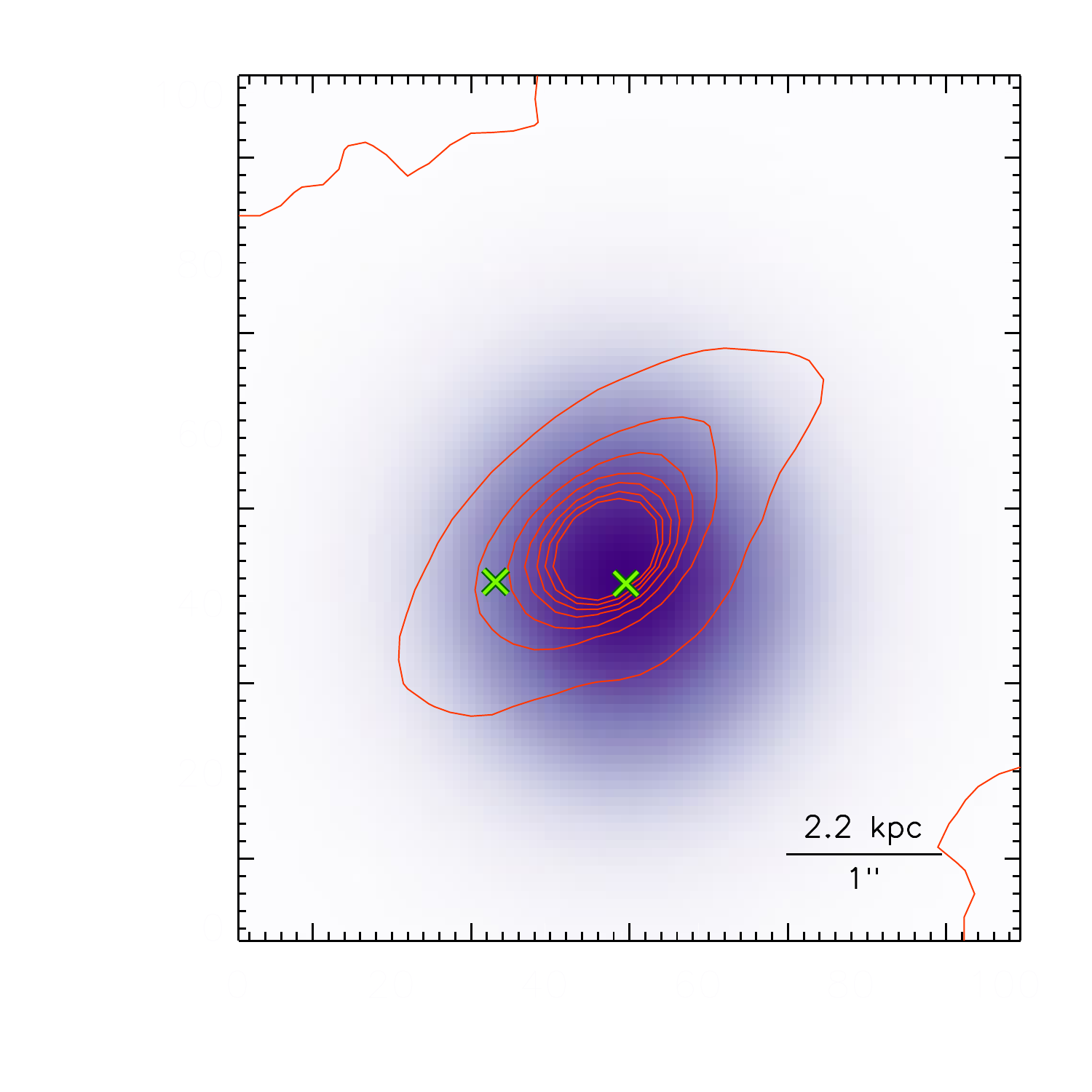}
\hspace{-.12in}
\raisebox{0.17\height}{\includegraphics[height=2.in]{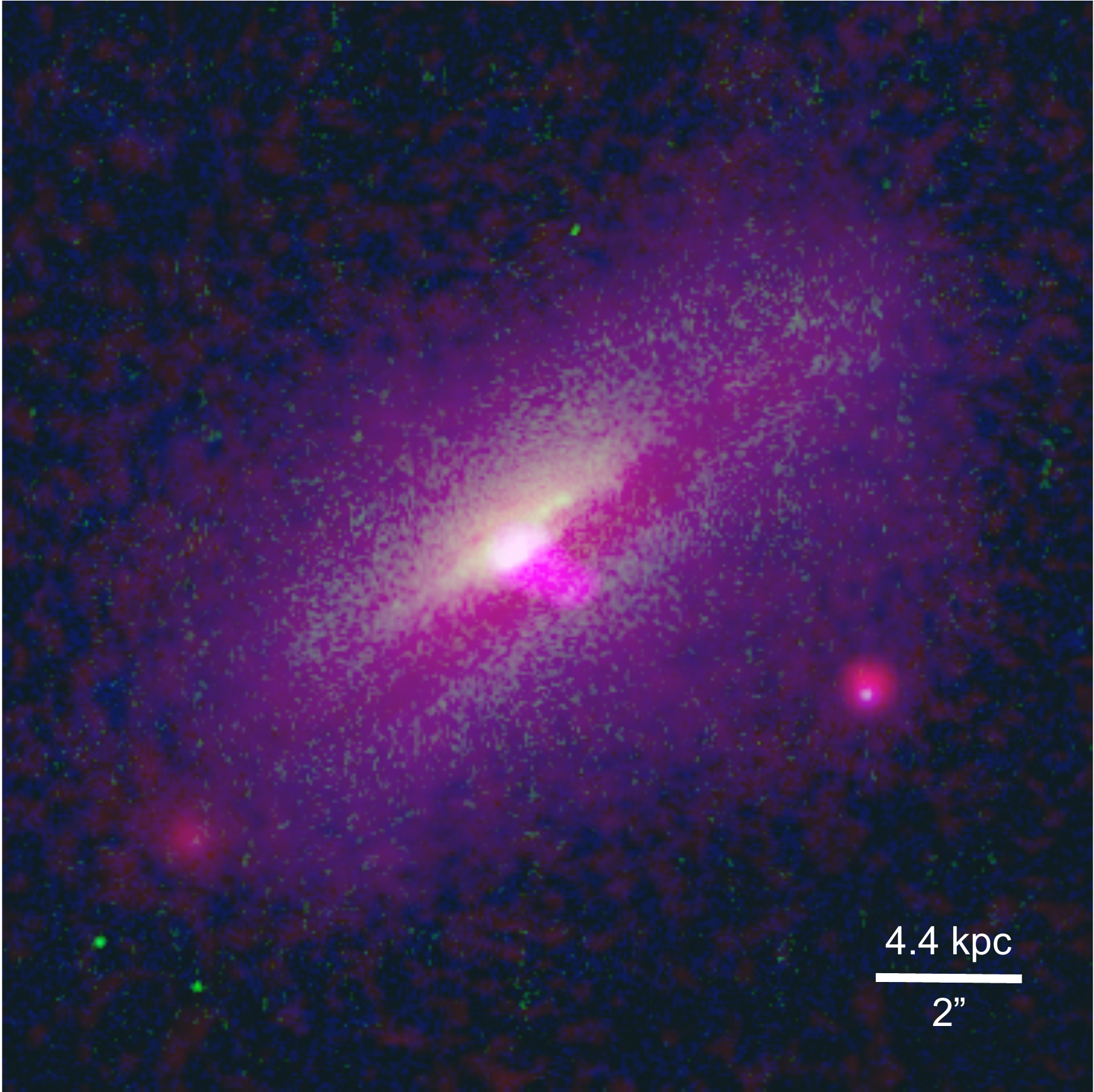}} \\
\vspace{-.4in}
\includegraphics[height=2.5in]{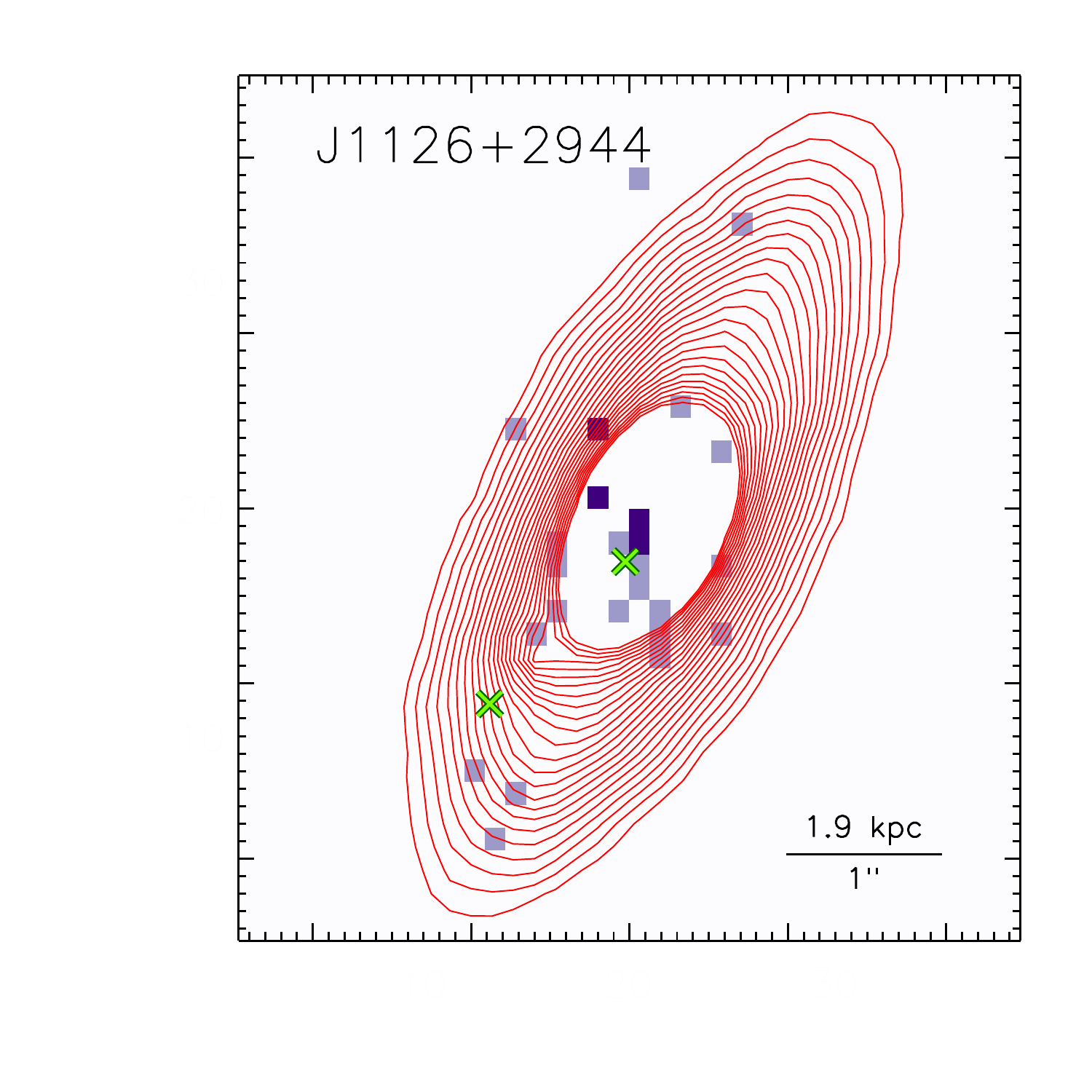}
\hspace{-.7in}
\includegraphics[height=2.5in]{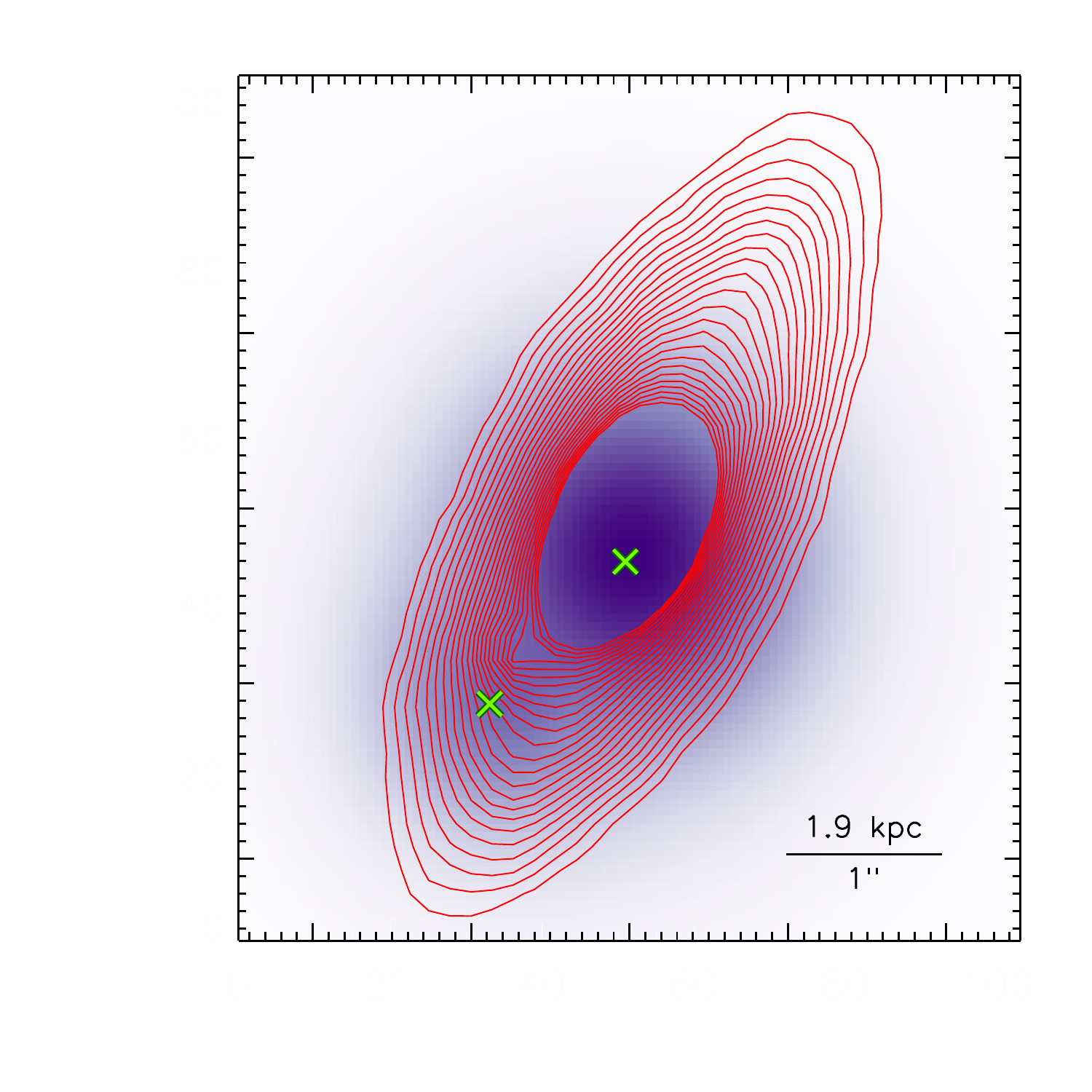}
\hspace{-.12in}
\raisebox{0.17\height}{\includegraphics[height=2.in]{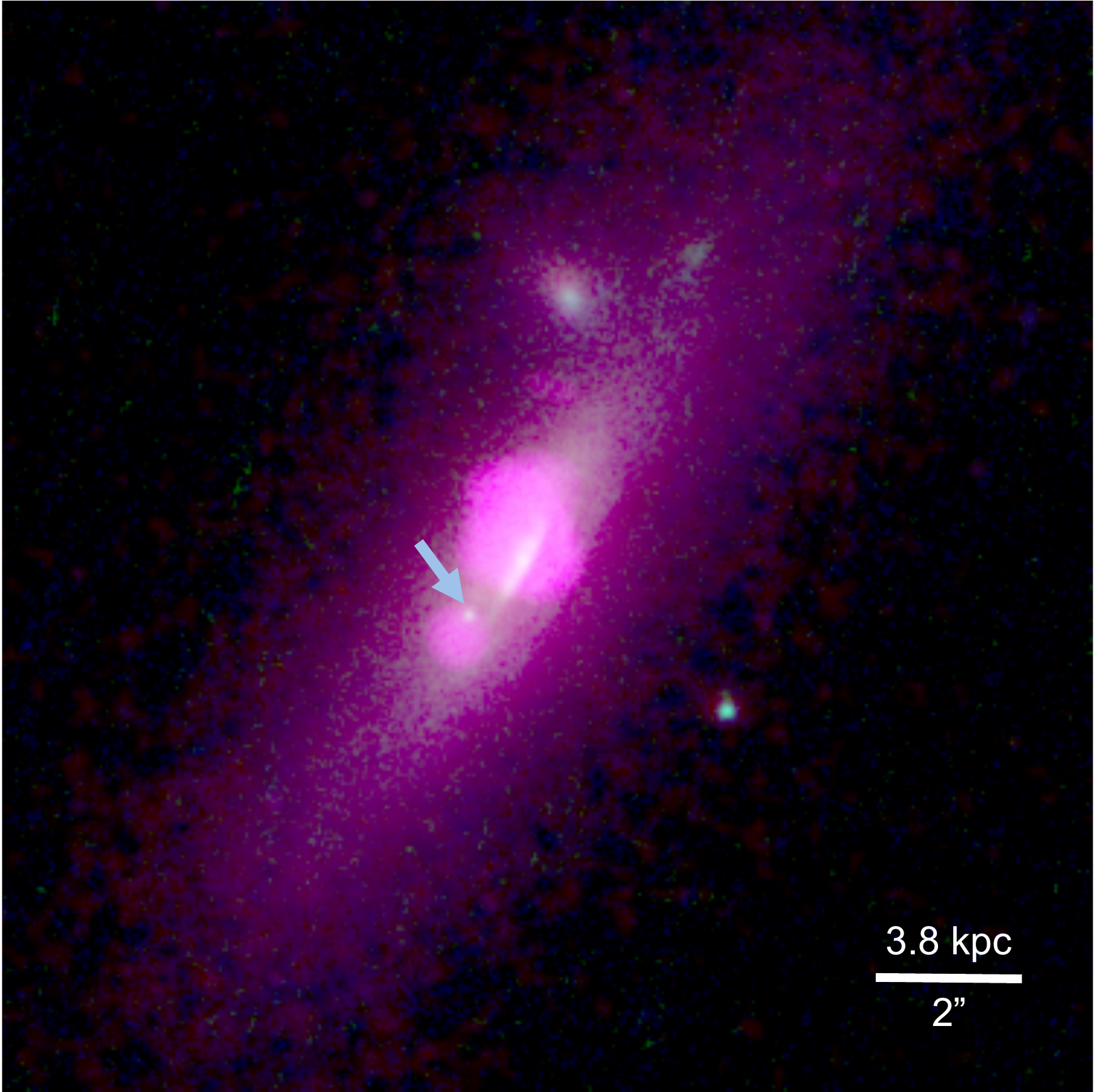}} \\
\vspace{-.4in}
\includegraphics[height=2.5in]{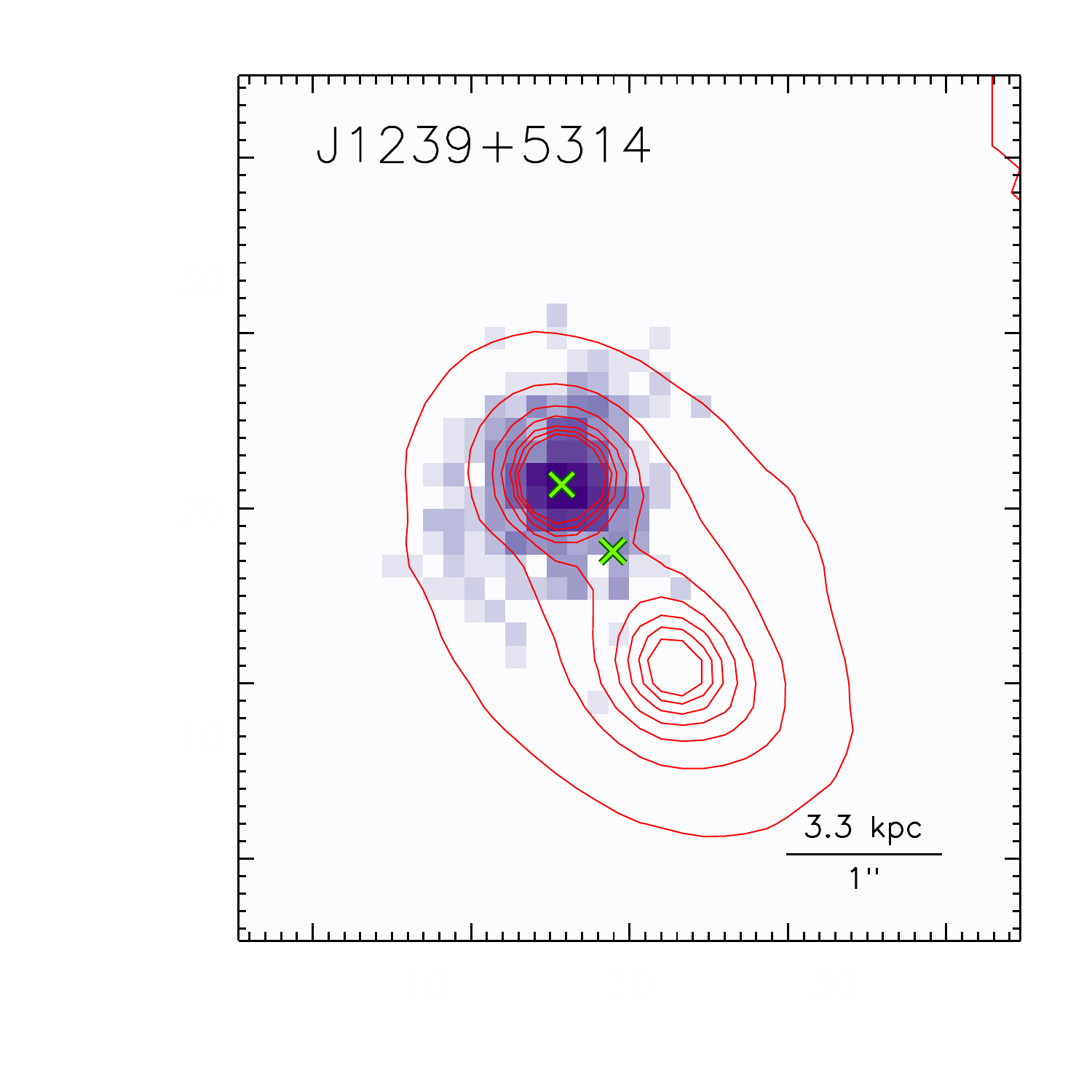}
\hspace{-.7in}
\includegraphics[height=2.5in]{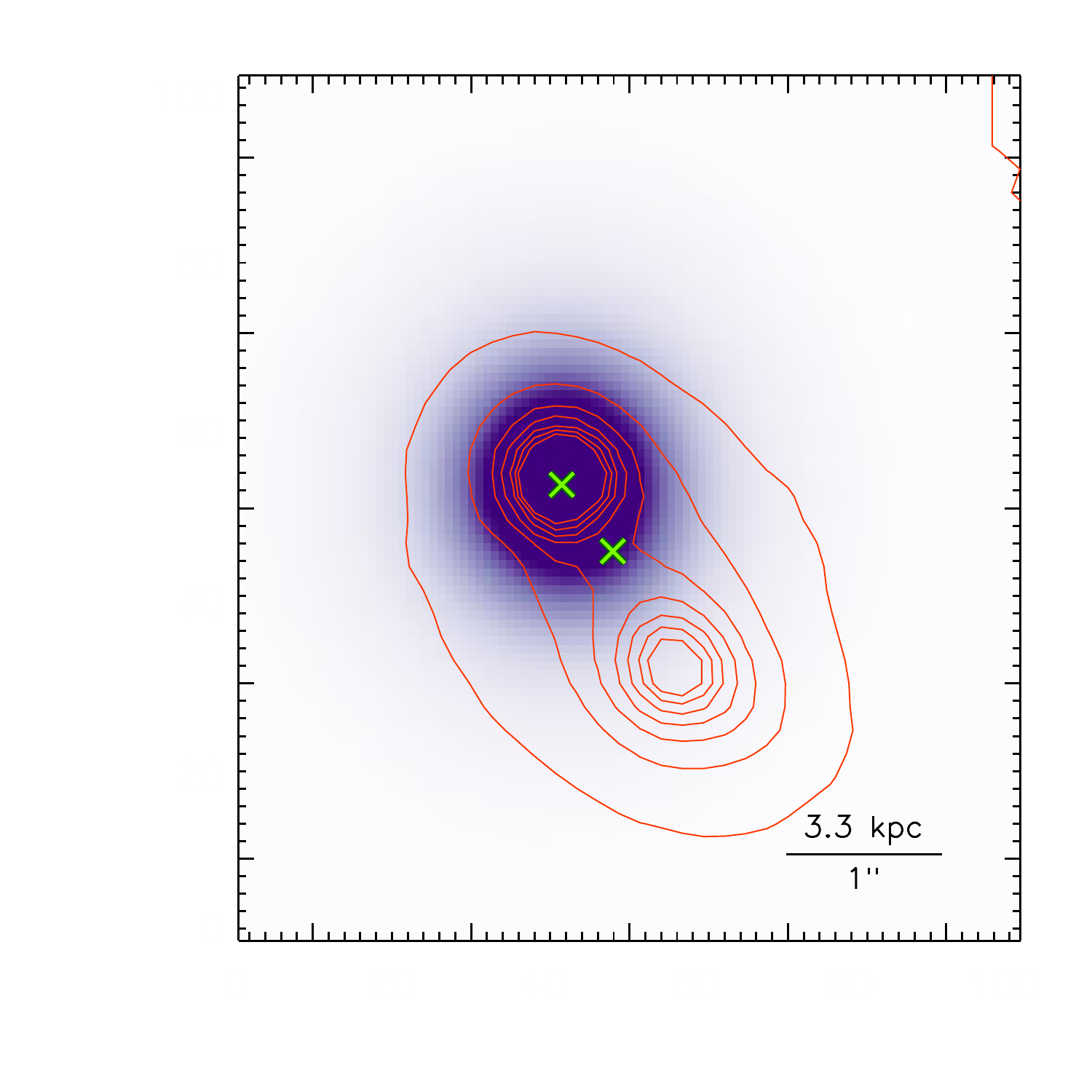}
\hspace{-.12in}
\raisebox{0.17\height}{\includegraphics[height=2.in]{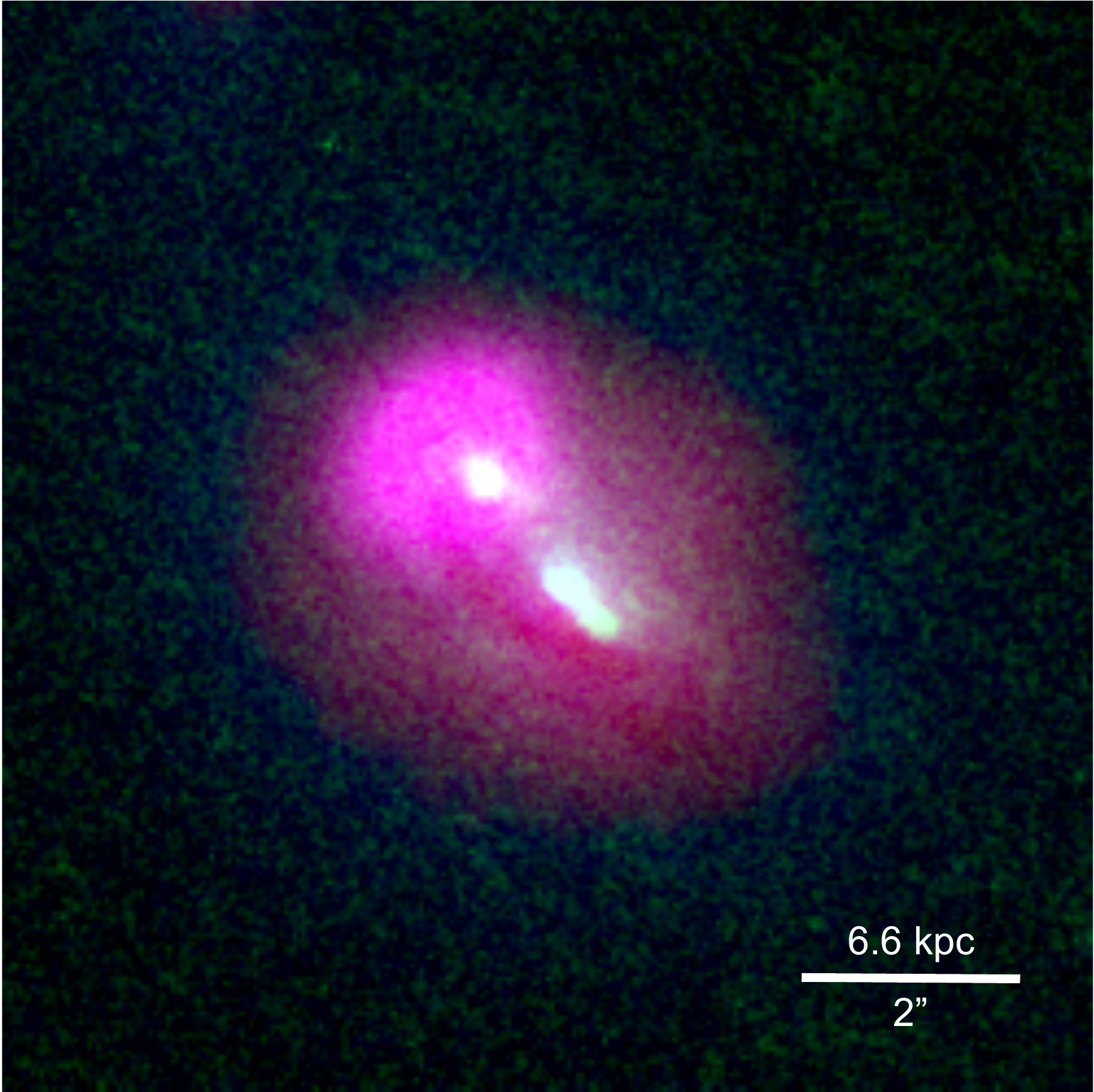}} \\
\vspace{-.4in}
\includegraphics[height=2.5in]{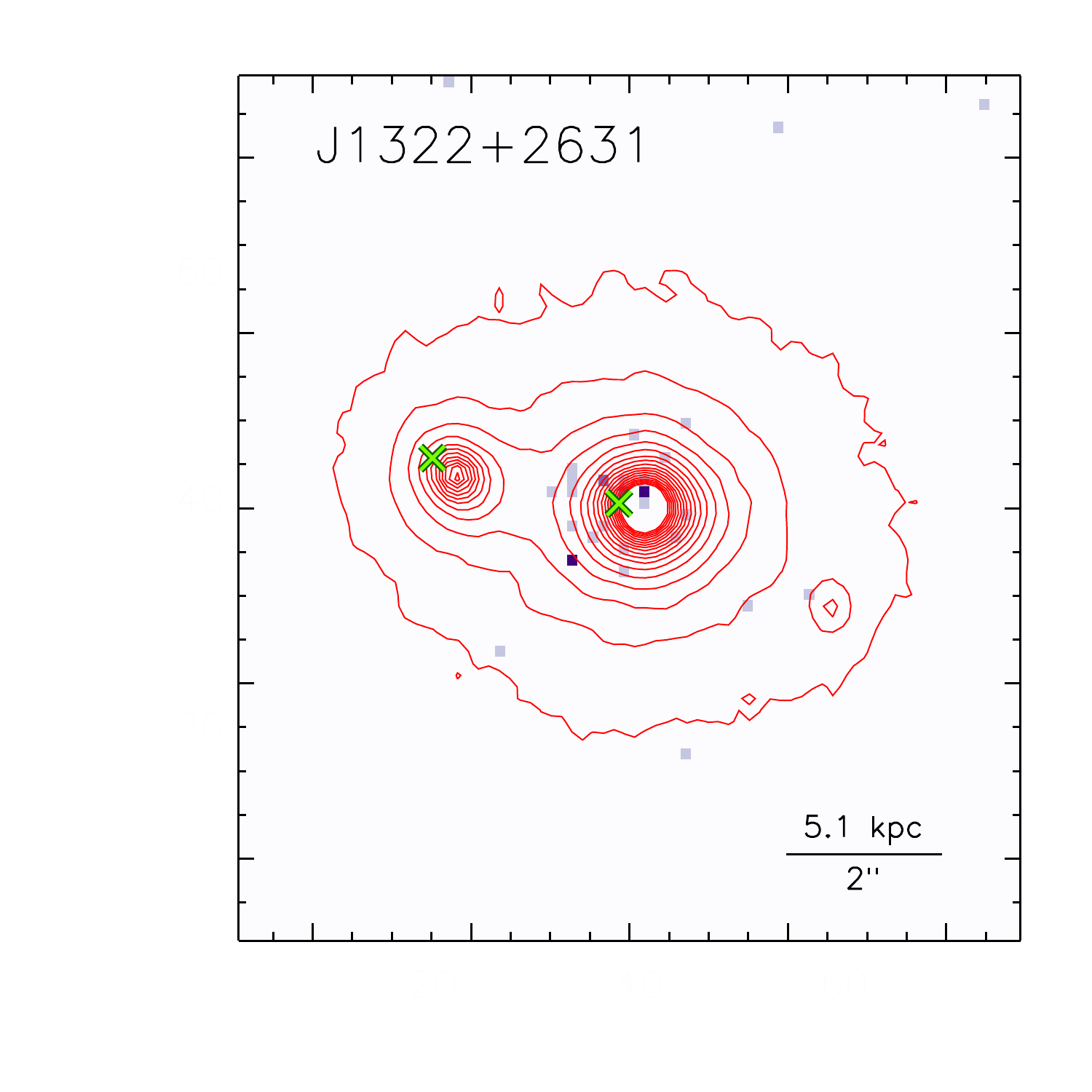}
\hspace{-.7in}
\includegraphics[height=2.5in]{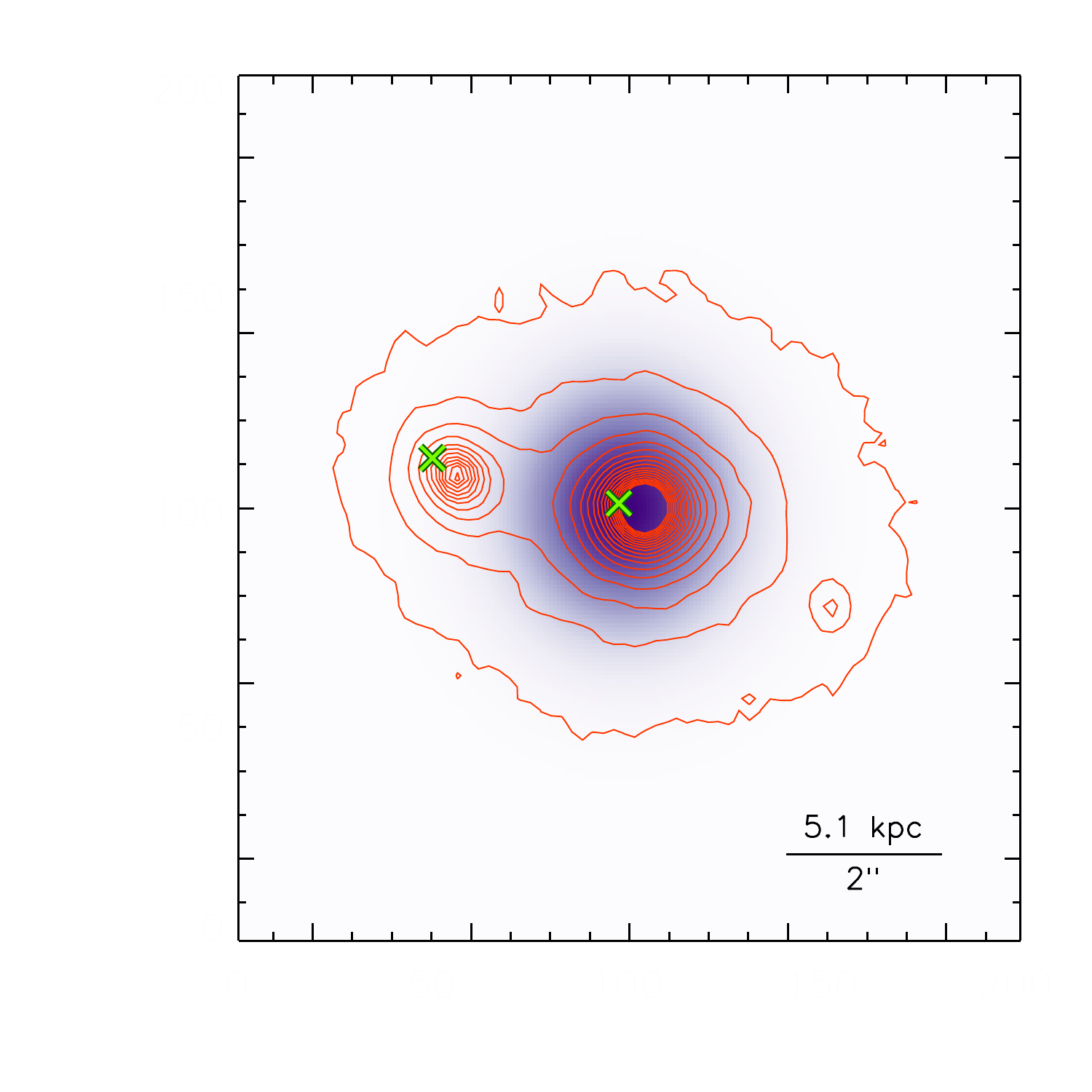}
\hspace{-.12in}
\raisebox{0.17\height}{\includegraphics[height=2.in]{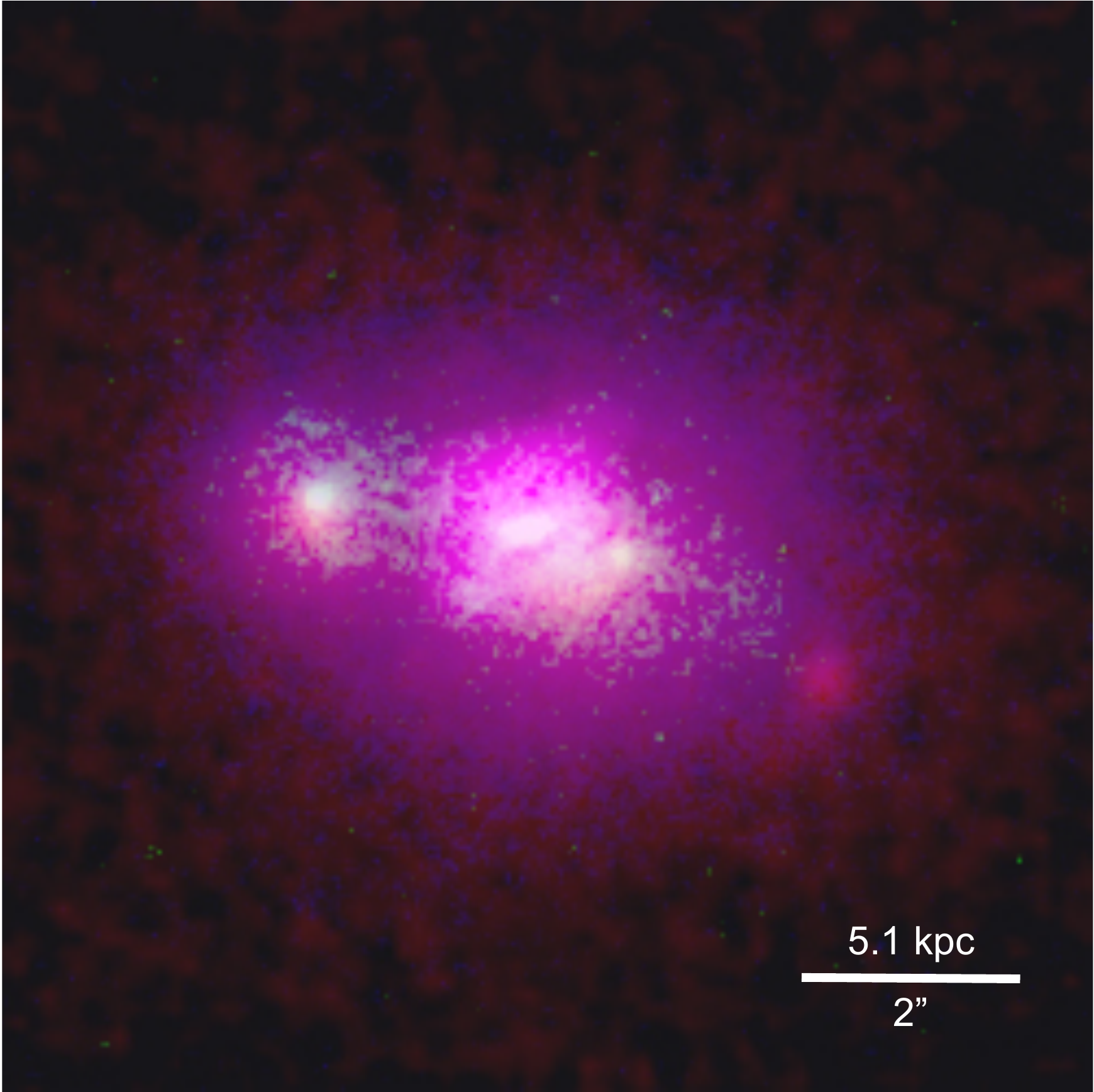}}
\end{center}
\end{figure*}

\begin{figure*}
\begin{center}
\includegraphics[height=2.5in]{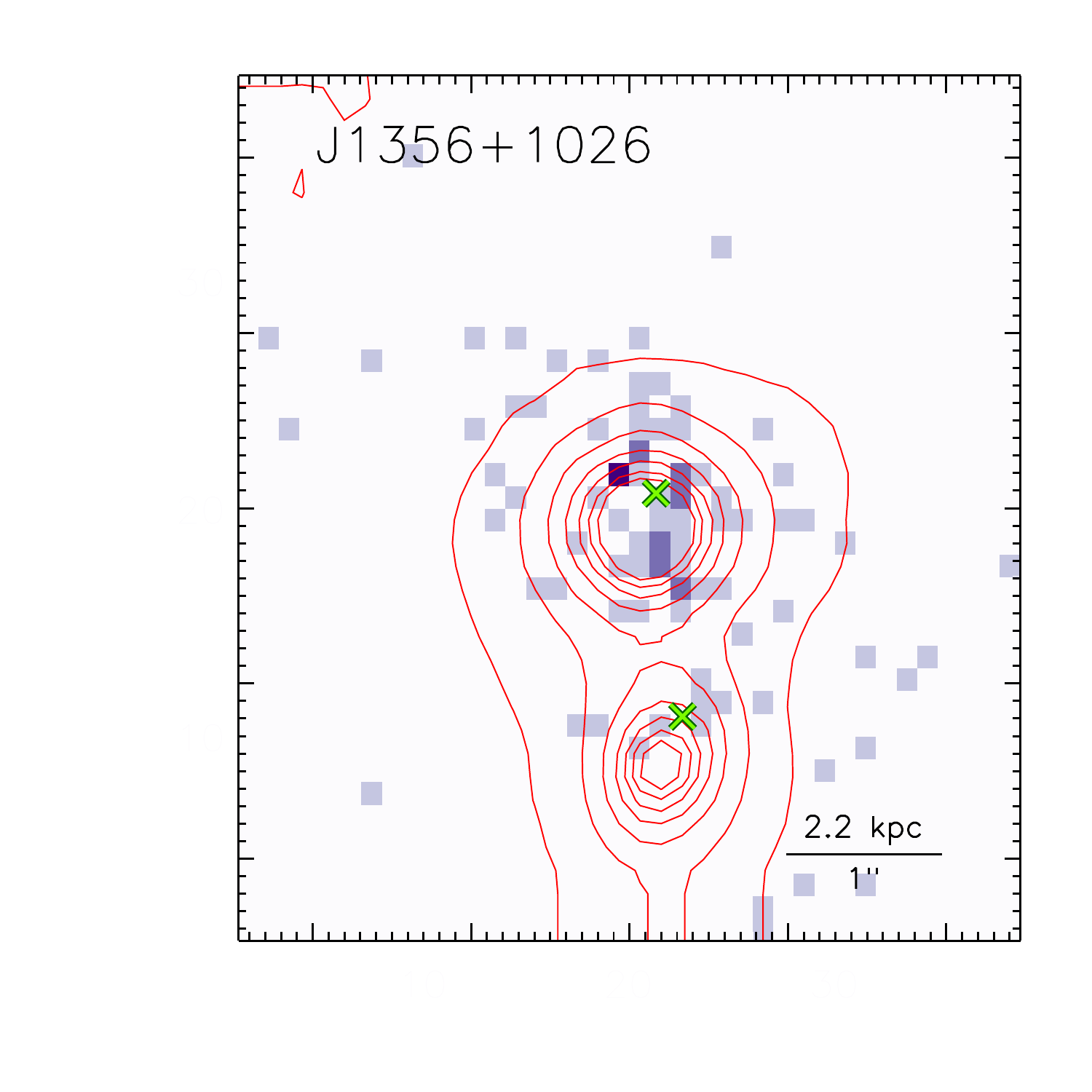}
\hspace{-.7in}
\includegraphics[height=2.5in]{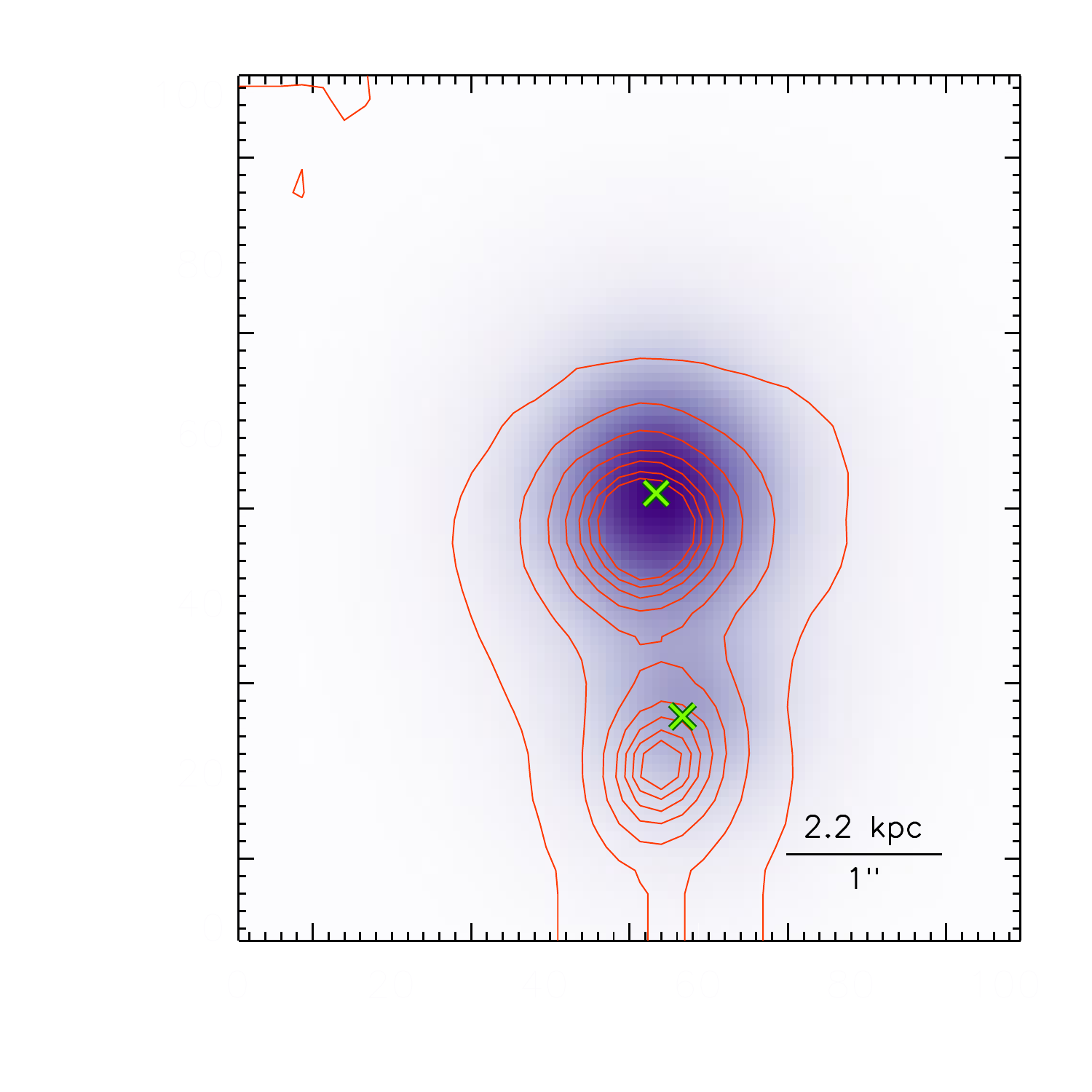}
\hspace{-.12in}
\raisebox{0.17\height}{\includegraphics[height=2.in]{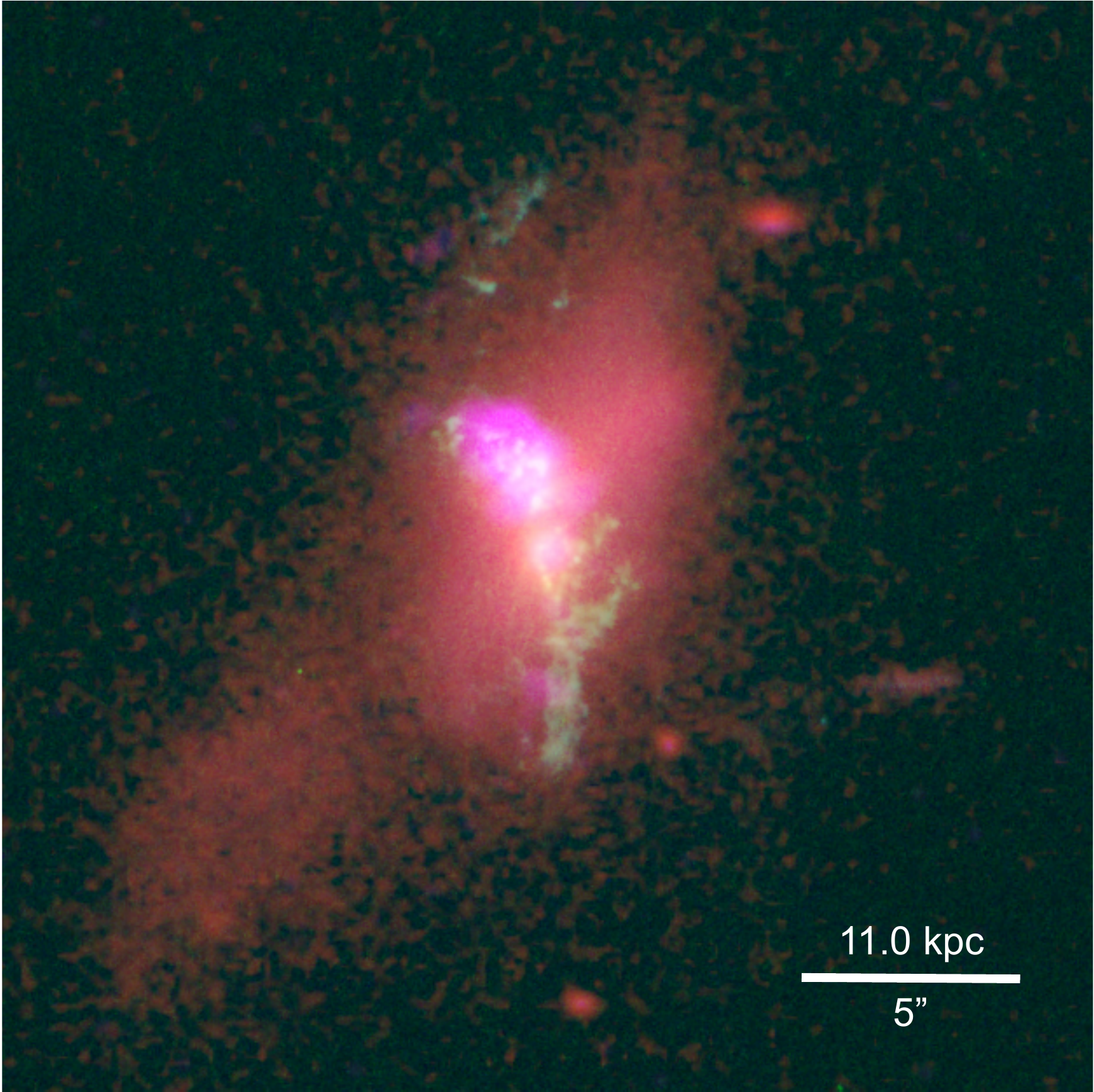}} \\
\vspace{-.4in}
\includegraphics[height=2.5in]{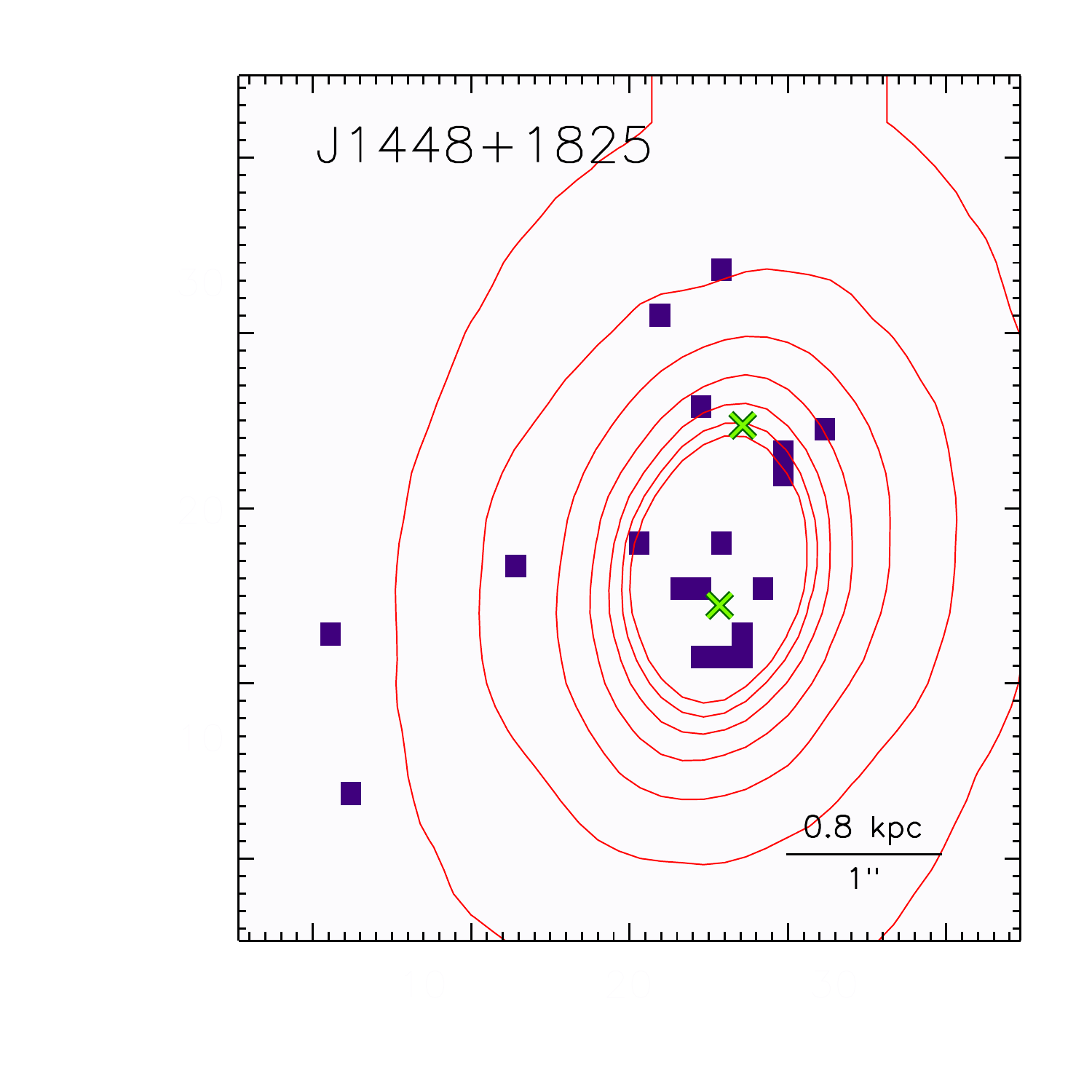}
\hspace{-.7in}
\includegraphics[height=2.5in]{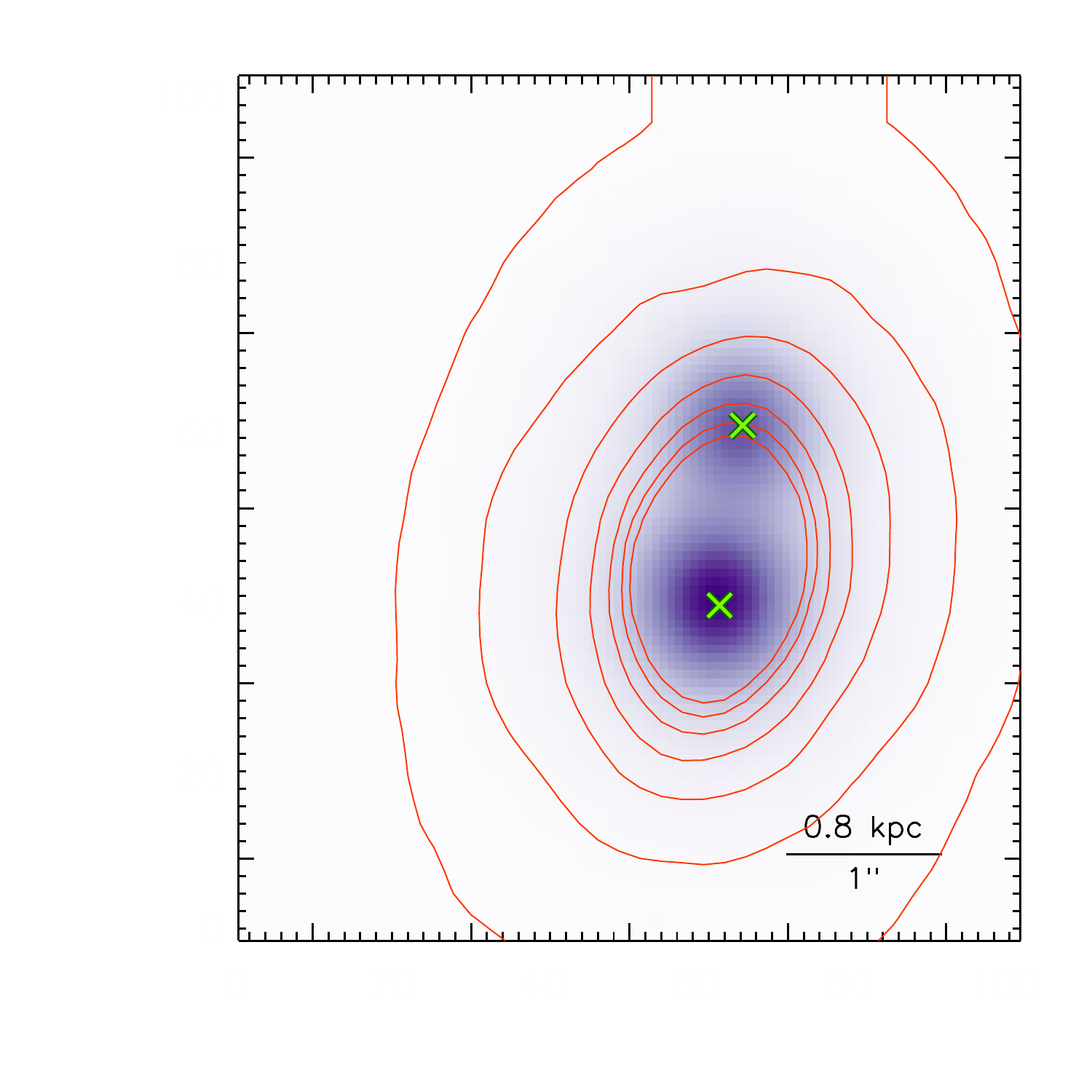}
\hspace{-.12in}
\raisebox{0.17\height}{\includegraphics[height=2.in]{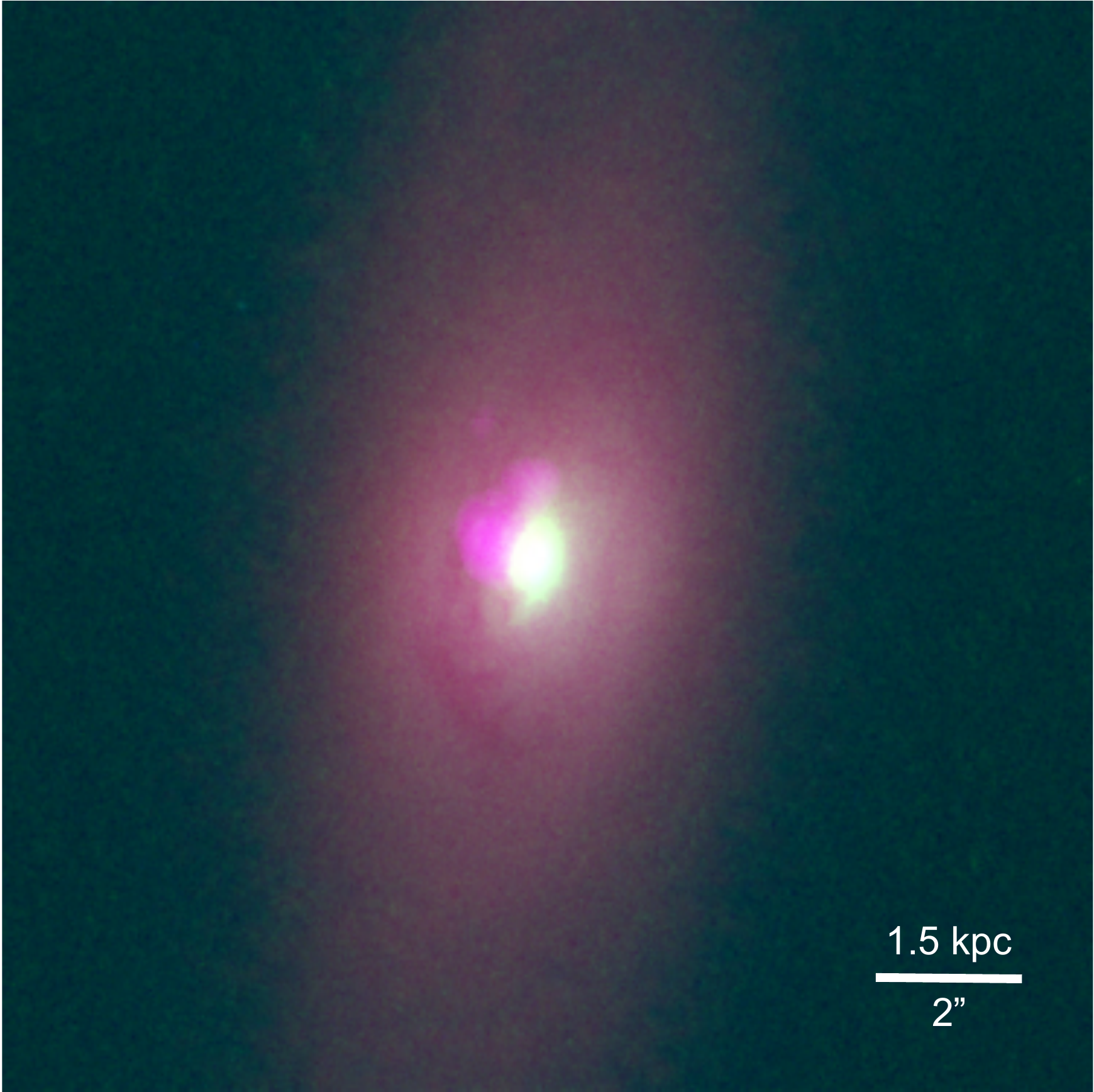}} \\
\end{center}
\vspace{-0.4in}
\caption{{\it Chandra} $0.3-8$ keV observations (left), model to the {\it Chandra} observations (middle), and combined {\it Chandra} and {\it HST} observations (right) of each of the dual AGN candidates.  In all panels, North is up and East is to the left.  The left and middle panels are centered on the coordinates of each SDSS spectrum.  The left and middle panels are all $5^{\prime\prime} \times 5^{\prime\prime}$ except for the J0841+0101 and J1322+2631 panels, which are $10^{\prime\prime} \times 10^{\prime\prime}$.  The left panels show one-fourth size {\it Chandra} pixels (purple), {\it HST} F160W contours (red), and best-fit locations of two X-ray sources (green crosses) that coincide within $3\sigma$ to the locations of two observed \oiiiw components. The middle panels show the model fits to the two X-ray sources (purple).  The right panels are four-color images of {\it HST} F160W (red), F814W (green), F438W (blue), and {\it Chandra} $0.3-8$ keV (purple, one-twelfth size pixels smoothed with a 16 pixel radius Gaussian kernel) observations of the dual AGN candidates.  The astrometric shifts described in Section~\ref{astrometry} have been applied to properly align the {\it Chandra} and {\it HST} images.}
\label{fig:images}
\end{figure*}

\begin{deluxetable*}{llllllll}
\tabletypesize{\scriptsize}
\tablewidth{0pt}
\tablecolumns{8}
\tablecaption{X-ray Spectral Fits to Dual AGN Candidates} 
\tablehead{
\colhead{SDSS Name} &
\colhead{$n_{H,exgal}$} & 
\colhead{$\Gamma$} &
\colhead{Reduced} &
\colhead{$F_{X, 0.5-8 \mathrm{keV}}$ (abs)} &
\colhead{$F_{X, 2-10 \mathrm{keV}}$ (abs)} &
\colhead{$F_{X, 0.5-8 \mathrm{keV}}$ (unabs)} &
\colhead{$F_{X, 2-10 \mathrm{keV}}$ (unabs)} \\
 & ($10^{20}$ cm$^{-2}$) & & C-stat & \tiny{($10^{-15}$ erg cm$^{-2}$ s$^{-1}$)} & \tiny{($10^{-15}$ erg cm$^{-2}$ s$^{-1}$)} & \tiny{($10^{-15}$ erg cm$^{-2}$ s$^{-1}$)} & \tiny{($10^{-15}$ erg cm$^{-2}$ s$^{-1}$)}  
 }
\startdata
J0142$-$0050 & $<0.009$ & $1.66^{+0.05}_{-0.07}$ & 0.902 & 313 & 240 & 323 & 240 \\    
\hline
J0752+2736 & $<0.072$ & $2.86^{+0.48}_{-0.43}$ & 0.195 & 5 & 1 & 5 & 1 \\
\hline
J0841+0101 & $<0.005$ & $1.17^{+0.07}_{-0.11}$ & 0.917 & 127 & 128 & 131 & 129 \\
  & $19.555^{+10.954}_{-6.363}$ & 1.7 (fixed)$^b$ & 0.816 & 116 & 164 & 494 & 358 \\
  & & $2.55^{+0.29}_{-0.29}$ & & 47 & 18  & 52 & 18 \\
\hline
J0854+5026 & $<0.318$ & $1.54^{+0.69}_{-0.68}$ & 0.143 & 7 & 8 & 10 & 8 \\
\hline
J0952+2552 & $<0.129$ & $0.19^{+0.25}_{-0.25}$ & 0.425 & 48 & 68 & 48 & 68 \\
 & $1.055^{+0.451}_{-0.342}$ & 1.7 (fixed)$^a$ & 0.467 & 26 & 26 & 37 & 27 \\
\hline
J1006+4647 & $<0.133$ & $1.55^{+0.57}_{-0.53}$ & 0.134 & 5 & 4 & 5 & 4 \\
\hline
J1126+2944 & $<0.031$ & $0.03^{+0.32}_{-0.33}$ & 0.325 & 33 & 49 & 33 & 49 \\ 
 &  $3.468^{+1.132}_{-0.966}$ & 1.7 (fixed)$^a$ & 0.385 & 28 & 34 & 58 & 42 \\ 
 &  $33.181^{+12.081}_{-9.671}$ &1.7 (fixed)$^b$ & 0.290 & 116 & 139 & 238 & 172 \\
  &  & $3.06^{+0.87}_{-0.83}$ & & 3 & 0.6 & 3 & 0.6 \\
\hline
J1239+5314 & $0.577^{+0.088}_{-0.084}$ & $1.44^{+0.10}_{-0.10}$ & 1.052 & 575 & 592 & 715 & 610 \\    
\hline
J1322+2631 &  $<0.295$ & $0.02^{+0.38}_{-0.30}$ & 0.333 & 38 & 57 & 38 & 57 \\ 
 & $1.223^{+0.528}_{-0.377}$ & 1.7 (fixed)$^a$ & 0.358 & 22 & 23 & 34 & 25 \\ 
\hline
J1356+1026 & $<0.016$ & $2.11^{+0.18}_{-0.22}$ & 0.436 & 26 & 15 & 27 & 15 \\    
\hline
J1448+1825 & $<0.049$ & $2.29^{+0.50}_{-0.47}$ & 0.169 & 6 & 3 & 6 & 3 \\   
\hline
J1604+5009 & $<0.166$ & $-0.21^{+0.24}_{-0.23}$ & 0.510 & 55 & 87 & 55 & 87 \\      
 &  $2.513^{+0.898}_{-0.678}$ & 1.7 (fixed)$^a$ & 0.562 & 36 & 41 & 65 & 47 
\enddata
\tablecomments{Column 2 shows the extragalactic column density. Column 3 shows the best-fit spectral index. Column 4 shows the reduced Cash statistic. Column 5 shows the absorbed soft X-ray flux (0.5-8 keV). Column 6 shows the absorbed hard X-ray flux (2-10 keV). Column 7 shows the unabsorbed soft X-ray flux (0.5-8 keV). Column 8 shows the unabsorbed hard X-ray flux (2-10 keV).}
\tablenotetext{a}{The best-fit spectrum was very hard and had little extragalactic absorption, so the fit was redone by freezing the spectral index to 1.7.}
\tablenotetext{b}{The best-fit spectrum was very hard and had little extragalactic absorption, and the fit with the spectral index frozen to 1.7 was also poor, so the fit was redone with two spectral indices.  This could be the case if one of the AGNs was heavily absorbed and the other was not.}
\label{tbl-3}
\end{deluxetable*}

\subsection{Chandra/ACIS X-ray Observations}
\label{chandra}

The 12 targets were observed with {\it Chandra}/ACIS over the course of two programs, GO1-12142X (PI: Gerke) and GO2-13130X (PI: Comerford).  We based the exposure times for each target on the \oiiiw flux of each peak in the target's double-peaked \oiiiw profile (measured in Section~\ref{sdss}).  We used the scaling relations between the optical \oiiiw flux and the hard X-ray ($2-10$ keV) flux, as determined for single AGNs by \cite{HE05.1}, to produce estimates of the hard X-ray flux of each peak in the double-peaked AGN.  Since these scaling relations have considerable scatter (standard deviations of 1.06 dex for Type 2 AGNs and 0.48 dex for Type 1 AGNs), we set the exposure times so that, even at 1$\sigma$ below the scaling relations, both peaks in each system should be observed with at least 15 counts.  The exposure times ranged from 20 ks to 30 ks (Table~\ref{tbl-1}).

The data were taken with the telescope aimpoint on the ACIS S3 chip in ``timed exposure'' mode and telemetered to the ground in ``faint'' mode.  We reduced the data with the latest {\it Chandra} software (CIAO\,4.6.1) and the most recent set of calibration files (CALDB\,4.6.2), and we reprocessed the data with the ``chandra\_repro'' script using the subpixel event repositioning algorithm of \cite{LI04.4}.  We searched for intervals of strong background flaring, but did not find any.

For each galaxy, we made a sky image of the field with one-tenth size {\it Chandra} pixels (0\farcs0492 per pixel) and events in the energy range of $0.3-8$ keV.  We then fit two-dimensional models to these images in Sherpa \citep{FR01.2} using modified \citet{CA79.1} statistics (``cstat'' in Sherpa).  To properly sample the parameter space, we first employed a Monte Carlo optimization method and then followed up with the \citet{NE65.1} optimization method (``simplex'' in Sherpa).  

To determine if the double \oiiiw emission components we observed in each target correspond to two X-ray sources, we constrained each X-ray model to contain two sources with a separation and orientation on the sky that were within $3\sigma$ of the measured separation and position angle of the two \oiiiw emission components on the sky (Section~\ref{longslit}).  Each model also had a fixed background component, based on a source-free region near the target.  For the source model we used a $\beta$ profile, which is a two-dimensional Lorentzian with a varying power law of the form $I(r) = A(1+(r/r_0)^2)^{-\alpha}$ and is a good match to the {\it Chandra} PSF.  We tied the power law index $\alpha$ to the $r_0$ parameter and required both components to have the same $r_0$, which was a model successfully used by \cite{PO09.1}. The components' positions and amplitudes were free parameters, and the best-fit amplitude of each component is given as a fraction of the total amplitude (Table~\ref{tbl-2}).  The best-fit positions and models of the two X-ray sources are shown for each system in Figure~\ref{fig:nohst} and Figure~\ref{fig:images}.

Next, we calculated the significance of each of the two sources in the model.  To do so, we set the amplitude of a source to zero and calculated the change in the fit statistic (``delta-cstat" in Sherpa).  We then calculated what confidence interval corresponded to that change in ``cstat".  These confidence intervals, in terms of equivalent Gaussian sigmas, are reported in Table~\ref{tbl-2}.

Out of a possible 24 individual X-ray source detections (two X-ray sources per target for 12 targets), we detected 16 X-ray sources at the $\geq1\sigma$ level.  Two sources are detected at $0.8\sigma$, one source is detected at $0.5\sigma$, and the remaining five sources are not detected and are given upper limits.  

Because the two X-ray sources in each galaxy have such small spatial separations, we cannot reliably extract spectra and make response files for each source separately.  Instead, we extracted a spectrum of both sources combined and fit the unbinned spectrum in Sherpa using cstat statistics and the simplex method.  The initial spectral model for each system was a simple absorbed power law; the absorption consisted of a fixed component with column density set to the Galactic value in the direction of each AGN \citep{DI90.1} as well as a free component with a redshift fixed at that of the host galaxy and column density allowed to float to the best-fit value; in most cases, this free component was negligible.  While this simple absorbed power-law model gave reasonable results in most cases, there were some instances where we needed to fix the spectral index $\Gamma$ at a certain value and/or add an additional component.  The results of the spectral fits are shown in Table~\ref{tbl-3}.  

For five of the systems (SDSS J0841+0101, SDSS J0952+2552, SDSS J1126+2944, SDSS J1322+2631, and SDSS J1604+5009), we found that the X-ray spectra are similarly well fit by several different models, including models with a frozen spectral index of $\Gamma=1.7$ or models with two spectral indices.  All of these different models are shown, for comparison, in Table~\ref{tbl-3}.  In these cases, we used the model with the smallest reduced Cash statistic in the analyses that follow.

We applied the results of our two-dimensional image fits to assign appropriate fractions of the total number of counts and the total flux to each X-ray component.  Then, we used the redshift to determine the distance to each system and convert the $2-10$ keV flux of each X-ray component to its $2-10$ keV luminosity (Table~\ref{tbl-4}). 

\begin{deluxetable*}{llllll}
\tabletypesize{\scriptsize}
\tablewidth{0pt}
\tablecolumns{5}
\tablecaption{Observed X-ray and \oiiiw Luminosities for Each Source} 
\tablehead{
\colhead{SDSS Name} &
\colhead{Counts} & 
\colhead{$L_{X, 2-10 \mathrm{keV}}$ (abs)} &
\colhead{$L_{X, 2-10 \mathrm{keV}}$ (unabs)} &
\colhead{Type} &
\colhead{$L_{\oiiiwn}$} \\ 
 & ($0.3-8$ keV) & ($10^{40}$ erg s$^{-1}$) & ($10^{40}$ erg s$^{-1}$) & & ($10^{40}$ erg s$^{-1}$) 
}
\startdata 
J0142$-$0050NW & $717.0^{+ 14.6}_{- 58.5}$ & $1101.3^{+ 22.5}_{- 89.9}$ & $1101.3^{+ 22.5}_{- 89.9}$ & 2 & $14.9 \pm 0.9$ \\    
J0142$-$0050SE &  $ 14.6^{+  7.3}_{-  7.3}$ & $ 22.5^{+ 11.2}_{- 11.2}$ & $  22.5^{+ 11.2}_{- 11.2}$ & 2 & $13.3 \pm 1.0$ \\    
\hline
J0752+2736NW & $ 24.8^{+  1.6}_{-  5.0}$ & $  1.4^{+  0.1}_{-  0.3}$ & $   1.4^{+  0.1}_{-  0.3}$ & 2 & $7.2 \pm 0.1$ \\
J0752+2736SE & $  1.6^{+  3.2}_{-  1.6}$ & $  0.1^{+  0.2}_{-  0.1}$ & $   0.1^{+  0.2}_{-  0.1}$ & 2 & $10.0 \pm 0.1$ \\
\hline
J0841+0101NE & $237.2^{+  0.0}_{- 14.2}$ & $576.9^{+  0.0}_{- 34.6}$ & $1192.1^{+  0.0}_{- 71.5}$ & 2 & $100 \pm 30$ \\
J0841+0101SW & $<  2.4$ & $<  5.8$ & $< 11.9$ & 2 & $100 \pm 30$ \\
\hline
J0854+5026NE & $  8.3^{+  0.0}_{-  3.2}$ & $ 17.5^{+  0.0}_{-  6.8}$ & $  19.0^{+  0.0}_{-  7.4}$ & 2 & $10.3 \pm 0.3$ \\
J0854+5026SW & $<  0.8$ & $<  1.7$ & $<  1.9$ & 2 & $10.2 \pm 0.3$ \\
\hline
J0952+2552NE  & $ 43.7^{+  0.9}_{- 12.9}$ & $2555.8^{+ 52.2}_{-756.3}$ & $2559.6^{+ 52.2}_{-757.4}$ & 1 & $110.1 \pm 20.4$ \\ 
J0952+2552SW & $  0.9^{+  2.2}_{-  0.9}$ & $ 52.2^{+130.4}_{- 52.2}$ & $  52.2^{+130.6}_{- 52.2}$ & 2 & $68.8 \pm 13.5$ \\ 
\hline
J1006+4647W & $ 10.2^{+  0.0}_{- 10.2}$ & $ 15.2^{+  0.0}_{- 15.2}$ & $  15.3^{+  0.0}_{- 15.3}$ & 2 & $14.2 \pm 0.5$ \\
J1006+4647E & $<  5.7$ & $<  8.5$ & $<  8.5$ & 2 & $17.8 \pm 0.7$ \\
\hline
J1126+2944NW & $ 23.7^{+  2.9}_{-  7.2}$ & $ 328.6^{+ 40.6}_{- 99.7}$ & $ 406.3^{+ 50.2}_{-123.3}$ & 2 & $18.5 \pm 0.3$ \\ 
J1126+2944SE & $  2.9^{+  2.9}_{-  1.6}$ & $ 40.6^{+ 40.6}_{- 22.2}$ & $  50.2^{+ 50.2}_{- 27.4}$ & 2 & $16.8 \pm 0.3$ \\ 
\hline
J1239+5314NE & $890.0^{+ 18.2}_{- 72.7}$ & $6742.4^{+137.6}_{-550.4}$ & $6947.4^{+141.8}_{-567.1}$ & n/a$^a$ & $22.6 \pm 1.9$ \\    
J1239+5314SW & $ 18.2^{+ 18.2}_{- 18.2}$ & $137.6^{+137.6}_{-137.6}$ & $ 141.8^{+141.8}_{-141.8}$ & n/a & $6.7 \pm 1.7$ \\     
\hline
J1322+2631SW &  $ 29.8^{+  0.0}_{-  5.1}$ & $316.7^{+  0.0}_{- 53.8}$ & $ 316.7^{+  0.0}_{- 53.8}$ & 2 & $20.0 \pm 0.6$ \\ 
J1322+2631NE & $<  0.6$ & $<  6.3$ & $<  6.3$ & 2 & $27.1 \pm 0.5$ \\ 
\hline
J1356+1026NE & $ 63.7^{+  9.4}_{-  7.9}$ & $  46.4^{+  6.9}_{-  5.7}$ & $  46.4^{+  6.9}_{-  5.7}$ & 2 & $387.3 \pm 0.5$ \\    
J1356+1026SW & $ 14.9^{+  5.5}_{-  4.7}$ & $ 10.9^{+  4.0}_{-  3.4}$ & $  10.9^{+  4.0}_{-  3.4}$ & 2 & $140.9 \pm 0.5$  \\ 
\hline
J1448+1825SE & $ 10.1^{+  5.7}_{-  4.7}$ & $   0.6^{+  0.3}_{-  0.3}$ & $   0.6^{+  0.3}_{-  0.3}$ & 2 & $3.4 \pm 0.1$ \\   
J1448+1825NW & $  5.7^{+  6.0}_{-  3.2}$ & $  0.3^{+  0.3}_{-  0.2}$ & $   0.3^{+  0.3}_{-  0.2}$ & 2 & $2.5 \pm 0.1$ \\   
\hline
J1604+5009NW & $ 63.6^{+  0.0}_{- 14.6}$ & $500.8^{+  0.0}_{-115.2}$ & $ 500.8^{+  0.0}_{-115.2}$ & 2 & $17.2 \pm 0.7$  \\      
J1604+5009SE & $<  0.6$ & $<  5.0$ & $<  5.0$ & 2 &  $16.7 \pm 0.6$
\enddata
\tablecomments{Column 2 shows the total $0.3-8$ keV counts in the model fit to each X-ray source.  Column 3 shows the absorbed hard X-ray luminosity (2-10 keV). Column 4 shows the unabsorbed hard X-ray luminosity (2-10 keV). Column 5 shows the AGN type, based on the optical spectra.  Column 6 shows the observed luminosity of \oiiiwn, as measured from the optical spectra.}
\tablenotetext{a}{The SDSS spectrum is classified as Type 1, but the type of each individual component is not known.}
\label{tbl-4}
\end{deluxetable*}

\subsection{HST/WFC3 F160W, F814W, and F438W  Observations}
\label{hst}

For each of the 10 systems observed with {\it Chandra} in GO2-13130X, we also obtained {\it HST}/WFC3 observations (GO 12754, PI:Comerford).  Each galaxy was observed for one orbit, and the observations covered three bands: UVIS/F438W ($B$ band), UVIS/F814W ($I$ band), and IR/F160W ($H$ band).  The exposure times are summarized in Table~\ref{tbl-1}, and a full analysis of the {\it HST} data will be presented in a forthcoming paper.  Here, we focus primarily on the stellar bulges detected in the F160W observations.

We used GALFIT V3.0 \citep{PE10.1} to model the {\it HST} images for the purposes of locating the positions of the central nuclei and for estimating flux ratios in systems with multiple nuclei.  To avoid modeling complex and irregular gas kinematics, we only ran GALFIT models on the F160W images because they do not contain significant line emission from ionized gas and are sensitive to the central stellar bulges, which is the primary component of interest.  For one exception, SDSS J1322+2631, we also ran a double-S\'ersic model on the F814W image to estimate the projected on-sky separation and position angle between H$\alpha$ emission components (see Section~\ref{longslit}). 

For each galaxy, we fit models that consist of S\'ersic profiles, AGN point sources (represented by PSFs), and a uniform sky background component.  We used TINYTIM \citep{KR11.1} to generate the PSF images used for input to GALFIT, and we re-binned them to the same pixel scale as the {\it HST}/F160W images.  In all cases we adopted the simplest model that provided a satisfactory reduced $\chi^{2}$ statistic and that successfully converged on the central components.  GALFIT outputs the positions of the sources and their integrated magnitudes, which we converted to a luminosity ratio ($L_{*,1}/L_{*,2}$) for the systems with two stellar bulges.  We then use this luminosity ratio as a proxy for the merger ratio (Section~\ref{major_mergers}).

In cases where a galaxy merger has clearly occurred, the observed morphologies may differ significantly from the idealized profiles of the components available in GALFIT.  However, our goals of this modeling are to estimate the positions of the stellar bulges and the integrated fluxes of the stellar bulges, both of which are successfully accomplished using GALFIT.  In all but two cases, S\'ersic components alone were sufficient to converge on the central stellar bulges, and no AGN point sources were required.  The exceptions are SDSS J0841+0101SW (the S\'ersic component is offset due to an extended tidal tail) and SDSS J1126+2944SE (too faint and compact for a S\'ersic component).   

For errors on the positions of the stellar bulges, we used the uncertainties generated by GALFIT.  As discussed in \cite{PE10.1}, the uncertainties are purely statistical and represent lower limits on the errors since they do not account for deviations between the galaxy morphologies and the idealized profiles.

\subsection{Astrometry}
\label{astrometry}

In order to identify whether any {\it Chandra} sources align with the stellar bulges seen in the {\it HST} data, we need an understanding of the astrometric errors.  First, we searched for strongly-detected, off-nuclear sources common in both the {\it HST}/F160W and {\it Chandra}/ACIS fields of view (FOVs) to determine the astrometric correction between the two images and the corresponding astrometric uncertainty.  We used the F160W images because they have the largest FOV ($70^{\prime\prime} \times 62^{\prime\prime}$) of the three {\it HST} filters that we used, which optimizes the number of sources available for registration.  To detect strong sources and find their positional centroids in the {\it HST} images, we used Source Extractor \citep{BE96.1} with a detection threshold of $3\sigma$.  The uncertainties associated with the Source Extractor positions are based on the rms errors after background subtraction.  

To detect and measure positions of the X-ray sources, we used ``wavdetect" to find the positions for all sources in the {\it Chandra}/ACIS images with a detection threshold of $10^{-8}$ (``sigthresh").  The positional errors are $1\sigma$ uncertainties based on the inverse variance of the flux.  The {\it Chandra}/ACIS FOV is $504^{\prime\prime} \times 504^{\prime\prime}$, and we required that X-ray sources used for registration be within $3'.0$ of the aimpoint, where the 99\% absolute uncertainty radius is $\lesssim 0\farcs8$.  For one exception, SDSS J1006+4647, it was necessary to expand the radius of {\it Chandra}/ACIS detections to $4.'5$ to find any matches.  We visually inspected all optical and X-ray matches to ensure that they are associated with the same source.  
 
Due to the small number of detected {\it Chandra}/ACIS sources and the relatively small {\it HST}/F160W FOV, we were able to directly register the {\it HST}/F160W and {\it Chandra}/ACIS images for only two of the targets (SDSS J0841+0101 and SDSS J1322+2631).  For the remaining eight targets, we used the SDSS fields as intermediate frames because the SDSS FOV ($810^{\prime\prime} \times 560^{\prime\prime}$) is comparable to that of {\it Chandra}/ACIS and significantly increases the number of matched sources.  We measured positions of the sources in the SDSS frames using Source Extractor with the same parameters that we used on the {\it HST}/F160W frames.  We then determined astrometric corrections from the {\it HST} to SDSS images and astrometric corrections from the SDSS images to the {\it Chandra} images.  The difference between these two measurements yields the astrometric correction between {\it HST} and {\it Chandra}.     

The errors in the spatial offsets between the positions of a matched source in two images were determined by adding in quadrature the positional errors from the detections in each image.  Once all of the matched sources were identified and their individual offsets measured, 
we removed any sources that had offsets $>1\sigma$ from the mean offset.  We then determined the final astrometric correction between any two images by taking the error-weighted average of all of the remaining offsets.  To further reduce the astrometric uncertainty, we determined these corrections separately using three SDSS filters with the highest sensitivities ($g$, $r$, and $i$) and adopted the error-weighted average of those three corrections as the final correction.  Table~\ref{tbl-5} shows the corresponding astrometric uncertainty for each object.

\begin{deluxetable}{lllll} 
\tabletypesize{\scriptsize}
\tablewidth{0pt}
\tablecolumns{5}
\tablecaption{Astrometry Measurements} 
\tablehead{
\colhead{SDSS Name} &
\colhead{$n_{CS}$} & 
\colhead{$n_{HS}$} & 
\colhead{$n_{HC}$} & 
\colhead{$\Delta\theta_{ast}$ ($\prime\prime$)} 
 }
\startdata  
J0142$-$0050 & 2 & 4 & 0 & 0.0771 \\    
J0841+0101 & 3 & 5 & 1 & 0.5882  \\ 
J0854+5026 &  3 & 2 & 0 & 0.2307 \\
J0952+2552 & 3 & 1 & 0 & 0.0968 \\
J1006+4647 &  1 & 1 & 0 & 0.3278 \\
J1126+2944 &  1 & 1 & 0 & 0.3018 \\ 
J1239+5314 & 3 & 3 & 0 & 0.1868 \\    
J1322+2631 & 2 & 1 & 1 & 0.7779 \\ 
J1356+1026 & 1 & 1 & 0 & 0.4164 \\    
J1448+1825 &  2 & 4 & 0 & 0.1192
\enddata
\tablecomments{Column 2 shows the number of sources matched between {\it Chandra} and SDSS $gri$ images. Column 3 shows the number of sources matched between {\it HST}/F160W and SDSS $gri$ images.  Column 4 shows the number of sources matched between {\it HST}/F160W and {\it Chandra} images.  Column 5 shows the astrometric accuracy measurement based on matching these sources.}
\label{tbl-5}
\end{deluxetable}

\subsection{WISE Observations}
\label{wise}

\begin{figure}
\begin{center}
\includegraphics[width=3.5in]{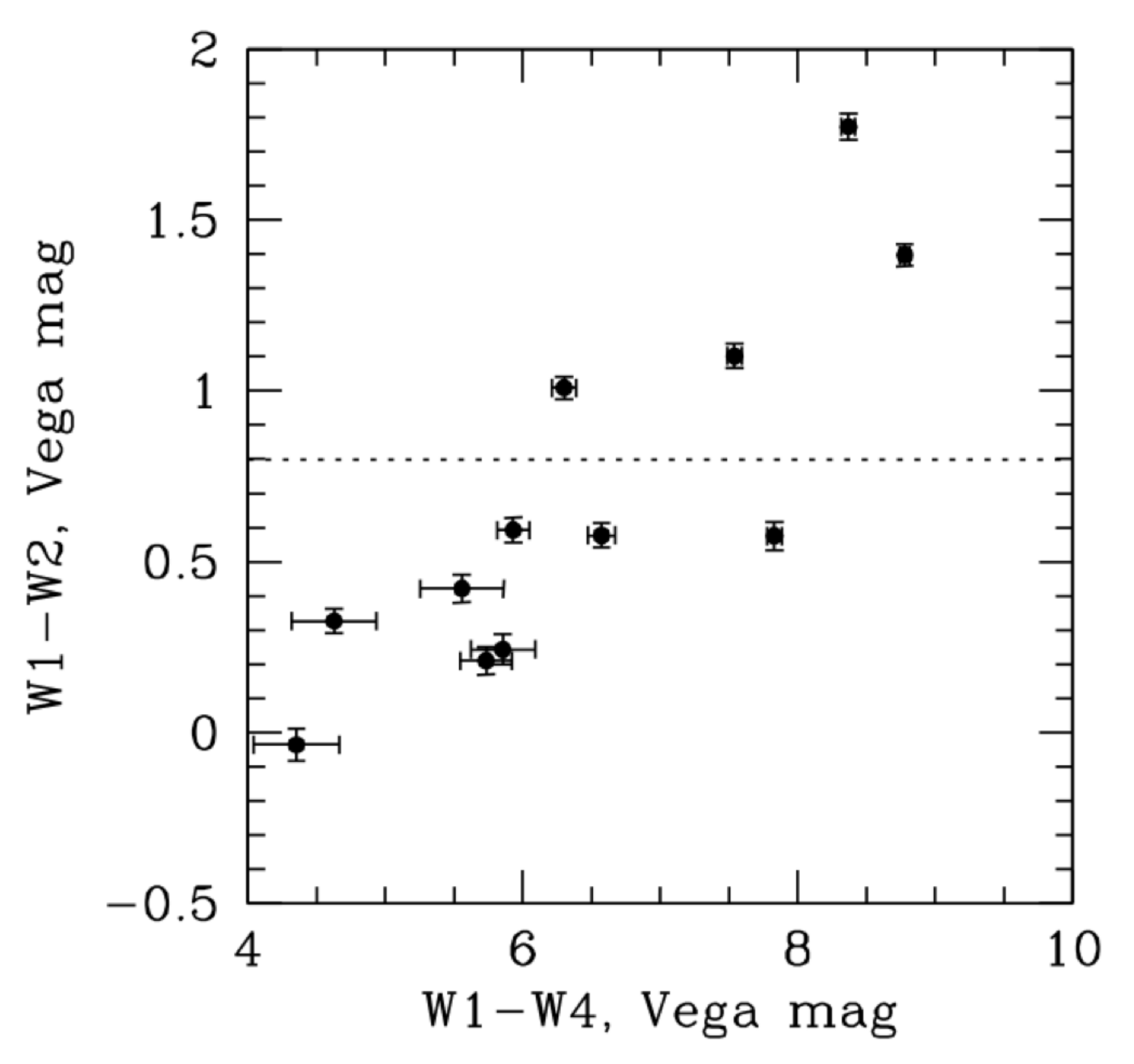}
\end{center}
\vspace{-.1in}
\caption{Mid-infrared {\it WISE} color-color plot of $W1-W2$ ([3.4]-[4.6]) against $W1-W4$ ([3.4]-[22]) for our 12 targets.  The dashed horizontal line indicates the AGN criterion of \cite{ST12.2}.   The four targets above the dashed line have mid-infrared fluxes dominated by AGN emission, while the eight targets below the dashed line have mid-infrared fluxes dominated by star formation.}
\label{fig:wise}
\end{figure}

We also examine the mid-infrared colors from the {\it Wide-field Infrared Survey Explorer} ({\it WISE}) for all of our targets to determine the relative contributions of AGN activity and star formation to our observed X-ray luminosities.   {\it WISE} mapped the full sky in four bands at 3.4 $\mu$m, 4.6 $\mu$m, 12 $\mu$m, and 22 $\mu$m ($W1$, $W2$, $W3$, and $W4$, respectively), and the {\it WISE} colors are useful because they can identify even heavily obscured AGNs.  AGNs are identified by $W1-W2 \geq 0.8$, as determined by comparing the {\it Spitzer} mid-infrared color-color diagnostics for AGNs \citep{ST05.1} to the {\it WISE} colors of COSMOS galaxies \citep{ST12.2} and combining the {\it WISE} data with SDSS optical observations \citep{YA13.1}.  While the $W1-W2$ selection of AGNs is clean (i.e., few false positives), it is also very incomplete (e.g., \citealt{MA13.2}).  Since the most obscured AGNs may be best identified at longer wavelengths such as $W4$, we plot $W1-W2$ against $W1-W4$ in Figure~\ref{fig:wise}.

As shown in Figure~\ref{fig:wise}, we find that four of our systems (SDSS J0841+0101, SDSS J0952+2552, SDSS J1322+2631, and SDSS J1356+1026) have AGN-dominated mid-infrared fluxes.  Of the eight objects with mid-infrared fluxes that are consistent with significant contributions from star formation, the two objects with the reddest $W1-W4$ colors (SDSS J0752+2736 and SDSS J1006+4647) may be highly obscured AGNs.  For all eight systems with $W1-W2 < 0.8$, we estimate the contribution to the observed X-ray emission from star formation.  First, we calculate the rest-frame 12 $\mu$m monochromatic luminosity for each system, and then convert this luminosity to the total infrared luminosity from star formation \citep{CA06.1, CH01.2, EL02.1}.  Then, we use the scalings of \cite{BE03.4} to convert the infrared luminosity due to star formation to a star formation rate.  Finally, we determine the expected $0.5-8$ keV X-ray luminosity for this star formation rate \citep{MI14.1}.

The results are given in Table~\ref{tbl-6}.  We find that only two systems, SDSS J0752+2736 and SDSS J1006+4647, could have a significant contribution ($40-50\%$) of star formation to the observed {\it Chandra} X-ray luminosities.

\begin{deluxetable*}{lllll} 
\tabletypesize{\scriptsize}
\tablewidth{0pt}
\tablecolumns{5}
\tablecaption{Star Formation Contributions to X-ray Observations} 
\tablehead{
\colhead{SDSS Name} &
\colhead{Interpretation of} & 
\colhead{SFR} & 
\colhead{$L_{X, \mathrm{SF}, 0.5-8 \mathrm{keV}}$ (unabs)} & 
\colhead{$L_{X, \mathrm{SF}, 0.5-8 \mathrm{keV}}$ (unabs) / } \\
\colhead{} &
\colhead{{\it WISE} colors} &
\colhead{(M$_{\odot}$ yr$^{-1}$)} &
\colhead{($10^{40}$ erg s$^{-1}$)} &
\colhead{$L_{X, \mathrm{observed}, 0.5-8 \mathrm{keV}}$ (unabs) }
 }
\startdata  
J0142$-$0050 & SF & 9.92 & 4.0 & 0.003 \\    
J0752+2736 & SF & 6.73 & 2.7 & 0.47 \\
J0841+0101 & AGN & & & \\ 
J0854+5026 & SF & 2.25 & 0.9 & 0.04 \\
J0952+2552 & AGN & & & \\ 
J1006+4647 & SF & 20.65 & 8.1 & 0.41 \\
J1126+2944 & SF & 1.97 & 0.8 & 0.001 \\ 
J1239+5314 & SF & 52.06 & 20.9 & 0.003 \\    
J1322+2631 & AGN & & & \\ 
J1356+1026 & AGN & & & \\ 
J1448+1825 & SF & 0.59 & 0.2 & 0.12 \\
J1604+5009 & SF & 4.26 & 1.7 & 0.005
\enddata
\tablecomments{Column 2 shows whether the {\it WISE} colors are dominated by star formation (SF) or AGN. Column 3 shows the star formation rate (SFR) inferred from the infrared luminosity.  Column 4 shows the unabsorbed $0.5-8$ keV X-ray luminosity, predicted from the SFR.  Column 5 shows the fraction of the total unabsorbed $0.5-8$ keV X-ray luminosity observed with {\it Chandra} that could be due to star formation.}
\label{tbl-6}
\end{deluxetable*}

\subsection{Photoionization}
\label{photoionization}

Finally, we consider the contribution of photoionized gas to our observed X-ray luminosities.  We are searching for double X-ray sources corresponding to dual AGNs, but a single AGN photoionizing the surrounding gas could also produced extended soft X-ray emission (e.g., \citealt{BI06.1,WA11.3}).  The observational signatures of such photoionized gas include a soft X-ray spectrum and a small value for the ratio of soft X-ray luminosity to \oiiiw luminosity ($\sim$ 0.1 to 0.3; \citealt{BI06.1,WA11.3}).  We examine our dual AGN candidates for any clear examples of X-ray and \oiiiw emission sources that could be produced by photoionization (Section~\ref{nature}).

\section{Nature of the 12 Dual AGN Candidates}
\label{nature}

Here, we combine the \oiiiw long-slit spectra, {\it Chandra} observations, and {\it HST} observations for each of the 12 AGNs with double-peaked narrow emission lines to determine its physical nature.  We conclude that one is a dual AGN system, five are dual AGNs or offset AGNs, and six are most likely single AGNs, as discussed below.

\subsection{One Dual AGN System}
\label{dualagn}

We define dual AGNs using a combination of our \oiiiwn, {\it Chandra}, and {\it HST} observations of each candidate.  By this approach, a galaxy hosts dual AGNs if it contains two X-ray sources (each detected at $\geq 2\sigma$) that have the same spatial separation and orientation as the two \oiiiw emission components (to within $3\sigma$) and that coincide with two stellar bulges (to within $3\sigma$ in astrometric accuracy; this is given by $\Delta \theta/\Delta \theta_{ast}$ in Table~\ref{tbl-7}).  Here, our interpretation is that the two stellar bulges are the remnants of the merging galaxies and each hosts an AGN that is visible in \oiiiw and in X-ray.

We find two systems that fit these criteria: SDSS J1126+2944 and SDSS J1356+1026.  However, SDSS J1356+1026 has known soft X-rays associated with an outflowing bubble that complicate the identification of dual AGNs.  To be conservative, we categorize it as either dual AGNs or an offset AGN (Section~\ref{1356}).

This leaves SDSS J1126+2944 as the only confirmed dual AGN system, and we present a summary of its properties in Table~\ref{tbl-8}.  The line-of-sight velocity separations between the two AGNs' \oiiiw emission components are taken from \cite{CO12.1}. We measured three different projected spatial separations between the AGNs: the separation between \oiiiw emission components (Table~\ref{tbl-2}), the separation between X-ray sources (Section~\ref{chandra}), and the separation between the stellar bulges hosting the AGNs (Section~\ref{hst}).   In Table~\ref{tbl-8} we show the mean of these three separation measurements, weighted by their inverse variances.  

SDSS J1126+2944 is the fourth dual AGN system to be identified, via spatial resolution of two AGNs in X-ray or radio, from the parent sample of galaxies with double-peaked narrow AGN emission lines in their SDSS spectra.   
One dual AGN system at $z=0.39$ was confirmed by double radio cores observed by the Jansky Very Large Array, where the separation between the AGNs is $1\farcs4$ or 7.4 kpc \citep{FU11.3}.  Two additional dual AGNs were identified by {\it Chandra} observations of double X-ray sources: one system at $z=0.18$ with the AGN X-ray sources separated by $0\farcs82$ or 2.5 kpc, and one system at $z=0.13$ with the AGN X-ray sources separated by $3\farcs03$ or 7.0 kpc \citep{LI13.1}.  The new dual AGN system presented here, SDSS J1126+2944, is unique because it has a lower redshift ($z=0.102$) and a smaller separation (2.19 kpc) than these other three dual AGN systems.  It is also unique because one of the two stellar bulges in the merging system is much fainter than the other (Section~\ref{1126}).

For the special case of galaxies with SDSS spectra that exhibit double-peaked narrow AGN emission lines, where both \oiiiw peaks have fluxes $>8 \times 10^{-16}$ erg cm$^{-2}$ s$^{-1}$, and where two \oiiiw emission components are spatially separated by $>0\farcs75$ on the sky, we conclude that the dual AGN fraction is at least $8\%$ (1/12).

We discuss SDSS J1126+2944 individually in the subsection below.  

\begin{deluxetable*}{lllllllll}
\tabletypesize{\scriptsize}
\tablewidth{0pt}
\tablecolumns{8}
\tablecaption{Chandra and HST Positions of Each Source} 
\tablehead{
\colhead{SDSS Name} &
\colhead{RA$_{Chandra}^a$} & 
\colhead{DEC$_{Chandra}^a$} &
\colhead{RA$_{HST}$} &
\colhead{DEC$_{HST}$} &
\colhead{$\Delta \theta (^{\prime\prime})$} &
\colhead{$\Delta \theta / \Delta \theta_{ast}$} &
\colhead{Galaxy$_1$:} &
\colhead{Merger Ratio} \\
 & & & & & & & Galaxy$_2$ & ($L_{*,1}/L_{*,2}$)
}
\startdata
J0142$-$0050NW & 01:42:09.008 & $-$00:50:49.98 & 01:42:09.003 & $-$00:50:49.94 & 
     0.0850 & 1.10 & & \\    
J0142$-$0050SE & 01:42:09.034 & $-$00:50:50.59 & n/a & n/a & n/a & n/a & & \\ 
\hline
J0841+0101NE & 08:41:35.079 & +01:01:56.26 & 08:41:35.090 & +01:01:56.40 & 
      0.2164 & 0.37 & NE:SW & 1.4 \\
J0841+0101SW & 08:41:34.913 & +01:01:53.90 & 08:41:34.894 & +01:01:53.84 & 
      0.2912 & 0.50 & & \\
\hline
J0854+5026NE & 08:54:16.808 & +50:26:32.08 & 08:54:16.776 & +50:26:32.10 & 
      0.4804 & 2.08 & & \\
J0854+5026SW & 08:54:16.761 & +50:26:31.43 & n/a & n/a & n/a & n/a & & \\ 
\hline
J0952+2552NE & 09:52:07.608 & +25:52:57.29 & 09:52:07.611 & +25:52:57.23 & 
     0.0750 & 0.77 & NE:SW & 2.0 \\
J0952+2552SW & 09:52:07.597 & +25:52:56.27 & 09:52:07.604 & +25:52:56.24 & 
      0.1092 & 1.13 & & \\
\hline
J1006+4647W & 10:06:54.222 & +46:47:16.62 & 10:06:54.237 & +46:47:16.71 & 
      0.2423 & 0.74 & & \\
J1006+4647E & 10:06:54.300 & +46:47:16.63 & n/a & n/a & n/a & n/a & & \\ 
\hline
J1126+2944NW & 11:26:59.558 & +29:44:42.48 & 11:26:59.569 & +29:44:42.63 & 
      0.2230 & 0.74 & NW:SE & 460 \\
J1126+2944SE & 11:26:59.622 & +29:44:41.69 & 11:26:59.631 & +29:44:41.92 & 
      0.2667 & 0.88 & & \\
\hline
J1239+5314NE & 12:39:15.452 & +53:14:15.14 & 12:39:15.454 & +53:14:15.14 & 
     0.0300 & 0.16 & SW:NE & 1.1 \\ 
J1239+5314SW & 12:39:15.417 & +53:14:14.77 & 12:39:15.454 & +53:14:15.14 & 
      0.6670 & 3.57 & & \\  
\hline
J1322+2631SW & 13:22:31.803 & +26:31:58.72 & 13:22:31.796 & +26:31:58.65 & 
      0.1262 & 0.16 & SW:NE & 6.4 \\ 
J1322+2631NE & 13:22:31.976 & +26:31:59.23 & 13:22:31.968 & +26:31:59.02 & 
      0.2419 & 0.31 & & \\ 
\hline
J1356+1026NE & 13:56:46.115 & +10:26:08.83 & 13:56:46.128 & +10:26:08.61 & 
      0.2940 & 0.71 & SW:NE & 3.0 \\  
J1356+1026SW & 13:56:46.104 & +10:26:07.59 & 13:56:46.120 & +10:26:07.29 & 
      0.3842 & 0.92 & & \\
\hline
J1448+1825SE & 14:48:04.183 & +18:25:37.68 & 14:48:04.186 & +18:25:37.75 & 
     0.0832 & 0.70 & & \\
J1448+1825NW & 14:48:04.173 & +18:25:38.68 & n/a & n/a & n/a & n/a & & 
\enddata
\tablecomments{Columns 2 and 3 show coordinates measured from {\it Chandra}/ACIS observations. Columns 4 and 5 show coordinates measured from {\it HST}/WFC3/F160W observations. Column 6 shows the difference between the {\it Chandra} and {\it HST} positions $\Delta \theta$.  Column 7 shows the ratio of the difference between the {\it Chandra} and {\it HST} positions $\Delta \theta$ and the astrometric accuracy $\Delta \theta_{ast}$.  For the galaxies with two stellar bulges detected in {\it HST}/F160W observations, column 8 identifies the more luminous bulge (Galaxy$_1$) and the less luminous bulge (Galaxy$_2$).  Column 9 shows the merger ratio, which is estimated as the ratio of the stellar bulge luminosities.}
\tablenotetext{a}{The astrometric shifts described in Section~\ref{astrometry} have been applied to the {\it Chandra} source positions.}
\label{tbl-7}
\end{deluxetable*}

\subsubsection{SDSS J1126+2944: Dual AGNs}
\label{1126}

SDSS J1126+2944 is a Type 2 AGN with double-peaked emission lines \citep{WA09.1,LI10.1}, and follow-up observations showed that the double peaks are produced by two \oiiiw emission components separated by $0\farcs94$, or 1.76 kpc \citep{CO12.1}.  These two sources, J1126+2944NW and J1126+2944SE, spatially coincide with two components in the {\it HST}/F160W observations, to within $0.9$ times the astrometric error.  

J1126+2944NW is detected at the $>5\sigma$ level in $0.3-8$ keV X-rays, while J1126+2944SE is detected at $2.3\sigma$.  The best fit to the X-ray spectrum is a two-component spectral index model, with one spectral index fixed at $\Gamma=1.7$ and the other allowed to vary, yielding a best-fit value of $\Gamma=3.06$ (Table~\ref{tbl-3}).  This two-component spectral index model could be explained by the combined X-ray contribution of the two AGNs, where one is heavily absorbed and the other is not.  However, we note that there are so few X-ray counts that the spectrum is almost as well fit by an extremely hard spectrum ($\Gamma=0.03$) or a fixed single spectral index of $\Gamma=1.7$.  Deeper X-ray observations are needed to distinguish between these scenarios.

The fascinating component of this galaxy is J1126+2944SE.  J1126+2944SE is faint in the F160W image (460 times fainter than J1126+2944NW), and was not identified in a 120 second image taken in the near-infrared ($K_p$) with the Keck laser guide star adaptive optics system \citep{FU12.1}.  However, it is also detected in our F814W and F438W images.  We detect J1126+2944SE in the {\it HST}/F160W data with a significance of $2.4\sigma$ above the primary galaxy background, and we have marked its location with an arrow in Figure~\ref{fig:images}.  The merger ratio of J1126+2944NW to J1126+2944SE is 460:1.

J1126+2944SE is also pointlike.  In the F160W image, we fit it with a PSF model using GALFIT and found that 57\% of the total PSF intensity is contained in the central 4 pixels.  An upper limit on the half-light radius is therefore 1 pixel, which corresponds to 280 pc ($0\farcs13$).

It is possible that J1126+2944SE is a foreground or background source that is not physically associated with the main galaxy.  Since we observe an \oiiiw emission component at the location of J1126+2944SE that has a redshift close to systemic, we conclude that J1126+2944SE is most likely associated with the galaxy.

Since J1126+2944SE is pointlike, it can be classified as an ultraluminous X-ray source (ULX).  ULXs are variable off-nuclear X-ray sources, and some of them may be produced by intermediate mass black holes (e.g., \citealt{KA03.4}).  However, the observed \oiii luminosity ($[1.68 \pm 0.03] \times 10^{41}$ erg s$^{-1}$) and $2-10$ keV luminosity ($4.06^{+4.06}_{-2.22} \times 10^{41}$ erg s$^{-1}$) are significantly above what are typically measured in ULXs (e.g., \citealt{PO10.1, CS11.1, WA11.2}).  The $L_{X, 2-10 \mathrm{keV}} / L_{\oiiiwn}$ ratio for J1126+2944SE is also within the scatter observed for Type 2 AGNs \citep{HE05.1}.

J1126+2944SE bears a striking resemblance to HLX1, a hyperluminous X-ray source in the galaxy ESO243-49.  HLX1 is a bright, pointlike X-ray source with a pointlike optical counterpart, and it is spatially offset from the center of the host galaxy \citep{FA09.1,GO09.1,WE10.1}.  The variable $0.2-10$ keV X-ray luminosity reaches up to $1.2 \times 10^{42}$ erg s$^{-1}$, which implies a black hole mass $> 500 \, M_\odot$.  HLX1 is seen in an edge-on disk galaxy, just as SDSS J1126+2944 is an almost edge-on disk galaxy.  It has been suggested that HLX1 is the central intermediate mass black hole of a dwarf galaxy that was consumed by ESO243-49. Similarly, J1126+2944SE could be the remnant black hole from a minor merger interaction with SDSS J1126+2944.  In fact, SDSS J1126+2944 shows evidence of tidal disruption in the northwest region of the galaxy.

\begin{deluxetable*}{llllllllll}
\tabletypesize{\scriptsize}
\tablewidth{0pt}
\tablecolumns{10}
\tablecaption{Summary of the One Confirmed Dual AGN System and Five Dual/Offset AGNs} 
\tablehead{
\colhead{SDSS Name} &
\colhead{Redshift} & 
\colhead{Type} & 
\colhead{$\Delta v$ (km s$^{-1}$)} & 
\colhead{$\Delta x$ (kpc)} &
\colhead{Galaxy$_1$:} &
\colhead{Merger Ratio} &
\colhead{$L_{bol,1}/$} &
\colhead{$f_{Edd,1}/$} &
\colhead{Classification} \\
 & & & & & Galaxy$_2$ & $(L_{*,1}/L_{*,2})$ & $L_{bol,2}$ & $f_{Edd,2}$
 }
\startdata
J1126+2944 & 0.102 & 2 & $310 \pm 5$ & $2.19 \pm 0.02$ & NW:SE & 460 & 1.1 & 0.002 & Dual AGNs \\
\hline
J0841+0101 & 0.111 & 2 & 0 & $8.00 \pm 0.01$ & NE:SW & 1.4 & 1.0 & 0.726 & Dual/Offset AGN \\
J0952+2552 & 0.339 & 1/2$^a$ & $438 \pm 18$ & $4.82 \pm 0.05$ & NE:SW & 2.0 & 1.6 & 0.811 & Dual/Offset AGN \\
J1239+5314 & 0.201 & n/a$^b$ & $359 \pm 23$ & $4.19 \pm 0.01$ & SW:NE & 1.1 & 0.3 & 0.265 & Dual/Offset AGN \\    
J1322+2631 & 0.144 & 2 & $401 \pm 5$ & $5.99 \pm 0.01$ & SW:NE & 6.4 & 0.7 & 0.116 & Dual/Offset AGN \\
J1356+1026 & 0.123 & 2 & $454 \pm 12$ & $2.93 \pm 0.02$ & SW:NE & 3.0 & 0.4 & 0.121 & Dual/Offset AGN 
\enddata
\tablecomments{Column 3 shows the AGN type, based on the optical spectra. Column 4 shows the line-of-sight velocity separation between the dual AGNs, given as the velocity separation between \oiiiw peaks.  Column 5 shows the projected spatial separation on the sky.  For the dual AGNs, this is given as the weighted mean of the spatial separations that we measured between \oiiiw emission components, X-ray sources, and stellar bulges.  For the dual/offsetAGNs, this is given as the separation between the stellar bulges.  Column 6 identifies the more luminous bulge (Galaxy$_1$) and the less luminous bulge (Galaxy$_2$) in the system.   Column 7 shows the merger ratio of the two stellar bulges observed in {\it HST}/F160W, as determined by their luminosity ratio.  Column 8 shows the ratio of the AGN bolometric luminosities, while column 9 shows the ratio of the Eddington fractions.  Column 10 shows our classification of each system, where a dual/offset AGN is a system that is either dual AGNs or an offset AGN.}
\tablenotetext{a}{J0952+2552NE is a Type 1 AGN, and J0952+2552SW is a Type 2 AGN (Section~\ref{0952}).}
\tablenotetext{b}{J1239+5314NE is likely a Type 1 AGN, and J1239+5314 is a Type 1 or a Type 2 AGN (Section~\ref{1239}).}
\label{tbl-8}
\end{deluxetable*}

Another analogy to J1126+2944SE may be M60-UCD1, which is an ultra-compact dwarf galaxy hosting a black hole of mass $2.1 \times 10^7$ $M_\odot$ \citep{SE14.1}. The high black hole mass fraction suggests that M60-UCD1 may be the remnant of a galaxy that was once more massive, but was tidally stripped during an encounter with the galaxy M60.  Likewise, J1126+2944SE may be a supermassive black hole in the remnant stellar nucleus of a galaxy that was tidally stripped by SDSS J1126+2944.  Ultra-compact dwarf galaxies such as M60-UCD1 have half-light radii of $3 - 50$ pc.  We measure an upper limit of 280 pc for the half-light radius of J1126+2944SE, which indicates that its actual size may fall in the regime of ultra-compact dwarf galaxies.

The question of interest is the mass of the black hole in J1126+2944SE.  Is it an intermediate mass black hole that originated in a dwarf galaxy, or is it a supermassive black hole that originated in a more massive galaxy?  Both scenarios require that the progenitor galaxy lost much of its mass to tidal stripping, and SDSS J1126+2944 does show signs of tidal disruption itself.  Coincident measurements of J1126+2944SE's X-ray and radio fluxes (e.g., \citealt{ME03.5,GU09.1}) would enable the black hole mass measurement necessary to decipher the nature of this peculiar source.

\subsection{Five Dual/Offset AGNs}

We define the ``dual/offset AGNs" as the galaxies that are either dual AGNs or offset AGNs, where offset AGNs are galaxy mergers where only one of the two merging stellar bulges hosts an AGN (e.g., \citealt{BA08.1,CO14.1}). Dual/offset AGN systems have two \oiiiw emission components with the same spatial separation and orientation on the sky as two stellar bulges (to within $3\sigma$), but two coincident X-ray sources are not detected at the $\geq 2\sigma$ level.  The four dual/offset AGNs that fit these criteria are SDSS J0841+0101, SDSS J0952+2552, SDSS J1239+5314, and SDSS J1322+2631, and in all four cases one stellar bulge has an X-ray source detected at $>5\sigma$ but the other stellar bulge does not.  We also classify SDSS J1356+1026 as a dual/offset AGN because of the difficulty in identifying a secondary AGN amongst the soft X-rays associated with the outflow from the primary AGN.

Further observations are needed to distinguish whether each of these galaxies is in fact a dual AGN system or an offset AGN system. If deeper X-ray observations reveal an AGN in the second stellar bulge of one of these systems, then it is a dual AGN system.  On the other hand, if there is no AGN present the second stellar bulge of one of these systems, it is an offset AGN system.    In the offset AGN case, the \oiiiw emission observed in the second stellar bulge can be explained if, e.g., the gas there is photoionized by the AGN in the other stellar bulge. 

Table~\ref{tbl-8} provides a summary of the properties of each dual/offset AGN.  The line-of-sight velocity separation between the two AGNs' \oiiiw emission components for SDSS J0841+0101 was measured in \cite{GR11.1}, the velocity separation for SDSS J0952+2552 is from \cite{CO12.1}, and the velocity separations for SDSS J1239+5314, SDSS J1322+2631, and SDSS J1356+1026 were measured in Section~\ref{sdss}.  The quoted spatial separations are those we measured between the stellar bulges (Section~\ref{hst}).

We discuss each dual/offset AGN in detail below.

\subsubsection{SDSS J0841+0101: Dual/Offset AGN}
\label{0841}

SDSS J0841+0101 was recognized as a Type 2 dual AGN candidate by \cite{GR11.1}.  The system has two \oiiiw emission sources with AGN-like line ratios, nearly identical luminosities of $L_{\oiiiwn}=10^{42}$ erg s$^{-1}$, and a separation of $3\farcs60$ (7.28 kpc).  The two \oiiiw emission components spatially coincide with the locations of two stellar bulges, J0841+0101NE and J0841+0101SW, to within 0.5 times the astrometric error.  The merger ratio is 1.4:1, indicating that this system is undergoing a major merger.  Our current {\it Chandra} observations show an X-ray AGN in the more massive stellar bulge, J0841+0101NE ($237.2^{+0.0}_{-14.2}$ counts at 0.3-8 keV, for a $>5\sigma$ detection), but an upper limit of only 2.4 counts at 0.3-8 keV were detected in the less massive stellar bulge, J0841+0101SW.  

In the {\it HST} F814W and F438W images, we also detect spatially extended emission between the two stellar bulges.  This emission, which is coincident with extended soft X-rays in the {\it Chandra} data, could originate from photoionized or shocked gas in the merger. 

\subsubsection{SDSS J0952+2552: Dual/Offset AGN}
\label{0952}

SDSS J0952+2552 was selected as a Type 1 double-peaked AGN in \cite{SM10.1}, and follow-up observations revealed two \oiiiw emission components separated by $1\farcs00$ (4.85 kpc; \citealt{MC11.1,CO12.1,FU12.1}) that spatially coincide with two stellar bulges seen in high resolution near-infrared imaging \citep{RO11.1,FU12.1}.  We also observe these stellar bulges in our {\it HST}/F160W data.  Based on extracted spectra for each individual source, \cite{MC11.1} concluded that J0952+2552NE is a Type 1 AGN while J0952+2552SW is a Type 2 AGN.

We make $0.3-8$ keV X-ray detections of J0952+2552NE at $>5\sigma$ and J0952+2552SW at $0.5\sigma$, and we find that the X-ray source positions agree with those of the stellar bulges (observed with {\it HST}/F160W) to within $1.2$ times the astrometric error.  The 2.0:1 merger ratio indicates an ongoing major merger. 

\subsubsection{SDSS J1239+5314: Dual/Offset AGN}
\label{1239}

This system was identified as a Type 1 double-peaked AGN in \cite{SM10.1}, and follow-up long-slit observations revealed that this system is much more complex than its SDSS spectrum alone suggests.  The long-slit spectra show four emission components that are distinct both spatially and in line-of-sight velocity \citep{CO12.1}.  The double-peaked profile of the SDSS spectrum corresponds to two \oiiiw emission components separated by only $0\farcs39$ (1.3 kpc), and our {\it HST} F160W observations show a stellar bulge coincident with the blueshifted emission component (J1239+5314NE) but do not show a stellar bulge coincident with the redshifted emission component.  Instead, there is a third emission component (which has a redshift between the blueshifted and redshifted components of the main double peak) at a larger distance ($1\farcs20$, or 3.98 kpc; also reported in \citealt{FU12.1}) that does coincide with a stellar bulge (J1239+5314SW), seen both in $K_p$ imaging \citep{FU12.1} and in our {\it HST}/F160W imaging (Figure~\ref{fig:images}).

This third emission component would be contained within the $3^{\prime\prime}$ diameter SDSS fiber, so that the resultant SDSS spectrum contains contributions from three spatially-distinct emission components.  What makes this system interesting is that it was selected as a dual AGN candidate because of its double-peaked \oiiiw emission lines, but the candidate dual AGNs we find are not responsible for the main double peaks.  Instead, we find that the candidate dual AGNs produce a more subtle double-peaked profile that was overlooked in the initial target selection (\citealt{SM10.1}; see also Figure 8 of \citealt{CO12.1} for spectra of the two emission components, labeled 'A' and 'B', that correspond to the candidate dual AGNs, while the main double-peaked profile is produced by 'A' and 'C').   

J1239+5314 has a combined SDSS spectrum that is Type 1, but it is unclear whether one or both of the sources (J1239+5314NE and J1239+5314SW) are Type 1 AGNs.  J1239+5314NE lies above the rms scatter of the mean value of $\log (L_{X, 2-10 \mathrm{keV}} / L_{\oiiiwn})$ for Type 1 single AGNs, and much further above the mean value for Type 2 single AGNs (\citealt{HE05.1}; Figure~\ref{fig:Lx_Loiii}).  J1239+5314SW lies within the rms scatter of both the mean value of $\log (L_{X, 2-10 \mathrm{keV}} / L_{\oiiiwn})$ for Type 1 single AGNs and that for Type 2 single AGNs.   As a result, we conclude that J1239+5314NE is likely a Type 1 AGN while J1239+5314SW is either a Type 1 or Type 2 AGN.

We detect two $0.3-8$ keV X-ray sources, at $>5\sigma$ and $1.0\sigma$ significances, that have a similar spatial separation and orientation on the sky as the two \oiiiw emission components, to within $3\sigma$.  Since the X-ray spectrum is not soft and each source has a high soft X-ray to \oiiiw luminosity ratio (Table~\ref{tbl-3}; Table~\ref{tbl-4}), we conclude that photoionization is not likely producing these sources.

While one X-ray source spatially coincides ($0.16\sigma$ difference in spatial position) with the less massive stellar bulge, J1239+5314NE, the other X-ray source is located $3.57\sigma$ away from the more massive stellar bulge, J1239+5314SW.  Because we do not detect significant X-ray emission associated with the J1239+5314SW stellar bulge, it remains an AGN candidate.  The mass ratio of the stellar bulges is 1.1:1, indicating an ongoing major merger.

\subsubsection{SDSS J1322+2631: Dual/Offset AGN}
\label{1322}

SDSS J1322+2631 is one of the Type 2 double-peaked AGNs selected in \cite{LI10.1}, and the follow-up long-slit observations at a position angle of $79^\circ$ East of North showed two \oiiiw emission components separated by $2\farcs1$ \citep{SH11.1}.  Since the system was not reobserved at a second position angle, $2\farcs1$ is a lower limit on the true spatial separation between the emission components on the sky, and we measured this true separation ($2\farcs35$, or 5.94 kpc) via the two \ha emission components apparent in the {\it HST}/F814W observations of this system (Section~\ref{hst}).

This system is a minor merger with a mass ratio of 6.4:1 between J1322+2631SW and J1322+2631NE.  Two X-ray sources spatially coincide with the two stellar bulges to within 0.31 times the astrometric error.  We detect $29.8^{+0.0}_{-5.1}$ counts (for $>5\sigma$ significance detection) in the more massive stellar bulge, J1322+2631SW, but an upper limit of only 0.6 counts in the less massive stellar bulge, J1322+2631NE.  We also note that there is a faint, third stellar bulge that is located $2\farcs59$ southwest of J1322+2631SW (the mass ratio of J1322+2631SW to the third stellar bulge is 36.3:1), but we see no evidence for an AGN in this third stellar bulge.

\subsubsection{SDSS J1356+1026: Dual/Offset AGN}
\label{1356}

SDSS J1356+1026 is a luminous Type 2 quasar with double-peaked emission lines in its SDSS spectrum \citep{LI10.1}, and follow-up observations have shown that this system is incredibly complex and interesting.   Two \oiiiw emission knots seen in long-slit observations \citep{GR11.1} that correspond to two stellar nuclei seen in near-infrared ($K_s$-band) imaging led \cite{SH11.1} to conclude that the double peaks in J1356+1026 are likely produced by dual AGNs.  The two emission sources, J1356+1026NE and J1356+1026SW, are separated by $1\farcs31$ (2.89 kpc).  However, \cite{FU12.1} observed extended \oiiiw emission in integral-field spectroscopy, in addition to the double stellar nuclei in near-infrared ($K_p$) imaging, that caused them to argue that the double peaks are produced by an extended narrow-line region powered by a single AGN, and not dual AGNs.

Further clarification of the complex nature of SDSS J1356+1026 came from observations of the extended \oiiiw emission, which \cite{GR12.2} attributed to a spectacular $\sim20$ kpc scale outflow driven by radio-quiet quasar feedback originating from the quasar in J1356+1026NE.  Analysis of the {\it Chandra} data that is also presented here showed extended soft X-ray gas located South of J1356+1026SW, where an extended ionized gas outflow is also present \citep{GR14.1,SU14.1}.  The soft X-ray gas may be due to shocks within a quasar-driven wind or a by-product of photoionization by the luminous buried quasar \citep{GR14.1}.

We detect both J1356+1026NE and J1356+1026SW in $0.3-8$ keV X-rays, with significances of $>5\sigma$ and $4.4\sigma$, respectively.  The X-ray source positions are in agreement with the two stellar bulges observed in {\it HST}/F160W, to within $1.0$ times the astrometric errors, and the X-ray spectrum is well-fit by a spectral index $\Gamma=2.11$, which indicates a very soft spectrum.  This, when combined with the low soft X-ray to \oiiiw luminosity ratio of each source (Table~\ref{tbl-3}; Table~\ref{tbl-4}), indicates that photoionization may be contributing to the observed flux of soft X-rays.

Although we make a firm detection of J1356+1026NE in $0.3-8$ keV X-rays ($>5\sigma$ significance), our $4.4\sigma$ detection of J1356+1026SW is complicated by the soft X-rays in the region that are associated with the outflow.  For this reason, we await upcoming deeper observations with {\it Chandra} (PI: Greene) before issuing a verdict on whether an AGN is present in J1356+1026SW, which would make SDSS J1356+1026 a dual AGN system.

\begin{figure*}
\begin{center}
\includegraphics[width=7in]{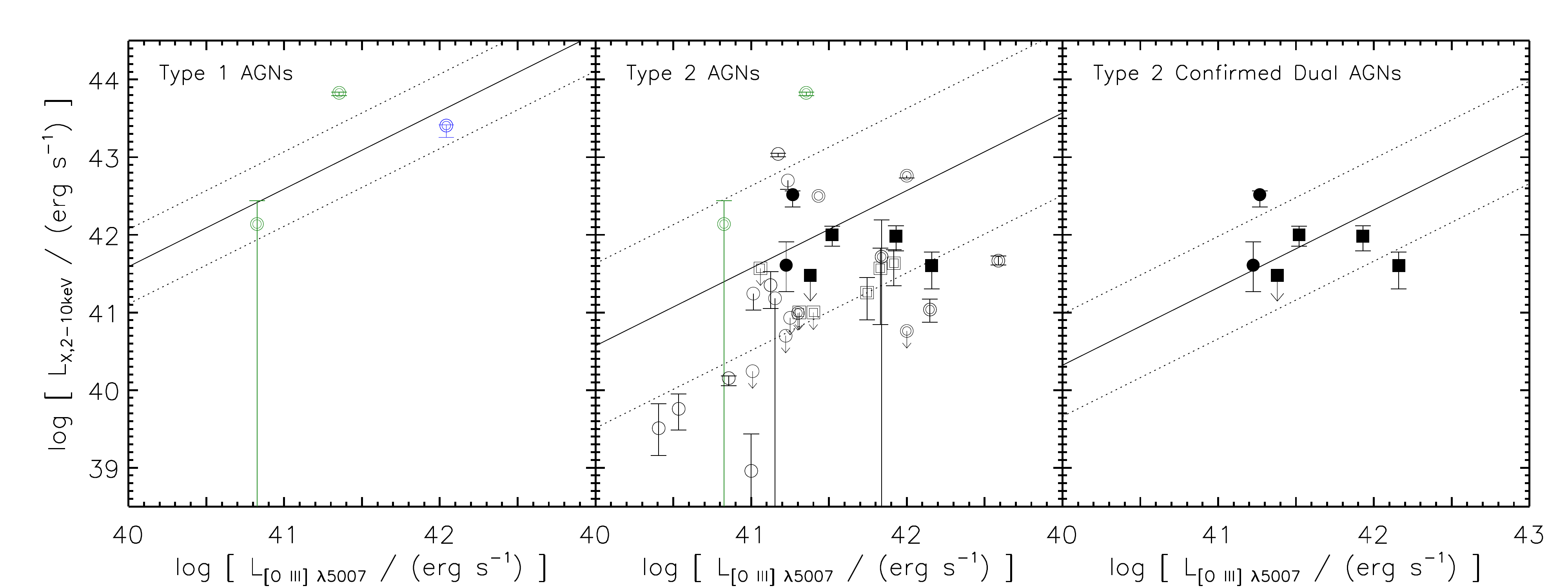}
\end{center}
\caption{Observed hard X-ray luminosity (2-10 kev) vs. observed \oiiiw luminosity for confirmed dual AGNs (filled symbols), dual/offset AGNs (double symbols), and the likely single AGNs (open symbols).  Left: a Type 1 AGN (blue circle) and two AGNs that could be Type 1 or Type 2 (green circles; Section~\ref{1239}).   The observed relation for optically-selected Type 1 single AGNs, log($L_{X,2-10\mathrm{keV}}/L_{\oiiiwn})=1.59 \pm 0.48$ \citep{HE05.1}, is shown as the solid line, and the dotted lines illustrate the rms scatter.  The Type 1 AGN that is part of a dual /offset AGN system falls below this relation but within the errors.  Middle: Type 2 AGNs (black circles) and Type 1 or 2 AGNs (green circles; Section~\ref{1239}) from this paper, and Type 2 AGNs from observations of SDSS double-peaked AGNs from \cite{CO11.2} and \cite{LI13.1} (black squares).   The observed relation for optically-selected Type 2 single AGNs, log($L_{X,2-10\mathrm{keV}}/L_{\oiiiwn})=0.57 \pm 1.06$ \citep{HE05.1}, is shown as the solid line, and the dotted lines illustrate the rms scatter.  The Type 2 AGNs in our sample fall systematically below this relation.  Right: confirmed dual AGNs that are Type 2 AGNs.  We find that these confirmed dual AGNs are fit by log($L_{X,2-10\mathrm{keV}}/L_{\oiiiwn})=0.32 \pm 0.66$, which is shown by the solid line.  The dotted lines illustrate the rms scatter.}
\label{fig:Lx_Loiii}
\end{figure*}

\subsection{Six Likely Single AGNs}
\label{single}

The remaining six galaxies (SDSS J0142$-$0050, SDSS J0752+2736, SDSS J0854+5026, SDSS J1006+4647, SDSS J1448+1825, and SDSS J1604+5009) are likely single AGNs where the double-peaked narrow emission lines are the result of outflows, jets, or disk rotation.  {\it HST}/IR observations are useful for revealing the stellar structures of these six galaxies.  Four of them were observed with {\it HST}/F160W for this program, one of them has archival {\it HST}/F105W imaging (J1604+5009; GO 12521, PI: Liu), and one of them has no published high-resolution imaging (J0752+2736).  We cannot be sure of the exact morphology of J0752+2736 (in SDSS imaging it appears almost edge-on, with no obvious substructure), but for the five galaxies with {\it HST} imaging we do not find double stellar bulges associated with the double \oiiiw emission components.

Two of the galaxies, SDSS J0752+2736 and SDSS J1006+4647, may have significant contributions from star formation to the observed X-ray luminosities (Section~\ref{wise}).  In SDSS J0752+2736, 47\% of the $0.5-8$ keV luminosity may be due to star formation, while in SDSS J1006+4647, 41\% of the $0.5-8$ keV luminosity may be attributed to star formation.  These two galaxies also have the reddest colors ($W1-W4 > 6.5$; Figure~\ref{fig:wise}) of the eight galaxies whose mid-infrared fluxes are dominated by star formation, which suggests that these two galaxies may host highly obscured AGNs.

We also examine the role of photoionization in these systems.  In SDSS J0142$-$0050, the X-ray spectrum is not soft and each source has a high soft X-ray to \oiiiw luminosity ratio (Table~\ref{tbl-3}; Table~\ref{tbl-4}), indicating that photoionization is not producing the observed X-ray and \oiiiw sources.  However, photoionized gas likely contributes to the observed fluxes of SDSS J0752+2736 and SDSS J1448+1825, given their soft X-ray spectra and low ratios of soft X-ray to \oiiiw luminosities.  In each of these two systems, the two observed X-ray and \oiiiw emission sources could be produced by an AGN and a photoionized region, or a buried AGN with a biconical photoionization region.  Given the evidence from the mid-infrared colors for an obscured AGN in SDSS J0752+2736, it most likely hosts a buried AGN with biconical photoionization regions producing the observed X-ray and \oiiiw sources.

\section{Results}

\subsection{Dual AGNs Have Systematically Lower Hard X-ray Luminosities, at Fixed \oiiiw Luminosity, Than Single AGNs}
\label{Lx-Loiii}

Because of their multiwavelength nature, AGNs can be selected in many different wavebands via many different approaches.  Perhaps the most common selection of AGNs is via their optical emission lines (e.g., \citealt{BA81.1,KE06.1}), but this selection is imperfect because the observed lines can suffer obscuration by dust and confusion with star formation.  A cleaner selection of AGNs is via their hard X-rays, which can penetrate the dust that obscures optical emission lines. Hence, a scaling relation between hard X-ray luminosity $L_{X, 2-10 \mathrm{keV}}$ and optical emission line luminosity (in this case, \oiiiw luminosity $L_{\oiiiwn}$) has many applications.  

While such relations have been well-studied for single AGNs, it is unclear whether dual AGNs are described by the same or different relations.  From a sample of 20 Type 1 AGNs, selected to have $z<0.2$ and \oiiiw fluxes greater than $2.5 \times 10^{-13}$ erg cm$^{-2}$ s$^{-1}$,  \cite{HE05.1} find $\log (L_{X, 2-10 \mathrm{keV}} / L_{\oiiiwn}) = 1.59 \pm 0.48$.  We note that the luminosities in this relation are {\it observed} luminosities that are not corrected for intrinsic absorption.  Our sample includes one Type 1 AGN that is a member of a dual/offset AGN system, J0952+2552NE, and it lies below this relation but still within the rms scatter (Figure~\ref{fig:Lx_Loiii}, left).  

For a sample of 29 Type 2 AGNs selected with the same redshift and \oiiiw flux criteria as for the Type 1 AGNs, \cite{HE05.1} find $\log (L_{X, 2-10 \mathrm{keV}} / L_{\oiiiwn}) = 0.57 \pm 1.06$.  As for the Type 1 AGN relation, these luminosities are observed luminosities, but the Type 2 relation has a smaller amplitude and larger scatter than the relation for Type 1 AGNs.  A sample of eight optically selected Type 2 quasars at $0.39 < z < 0.73$ also follow roughly the same relation as the Type 2 Seyferts \citep{PT06.1}.   For comparison, we find that Type 2 AGNs that are confirmed dual AGNs and dual/offset AGNs lie systematically below the relation for Type 2 single AGNs (Figure~\ref{fig:Lx_Loiii}, middle).  

From a sample of four optically selected Type 2 double-peaked AGNs from SDSS, \cite{LI13.1} find a hint that these objects fall systematically below the $L_{X, 2-10 \mathrm{keV}}$ -- $L_{\oiiiwn}$ relation for single Type 2 AGNs.  For their two confirmed dual AGNs and two candidate dual AGNs, \cite{LI13.1} find $\log (L_{X, 2-10 \mathrm{keV}} / L_{\oiiiwn}) = -0.2 \pm 0.40$.  Four of the eight AGNs in their sample had upper limits on their 2-10 keV luminosities, which they included as detections when measuring $\log (L_{X, 2-10 \mathrm{keV}} / L_{\oiiiwn})$.

Here, we measure $\log (L_{X, 2-10 \mathrm{keV}} / L_{\oiiiwn})$ for five firm detections of Type 2 AGNs in dual AGN systems.  We combine our detections of two Type 2 AGNs in a dual AGN system with the three firm detections of Type 2 AGNs in dual AGN systems in \cite{LI13.1}.  J1146+5110NE is part of the confirmed dual AGN system SDSS J1146+5110, but we do not include J1146+5110NE here since it has only an upper limit measurement of $L_{X, 2-10 \mathrm{keV}}$ \citep{LI13.1}.  We also do not include any of the dual/offset AGNs.  With the total sample of five Type 2 AGNs in dual AGN systems, which have double-peaked narrow AGN emission lines in their SDSS spectra at $0.1 < z < 0.2$, we find
\begin{equation}
\log \left( \frac{L_{X, 2-10 \mathrm{keV}}} {L_{\oiiiwn}} \right) = 0.32 \pm 0.66 \; ,
\end{equation} 
which is half what is measured for single Type 2 AGNs (Figure~\ref{fig:Lx_Loiii}, right).

This difference can be explained if the Type 2 dual AGNs are overluminous in \oiiiw and/or underluminous in hard X-ray luminosity, when compared to single Type 2 AGNs.  We find that the dual AGN system presented in this paper, SDSS J1126+2944, is overluminous in \oiiiw not only when compared to its hard X-ray luminosity, but also when compared to its mid-infrared (12 $\mu$m) luminosity $L_\mathrm{MIR}$.    SDSS J1126+2944 has the lowest ratio of mid-infrared to \oiiiw luminosities ($\log(L_\mathrm{MIR} / L_{[\mathrm{O \,III}]}$$)=1.2$) of our 12 targets, and it is also 1.4 dex below the mean luminosity ratio for a sample of optically-selected Type 2 AGNs in SDSS ($\log(L_\mathrm{MIR} / L_{[\mathrm{O \,III}]}$$)=2.6$; standard deviation 0.4; \citealt{LA10.1}).  The higher \oiiiw luminosity of the dual AGN system could be a consequence of excess gas that is available due to the ongoing galaxy merger.
 
On the other hand, the low values of $\log (L_{X, 2-10 \mathrm{keV}} / L_{\oiiiwn})$ for Type 2 dual AGNs could also be explained by galaxy mergers that produce systematically lower hard X-ray luminosities in dual AGNs, as is suggested by \cite{LI13.1}.   Galaxy mergers have been linked to both star formation and to AGN activity, as tidal torques induced by mergers funnel gas towards the galaxy centers (e.g., \citealt{HE84.1,SA88.2,BA92.2,HO08.3}).  The excess of gas at the galaxy center absorbs an AGN's emitted X-ray flux, which is produced very close in to the AGN (photons from the accretion disk are Compton scattered by hot electrons in the corona above the accretion disk), to a greater extent than it absorbs the \oiiiw flux, which is emitted by low-density gas on the much larger scales of the NLR (e.g., \citealt{PO89.1}).  Dual AGNs are in ongoing galaxy mergers, and the higher nuclear gas column in mergers could explain why dual AGNs have systematically lower hard X-ray to \oiiiw luminosity ratios than single AGNs.

\begin{figure}
\begin{center}
\includegraphics[width=3.5in]{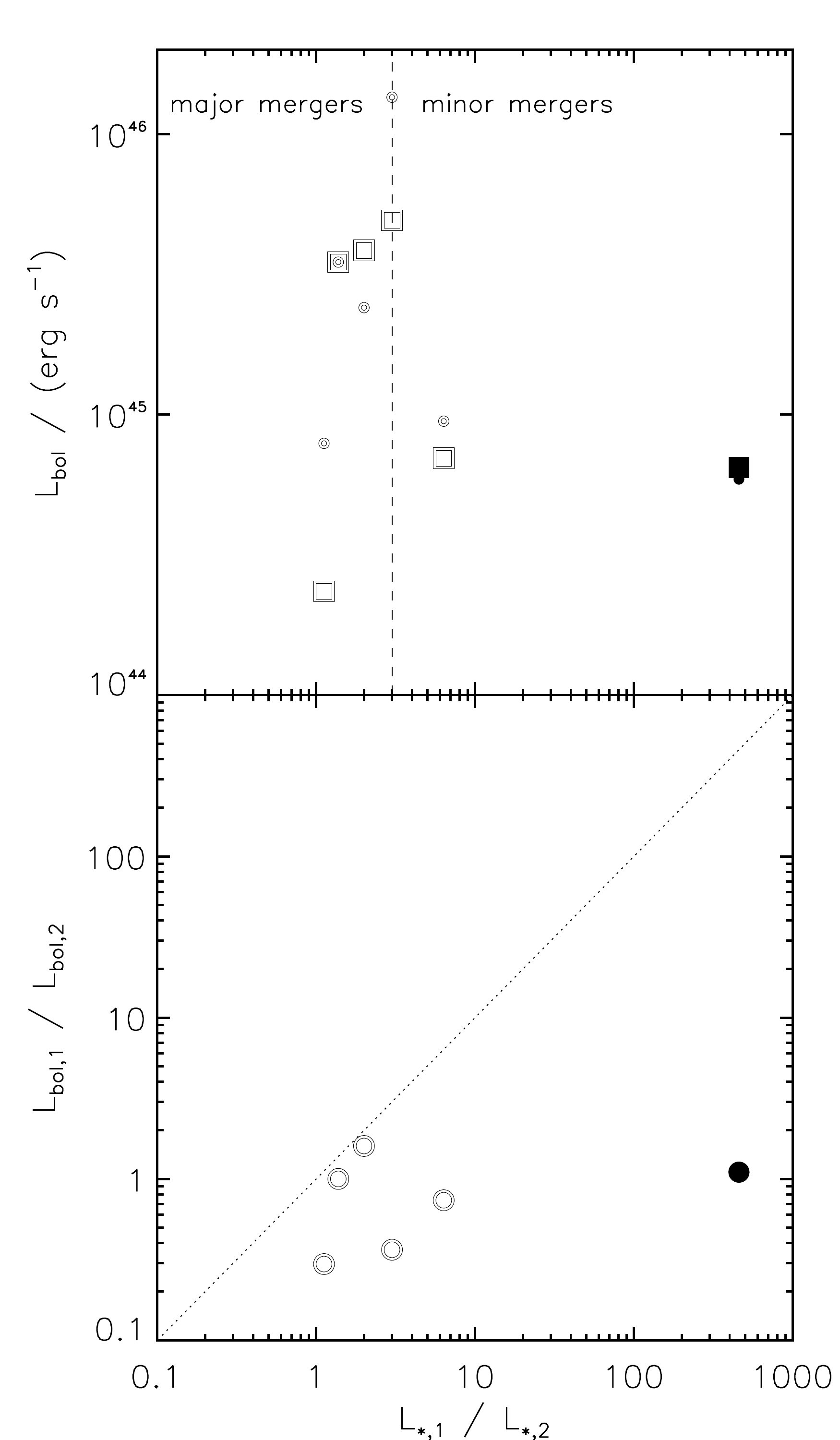}
\end{center}
\caption{Luminosities of the dual AGNs and dual/offset AGNs in comparison to the mass ratios of the galaxy mergers, measured as the ratio of the stellar bulge luminosities $L_{*,1}/L_{*,2}$.  Filled symbols illustrate the dual AGNs, while double symbols illustrate the dual/offset AGNs.  Top: Bolometric luminosity of each AGN, with the luminosity of the AGN in the more massive stellar bulge shown as a square and the luminosity of the AGN in the less massive stellar bulge shown as a circle.  In the dual AGN system the more massive stellar bulge hosts the more luminous AGN.  The dashed vertical line shows the $L_{*,1}/L_{*,2}=3$ dividing line between major and minor mergers.  Bottom: Ratios of AGN bolometric luminosities for the dual AGNs and dual/offset AGNs, illustrating the relation $L_{bol,1}/L_{bol,2}=(f_{Edd,1}/f_{Edd,2}) (L_{*,1}/L_{*,2})$.  The dotted line shows the case of $f_{Edd,1}/f_{Edd,2}=1$, and all of the galaxies lie below this line.  This demonstrates that the AGN in the less luminous stellar bulge has the higher accretion rate in all cases.}
\label{fig:merger_ratio}
\end{figure}

\cite{LI13.1} also suggest that the lower hard X-ray luminosities could be a selection effect not of dual AGNs intrinsically, but of dual AGNs that also produce double-peaked narrow emission lines.  The detection of dual AGNs via double-peaked narrow emission lines requires the two AGNs to be well-separated in line-of-sight velocity, and since dual AGNs orbit in the potential of the host galaxy, large line-of-sight velocity separations are more likely to be observed in edge-on galaxies.  Edge-on galaxies, in turn, have larger absorbing gas columns that could explain the lower observed hard X-ray luminosities.  However, the {\it HST} morphologies of the dual AGNs in \cite{LI13.1} do not suggest a preference for edge-on host galaxies.  Further, three other known dual AGNs with X-ray detections -- NGC 6240 \citep{KO03.1}, Mrk 463 \citep{BI08.1}, and Mrk 266 \citep{MA12.1} -- were not selected by double-peaked narrow emission lines, yet they still have lower hard X-ray luminosities than single AGNs \citep{LI13.1}.  

As a result, we interpret the lower hard X-ray to \oiiiw luminosity ratios of dual AGNs to most likely be a consequence of galaxy mergers.  Galaxy mergers drive excess gas onto the AGNs that increase their \oiiiw luminosities, and mergers also induce higher nuclear gas columns that suppress the hard X-ray luminosities of the AGNs.

\subsection{Four of the Six Dual and Offset AGNs Are Undergoing Major Mergers}
\label{major_mergers}

Galaxy merger simulations have shown that major mergers are capable of triggering dual AGNs with double-peaked emission line signatures such as those observed in our sample \citep{BL13.1}.  More generally, observations of AGN pairs have found that the AGN fraction is highest for major mergers \citep{WO07.2,EL11.1}. If we use the stellar bulge luminosity ratio as a proxy for the galaxy mass ratio, we find that four of our six dual and offset AGN systems are in agreement with this picture and are major mergers with mass ratios of 1:1 to 3:1.  

The dual AGN system SDSS J1126+2944 is in an extremely minor merger with a mass ratio of 460:1.  Our discovery of dual AGNs in a minor merger is in tension with theoretical expectations for dual AGNs.  A phenomenological model of dual AGNs argues that minor mergers are unlikely to produce dual AGNs with double-peaked emission lines that have a comparable luminosity in each peak, which are the types of line profiles typically selected in spectroscopic searches.  There has been one simulation of a minor merger, a 10:1 spiral-spiral merger \citep{VA12.1} in which the time period of dual AGN activity was shorter than for major mergers \citep{VA12.1}.  The enhanced tidal stripping and ram pressure on the less massive galaxy limited the time frame for efficient accretion onto its central SMBH.  Minor mergers need to be explored in more detail by simulations to place SDSS J1126+2944 in the broader context of how and how often such minor mergers produce dual AGNs.  We also note that SDSS J1126+2944 may have begun as a major merger, with extreme tidal stripping leading to the 460:1 mass ratio we presently observe.

Studies of galaxy pairs with separations $<80$ kpc have found that the AGN fraction is enhanced relative to field galaxies and that this enhancement is strongest for major mergers and for the more massive galaxy in the merger \citep{WO07.2,EL11.1}.  Other observations of galaxy pairs find that \oiiiw luminosities are higher in the less massive galaxy in a pair \citep{LI12.2}.  A phenomenological model of dual AGNs finds that $30\% - 40\%$ of the simulated dual AGNs have the brighter AGN in the smaller stellar bulge \citep{YU11.1}, while  galaxy merger simulations also find that which AGN is brighter changes with time as the merger progresses \citep{VA12.1}.  For the dual AGN system, we find that the more luminous AGN is hosted in the more luminous stellar bulge (Figure~\ref{fig:merger_ratio}). The dual/offset AGNs are a mixed bag, with three systems that have the higher \oiiiw luminosity in the less luminous stellar bulge, one system that has the higher \oiiiw luminosity in the more luminous stellar bulge, and one system that has equal \oiiiw luminosities in both stellar bulges.

We also examine the relative accretion rates of the two AGNs in the dual AGN system, where we denote the more luminous stellar bulge with ``1" and the less luminous stellar bulge with ``2".  If we combine the Eddington ratio $f_{Edd} \equiv L_{bol}/L_{Edd}$ and the Eddington luminosity $L_{Edd} = 4 \pi c G M_{BH} m_p / \sigma_T$ (where $m_p$ is the mass of a proton and $\sigma_T$ is the Thomson cross section for scattering), then we find that $L_{bol,1}/L_{bol,2}=(f_{Edd,1}/f_{Edd,2}) (M_{BH,1}/M_{BH,2})$.  If we assume that the black hole mass traces the host stellar bulge luminosity \citep{MA03.5,MC04.1,GR07.1}, such that $M_{BH,1}/M_{BH,2}=L_{*,1}/L_{*,2}$, then
\begin{equation}
\frac{L_{bol,1}}{L_{bol,2}} = \frac{f_{Edd,1}}{f_{Edd,2}} \frac{L_{*,1}}{L_{*,2}} \; .
\end{equation}

With this approach, we find that the dual AGN system has $f_{Edd,1}/f_{Edd,2} = 0.002$ (Table~\ref{tbl-8}, Figure~\ref{fig:merger_ratio}) and that the five dual/offset AGNs also have $f_{Edd,1}/f_{Edd,2} < 1$.  Our finding that every dual AGN and dual/offset AGN system has $f_{Edd,1}/f_{Edd,2} < 1$ means that the AGN in the less luminous stellar bulge is accreting at a higher Eddington ratio than the AGN in the more luminous stellar bulge.  

Our result is in agreement with hydrodynamical simulations of galaxy mergers, which find that the Eddington rate is higher for the AGN in the less massive of the two merging galaxies \citep{CA14.1}.  Other merger simulations have also produced situations where the less massive black hole accretes at a higher Eddington fraction until the less massive galaxy's gas is lost to ram pressure stripping \citep{VA12.1}.  The higher Eddington rates in the less massive galaxies can be understood if the less massive galaxies have more gas available to them (e.g., higher gas fractions) or if the gas accretion is more efficient in less massive galaxies (e.g., they experience stronger gravitational instabilities during mergers).  

\subsection{A Hint That Major Mergers Trigger Higher Luminosity AGNs}

While galaxy merger simulations have predicted that higher luminosity AGNs are more likely to be triggered in major galaxy mergers than lower luminosity AGNs (e.g., \citealt{HO09.1}), the observational results have not converged.  One study, which identifies mergers via visual morphology classifications, finds that the fraction of AGNs in mergers increases from $\sim20\%$ to $\sim60\%$ from AGN bolometric luminosities of $\sim10^{44}$ erg s$^{-1}$ to $\sim10^{46}$ erg s$^{-1}$ \citep{TR12.1}.  Two other studies, which quantify mergers visually \citep{KO12.2} or by the asymmetry of the galaxies \citep{VI14.1}, find no correlation between major mergers and AGN luminosity.  The discrepancy in these results might be attributed to the known difficulties in making complete and accurate identifications of mergers with these observational approaches (e.g., \citealt{CO03.2,LO11.1}).

Dual AGNs and offset AGNs present a solution to this observational problem, since they are by definition AGNs in galaxy mergers.  Here, we consider only the four dual/offset AGNs that are in major mergers (Section~\ref{major_mergers}), and we compare them to the six likely single AGNs (Section~\ref{single}).  The existing host galaxy stellar mass measurements based on fits to the SDSS photometry \citep{KA03.2,SA07.2} show that the mean stellar mass of the major mergers ($\log[M_* (M_\odot)]=11.1$) is similar to the mean stellar mass of the host galaxies of the single AGNs ($\log[M_* (M_\odot)]=10.9$).  We convert the AGNs' \oiiiw luminosities to bolometric luminosities \citep{HE04.3}, and we find that the four dual/offset AGNs have a mean AGN bolometric luminosity of $4.1 \times 10^{45}$ erg s$^{-1}$ (standard deviation $4.1 \times 10^{45}$ erg s$^{-1}$) while the six single AGNs have a mean AGN bolometric luminosity of $4.0 \times 10^{44}$ erg s$^{-1}$ (standard deviation $1.8 \times 10^{44}$ erg s$^{-1}$).  The dual/offset AGNs in major mergers are a factor of 10 times more luminous, on average, than the single AGNs.  

If we assume that the six single AGNs are not in mergers, then this suggests that AGNs in major mergers (the dual/offset AGNs in major mergers) have higher luminosities, in agreement with the trend seen in simulations.  However, so far this is only a hint, as a larger sample is needed to confirm the trend.  

\subsection{Dual AGN Occurrence May Depend on the Separation between Two Merging Galaxies}

Since galaxy mergers are efficient at driving central gas inflows (e.g., \citealt{BA91.1,SP05.2,HO09.1}), there is an expectation that the AGN fraction will increase as the merger progresses and the supermassive black hole pairs in a merger reach smaller separations.  In general, simulations have predicted that the AGN fraction peaks at black hole pair separations $<10$ kpc \citep{VA12.1,BL13.1}.

To date, observational studies of merger-driven AGN activity have focused either on $>10$ kpc separation AGN or quasar pairs (e.g., \citealt{HE06.1,HE10.1,EL11.1,LI11.3,SA14.1}) or single AGNs in galaxies with merger-like morphologies (e.g., \citealt{KO12.2,TR12.1,VI14.1}).  Our sample of dual and offset AGNs has separations $2 < \Delta x (\mathrm{kpc}) < 8$, which straddle these two regimes of pre-mergers and post-mergers.  Because of the small projected separations and line-of-sight velocities (Table~\ref{tbl-8}), the sample likely consists of merging systems and not flybys.  The $<8$ kpc separations between merging galaxies also enable us to measure the AGN fraction to smaller separations than ever before.  

Observations have shown that the AGN fraction increases as the separation between two merging galaxies decreases from 100 kpc to 10 kpc \citep{EL11.1,KO12.1}, and our data hint that the trend continues for separations $<10$ kpc.  Of the six merging galaxies in our sample, the smallest separation merger is the only confirmed dual AGN system ($\Delta x = 2.19$ kpc; SDSS J1126+2944) and the second-smallest separation merger is a likely dual AGN system ($\Delta x = 2.93$ kpc; SDSS J1356+1026; awaiting confirmation from upcoming {\it Chandra} observations).  This suggests that dual AGN activation could indeed be more common for merging galaxies with smaller separations, and this trend could be confirmed with a larger sample of dual AGNs.

\section{Conclusions}

We have presented {\it Chandra}/ACIS observations of 12 dual AGN candidates, including {\it HST}/WFC3 observations of 10 of them. These dual AGN candidates were selected by double-peaked narrow AGN emission lines in their SDSS spectra; two AGN-fueled \oiiiw emission components separated by $>0\farcs75$ in optical long-slit spectroscopy; and an estimated $2-10$ keV flux (estimated from the measured \oiiiw flux) for each AGN of $F_{X,2-10\mathrm{keV}} > 8 \times 10^{-15}$ erg cm$^{-2}$ s$^{-1}$.  The optical spectra alone are insufficient to identify whether these candidates are dual AGNs, but the criteria above were set up to enable {\it Chandra} detections of two X-ray sources corresponding to the two \oiiiw emission components, which would confirm dual AGNs.

The main results of our work are summarized below.

1.  We have discovered a dual AGN system in SDSS J1126+2944.  In this system, we detected two X-ray sources, at $\geq 2\sigma$ each; the projected separation and orientation on the sky of the two X-ray sources match those of the two observed \oiiiw emission components to within $3\sigma$; and the positions of the two X-ray sources match the positions of two stellar bulges observed in {\it HST}/F160W imaging, to within $3\sigma$ in astrometric accuracy.  The projected spatial separation of the dual AGNs is 2.19 kpc. If the secondary AGN is indeed associated with the faint secondary stellar bulge, then the merger ratio of the system is 460:1.  This suggests that the secondary source may be a dwarf galaxy hosting an intermediate mass black hole or the tidally-stripped remnant of a galaxy hosting a supermassive black hole.

2.  We find five systems that are either dual AGNs or offset AGNs: SDSS J0841+0101, SDSS J0952+2552, SDSS J1239+5314, SDSS J1322+2631, and SDSS J1356+1026.  SDSS J1356+1026 might be defined as dual AGNs, except that it has soft X-rays associated with an outflow that confound detection of the secondary AGN. The other four dual/offset AGN systems have two \oiiiw emission components with the same spatial separation and orientation on the sky as two stellar bulges (to within $3\sigma$), and X-ray sources detected at $>5\sigma$ in one stellar bulge but $<2\sigma$ in the other stellar bulge.  The projected spatial separations between the stellar bulges are $2 <\Delta x$ (kpc)$ < 8$.  Follow-up observations are needed to determine if each system hosts an AGN in the second stellar bulge (confirming dual AGNs) or not (confirming an offset AGN).

3.  We find six systems that are likely single AGNs: SDSS J0142$-$0050, SDSS J0752+2736, SDSS J0854+5026, SDSS J1006+4647, SDSS J1448+1825, and SDSS J1604+5009.  We do not detect double X-ray sources or double stellar bulges associated with the double \oiiiw emission components in these galaxies, and we conclude that the double-peaked narrow emission lines in these galaxies are the result of outflows, jets, or disk rotation.

4. Three of the galaxies show evidence for photoionized gas, via their soft X-ray spectra and low ratios of soft X-ray to \oiiiw luminosities.  One of these galaxies is either dual AGNs or an offset AGN (SDSS J1356+1026), and the other two are likely single AGNs (SDSS J0752+2736 and SDSS J1448+1825).  In these systems, one or both of the observed X-ray and \oiiiw emission sources could be produced by a photoionization region near a single AGN.

5. Type 2 dual AGNs have 2-10 keV hard X-ray luminosities that are two times lower, on average and for fixed \oiiiw luminosity, than for Type 2 single AGNs.  We interpret this as the likely consequence of dual AGNs residing in galaxy mergers, where the merger drives excess gas onto the AGNs that increase their \oiiiw luminosities.  The merger also acts to decrease the observed hard X-ray luminosities, since it produces high nuclear gas columns that absorb more X-ray flux (which is produced in close proximity to the AGN) than \oiiiw flux (which is produced on the larger scales of the NLR).

6.  Using the ratio of the stellar bulge luminosities as a proxy for the merger mass ratio, we find that four of the six dual and dual/offset AGN systems occur in major mergers with mass ratios of 3:1 to 1:1.  The dual AGN system SDSS J1126+2944 is in an extremely minor merger of mass ratio 460:1, which does not fit the common theoretical expectation that dual AGNs are produced in major mergers.  The dual AGN and dual/offset systems are mixed in terms of whether the more luminous AGN is hosted in the less or more luminous stellar bulge, which is consistent with predictions from simulations and phenomenological models.  In all of the dual AGN and dual/offset systems, the less luminous stellar bulge hosts the AGN with the higher Eddington ratio, which suggests that the less massive galaxies in mergers have higher gas fractions and/or more efficient gas accretion.  

7.  The dual/offset AGNs in major mergers have bolometric luminosities that are 10 times higher, on average, than those of the six single AGNs in our sample.   If we assume that the six single AGNs are not in major mergers, then this result hints that AGNs in major mergers are more luminous than those that are not.  However, a larger sample size is required to confirm this result.  

8.  The probability of dual AGN activation may increase as the separation between two merging galaxies decreases from $\sim10$ kpc to $\sim1$ kpc.  The galaxy mergers in our sample have separations that range from 8 kpc to 2 kpc, and the smallest separation system is the confirmed dual AGN system ($\Delta x = 2.19$ kpc; SDSS J1126+2944), while the next-smallest separation system is likely to be a dual AGN system ($\Delta x = 2.93$ kpc; SDSS J1356+1026).  Again, a larger sample of dual AGNs is needed for confirmation of this trend.

Our finding that dual AGNs have systematically lower hard X-ray luminosities than single AGNs underscores the need for deeper X-ray observations of dual AGN candidates.  When setting up this program, we relied on the \oiiiw to hard X-ray scaling relation for single AGNs to estimate the {\it Chandra} exposure times needed to make firm X-ray detections of the AGNs.  In reality the dual AGNs have significantly lower hard X-ray luminosities than predicted by the relation for single AGNs, which is why many galaxies in our sample have too few X-ray counts to confirm or refute them as dual AGNs.  The \oiiiw to hard X-ray scaling relation for dual AGNs, which we establish here, will be invaluable for setting up future X-ray programs to confirm dual AGNs.

\acknowledgements Support for this work was provided by NASA through Chandra Award Number GO2-13130A issued by the Chandra X-ray Observatory Center, which is operated by the Smithsonian Astrophysical Observatory for and on behalf of NASA under contract NAS8-03060.  Support for HST program number GO-12754 was provided by NASA through a grant from the Space Telescope Science Institute, which is operated by the Association of Universities for Research in Astronomy, Inc., under NASA contract NAS5-26555.

The scientific results reported in this article are based in part on observations made by the Chandra X-ray Observatory, and this research has made use of software provided by the Chandra X-ray Center in the application packages CIAO, ChIPS, and Sherpa.  The results reported here are also based on observations made with the NASA/ESA Hubble Space Telescope, obtained at the Space Telescope Science Institute, which is operated by the Association of Universities for Research in Astronomy, Inc., under NASA contract NAS 5-26555. These observations are associated with program number GO-12754.

{\it Facilities:} \facility{{\it CXO} (ACIS)}, \facility{{\it HST} (WFC3)}

\bibliographystyle{apj}

\end{document}